\documentclass[aps,prd,twocolumn,groupedaddress,showpacs,floatfix]{revtex4-1}
\usepackage[dvipdfm]{graphicx}
\usepackage{epsf}
\usepackage{amsmath,amssymb}

\newcommand{\beq}{\begin{equation}}
\newcommand{\eeq}{\end{equation}}
\newcommand{\beqn}{\begin{eqnarray}}
\newcommand{\eeqn}{\end{eqnarray}}

\begin{document}
\title{Gravitational waves from nonspinning black hole-neutron star
binaries: dependence on equations of state}
\author{Koutarou Kyutoku, Masaru Shibata}
\affiliation{Yukawa Institute for Theoretical Physics, Kyoto University,
Kyoto 606-8502, Japan}
\author{Keisuke Taniguchi}
\affiliation{Graduate School of Arts and Sciences, University of Tokyo,
Komaba, Meguro, Tokyo 153-8902, Japan}
\date{\today}

\begin{abstract}
 We report results of a numerical-relativity simulation for the merger
 of a black hole-neutron star binary with a variety of equations of
 state (EOSs) modeled by piecewise polytropes. We focus, in particular,
 on the dependence of the gravitational waveform at the merger stage on
 the EOSs. The initial conditions are computed in the moving-puncture
 framework, assuming that the black hole is nonspinning and the neutron
 star has an irrotational velocity field. For a small mass ratio of the
 binaries (e.g., $M_{\rm BH}/M_{\rm NS}=2$, where $M_{\rm BH}$ and
 $M_{\rm NS}$ are the masses of the black hole and neutron star,
 respectively), the neutron star is tidally disrupted before it is
 swallowed by the black hole irrespective of the EOS. Especially for
 less-compact neutron stars, the tidal disruption occurs at a more
 distant orbit. The tidal disruption is reflected in a cutoff frequency
 of the gravitational-wave spectrum, above which the spectrum amplitude
 exponentially decreases. A clear relation is found between the cutoff
 frequency of the gravitational-wave spectrum and the compactness of the
 neutron star. This relation also depends weakly on the stiffness of the
 EOS in the core region of the neutron star, suggesting that not only
 the compactness but also the EOS at high density is reflected in
 gravitational waveforms. The mass of the disk formed after the merger
 shows a similar correlation with the EOS, whereas the spin of the
 remnant black hole depends primarily on the mass ratio of the binary,
 and only weakly on the EOS. Properties of the remnant disks are also
 analyzed.
\end{abstract}
\pacs{04.25.D-, 04.30.-w, 04.40.Dg}

\maketitle

\section{introduction} \label{sec:intro}

Gravitational-wave observation is becoming one of the reliable tools for
observing our Universe. Current ground-based laser-interferometric
gravitational-wave detectors such as LIGO \cite{ligo2009} and VIRGO
\cite{virgo2008} have already achieved nontrivial scientific results;
e.g., upper limits on the amplitude of a stochastic gravitational-wave
background have been improved and we now know that gravitational waves
are not the main energy source of our Universe \cite{ligovirgo2009}.
Advanced gravitational-wave detectors such as advanced LIGO will be in
operation within the next several years and detect gravitational waves,
which can be used to explore the nature of strongly gravitating
phenomena. The most promising sources of gravitational waves are the
coalescing compact binaries composed of compact objects such as black
holes (BHs) and neutron stars (NSs). As illustrated in this paper, black
hole-neutron star (BH-NS) binaries are potential sources for exploring
the nature of the NSs and high-density nuclear matter.

According to a statistical study based on population synthesis
calculations, the detection rate of gravitational waves from BH-NS
binaries is estimated to be 0.5--50 events per year for advanced
gravitational-wave detectors \cite{kbkow,btkrb2007}. This suggests that
we will observe a variety of BH-NS binaries in the next decade. To
extract physical information of BH-NS binaries as well as the
information about the BH and NS themselves from gravitational waves,
theoretical templates of gravitational waves are necessary. This fact
motivates the numerical-relativity community to study in depth the
coalescence of BH-NS binaries, because numerical relativity is the
unique approach for accurately computing gravitational waves emitted
from the late inspiral and merger phases of such compact binaries.

Another astrophysical interest in BH-NS binaries is motivated by their
potential to be a progenitor of short-hard gamma-ray bursts (GRBs);
see~\cite{nakar,leeramirezruiz} and references therein for reviews.
According to a merger scenario of GRBs, a NS is tidally disrupted by a
low-mass BH before the orbit reaches an innermost stable circular orbit
(hereafter ISCO), resulting in a system consisting of a rotating BH and
a hot, massive accretion disk of mass $\gtrsim 0.01 M_\odot$ which could
become the central engine of a GRB. This BH-massive disk system could
subsequently radiate a large amount of energy $\gtrsim 10^{48}$ erg by
neutrino emission or by the so-called Blandford-Znajek process
\cite{blandfordznajek1977} in a short time scale $\lesssim 1$ s. Then,
neutrino-anti neutrino pair annihilation or electromagnetic Poynting
flux could drive a GRB. One of the key questions for the merger scenario
is whether the tidal disruption could lead to formation of the BH-disk
system. Numerical relativity is again the unique approach for answering
this question.

The equation of state (EOS) of NSs, which is still unknown, is the key
for determining gravitational waveforms emitted in the tidal-disruption
phase as well as for determining other properties of the BH-disk system
such as the mass and typical density of the disk. The EOS (specifically
its stiffness) determines the relation between mass and radius of a NS,
and hence, the relation between the tidal-disruption process and
associated gravitational waveforms. The reason is that the sensitivity
of a NS to the tidal force by the companion BH depends on its radius;
e.g., a NS of larger radius (with a stiffer EOS) will be disrupted at a
larger orbital separation (or a lower orbital frequency). If the tidal
disruption of a NS occurs at a larger distance, more material may be
spread around the companion BH, and consequently, a high-mass remnant
disk may be formed. Also, the gravitational-wave frequency at the tidal
disruption, which will be one of the characteristic frequencies, is
lower for NSs of larger radius. The EOS of nuclear matter beyond the
normal nuclear density is highly uncertain due to the lack of
constraints obtained from experiments. Gravitational-wave astronomy will
become a new and robust tool for determining or at least constraining
the EOS at such high densities through the observation of NSs
\cite{lindblom1992,vallisneri2000,rmsucf2009,fgp2010}. For this purpose,
we need theoretical templates of gravitational waves and it is necessary
to perform many simulations employing a wide variety of possible EOSs
for the NS matter.

In recent years, fully general relativistic studies of BH-NS binaries
have been performed both in calculations of quasiequilibrium states
\cite{grandclement2006,*grandclement2006e,tbfs2007,tbfs2008,fkpt2008,kst2009}
and in dynamical simulations of the mergers
\cite{shibatauryu2006,shibatauryu2007,shibatataniguchi2008,eflstb2008,dfkpst2008,elsb2009,skyt2009,dfkot2010}.
However, we have not yet understood the effect of the EOS on the merger
of BH-NS binaries in spite of its importance; in most of the previous
studies, NSs are modeled by simple and unrealistic $\Gamma=2$ polytropic
EOS (but see~\cite{dfkot2010}). One of the next goals in numerical
relativity is to clarify the effect of the EOS on the merger process of
BH-NS binaries and on resulting gravitational waveforms. For this
purpose, a systematic parametrization of possible EOSs by a small number
of parameters is quite useful.

In this paper, we report new results obtained by a simulation using a
wide variety of piecewise polytropic EOSs, which are shown to be useful
for parametrizing nuclear-theory-based EOSs in the cold
approximation~\cite{rlof2009,rmsucf2009,ozelpsaltis2009} \footnote{In
the original piecewise polytropic EOS, finite-temperature effects are
not taken into account. In our numerical simulation, a correction of
finite temperature induced by shock heating is taken into account; see
Sec. III A.}. We employ eight types of the piecewise polytropic EOSs,
ranging from highly stiff to soft ones \footnote{In this paper, the
stiffness is simply determined by the magnitude of pressure for the
nuclear-density region. We do not determine it by the adiabatic index.}.
We systematically choose the BH and NS masses in a realistic range of
interest. As a first step in this series of work, the BH is assumed to
be nonspinning. We track orbital evolutions of BH-NS binaries typically
for $\sim 5$ orbits so that the orbital eccentricity would not give a
serious error in gravitational waveforms at the onset of the merger
phase. We clarify the dependence of gravitational waveforms and merger
remnants on the EOS. In particular, we show that a gravitational-wave
spectrum contains valuable information on the EOS properties.

This paper is organized as follows. In Sec.~\ref{subsec:initial_method},
we summarize initial conditions employed in this paper.
Section~\ref{subsec:initial_pwp} describes the piecewise polytropic EOS
and the models adopted in this paper. Section \ref{sec:simulation}
describes the formulation and methods of numerical simulations. Section
\ref{sec:result} presents the numerical results and clarifies the effect
of the EOS on gravitational waveforms and merger remnants. Section
\ref{sec:summary} is devoted to a summary. Throughout this paper, we
adopt the geometrical units in which $G=c=1$, where $G$ and $c$ are the
gravitational constant and the speed of light, respectively. The
irreducible mass of the BH, gravitational mass of the NS in isolation,
circumferential radius of the NS in isolation, Arnowitt-Deser-Misner
(ADM) mass of the system, and sum of the BH and NS masses at infinite
separation are denoted by $M_{\rm BH}, M_{\rm NS}, R_{\rm NS}, M, m_0 =
M_{\rm BH} + M_{\rm NS}$, respectively. The mass ratio $Q$ is defined by
$Q \equiv M_{\rm BH} / M_{\rm NS}$ and the compactness of the NS
($\cal{C}$) is defined by ${\cal C} \equiv M_{\rm NS} / R_{\rm
NS}$. Latin and Greek indices denote spatial and spacetime components,
respectively.

\section{Initial condition} \label{sec:initial}

We employ BH-NS binaries in quasiequilibria for initial conditions of
numerical simulations as in \cite{shibatataniguchi2008,skyt2009}. The
quasiequilibrium state is computed in the moving-puncture framework
\cite{shibatauryu2006,shibatauryu2007,kst2009} with a piecewise
polytropic EOS \cite{rlof2009,rmsucf2009,ozelpsaltis2009}. Here, we
first summarize the formulation and numerical methods for computing the
quasiequilibrium state and then describe EOSs employed in this
paper. The details of the formulation and methods for computing initial
conditions are described in \cite{kst2009}, to which the reader may
refer. Computation of the quasiequilibrium state is performed using the
spectral-method library LORENE \cite{LORENE}.

\subsection{Formulation and methods} \label{subsec:initial_method}

We derive quasiequilibrium states of BH-NS binaries as solutions of the
initial value problem of general relativity \cite{cook}. When the
orbital separation of the binary is large enough, the time scale for the
gravitational-wave emission, $t_{\rm GW}$, is much longer than the
orbital period $P_{\rm orb}$, so that we can safely neglect the
radiation reaction of the gravitational-wave emission. In numerical
simulations, the orbital evolution has to be followed for $\agt 5$
orbits to derive a realistic waveform both for the late inspiral and
merger phases. For such a purpose, we have to choose the initial
separation of the binary which satisfies $t_{\rm GW} \gg P_{\rm orb}$,
and have to provide BH-NS binaries in a quasicircular orbit as the
initial condition, i.e., the binary is approximately in an equilibrium
state if it is observed in the comoving frame. To satisfy these two
conditions, we assume the presence of a helical Killing vector field
with the orbital angular velocity $\Omega$,
\begin{equation}
 \xi^\mu = (\partial_t)^\mu + \Omega (\partial_\varphi)^\mu ,
\end{equation}
and a hydrostatic equilibrium for the fluid configuration in the
comoving frame. In addition, we assume that the BH is nonspinning and
the NS has an irrotational velocity field. The irrotational velocity
field is believed to be an astrophysically (approximately) realistic
configuration~\cite{bildstencutler1992,kochanek1992}.

To compute the three-metric $\gamma_{ij}$, the extrinsic curvature
$K_{ij}$, the lapse function $\alpha$, and the shift vector $\beta^i$,
we employ a mixture of the conformal thin-sandwich approach and the
conformal transverse-traceless decomposition of Einstein's equation
\cite{cook}. We assume the conformal flatness of the three-metric
$\gamma_{ij} = \psi^4 \hat{\gamma}_{ij} = \psi^4 f_{ij}$, the
stationarity of the conformal three-metric $\partial_t \hat{\gamma}_{ij}
= 0$, and the maximal slicing condition for the trace part of the
extrinsic curvature $K = \gamma_{ij} K^{ij}$, i.e., $K = \partial_t K =
0$. Here, $f_{ij}$ denotes the flat spatial metric. Then, the basic
equations for the conformal factor $\psi$, the shift vector $\beta^i$,
and a weighted lapse function $\Phi \equiv \alpha \psi$ are derived from
the Hamiltonian constraint, the momentum constraint, and the maximal
slicing condition $\partial_t K = 0$ as
\begin{eqnarray}
 \Delta \psi &=& - 2 \pi \psi^5 \rho_H - \frac{1}{8} \psi^{-7}
 \hat{A}_{ij} \hat{A}^{ij} , \label{eq:ham} \\
 \Delta \beta^i &+& \frac{1}{3} \hat{\nabla}^i \hat{\nabla}_j \beta^j =
 16 \pi \Phi \psi^3 j^i + 2 \hat{A}^{ij} \hat{\nabla}_j (\Phi
 \psi^{-7}) , \\
 \Delta \Phi &=& 2 \pi \Phi \psi^4 (\rho_H + 2 S) + \frac{7}{8} \Phi
 \psi^{-8} \hat{A}_{ij} \hat{A}^{ij} \label{eq:max} ,
\end{eqnarray}
where $\hat{A}^{ij} \equiv \psi^{10} K^{ij}$, $\Delta \equiv f^{ij}
\hat{\nabla}_i \hat{\nabla}_j$, and $\hat{\nabla}_i$ denotes the
covariant derivative associated with $f_{ij}$. We assume an ideal fluid
for the matter field
\begin{equation}
T^{\mu \nu} = \rho h u^\mu u^\nu + P g^{\mu \nu}.
\end{equation}
where $\rho$ is the rest-mass density, $P$ is the pressure, $h \equiv 1
+ \varepsilon + P/\rho$ is the specific enthalpy, $\varepsilon$ is the
specific internal energy, and $u^\mu$ is the four-velocity of the fluid.
Then, the fluid quantities seen by the Eulerian observer are denoted by
\begin{eqnarray}
&& \rho_H = \rho h (\alpha u^t)^2 - P,\\
&& j^i = \rho h \alpha u^t u^{\mu} \gamma_{\mu}^{~i}, \\
&& S = \rho h [(\alpha u^t)^2 - 1] + 3P.
\end{eqnarray}
The EOS fully determines relations among the thermodynamical quantities
$\rho$, $\varepsilon$, $P$, and $h$. We describe the EOS adopted in this
work in Sec.~\ref{subsec:initial_pwp}.

In the moving-puncture framework, we set $\psi$ and $\Phi$ as
\begin{equation}
 \psi = 1 + \frac{M_{\rm P}}{2 r_{\rm BH}} + \phi \; , \; \Phi = 1 -
 \frac{M_\Phi}{r_{\rm BH}} + \eta ,
\end{equation}
where $M_{\rm P}$ and $M_\Phi$ are positive constants of mass dimension
and $r_{\rm BH} = |{x^i - x^i_{\rm P}}|$ is a coordinate distance from
the puncture located at $x^i_{\rm P}$. We numerically solve the
nonsingular parts $\phi$ and $\eta$ using Eqs.~(\ref{eq:ham}) and
(\ref{eq:max}) and adjusting the parameter $M_{\rm P}$ to achieve a
desired BH mass. The other parameter, $M_\Phi$, is determined by the
virial relation, i.e., the condition in which the ADM mass ($M_0$) and
the Komar mass agree, which holds for the stationary and
asymptotically flat spacetime \cite{beig1978,ashtekarashtekar1979},
\begin{equation}
 \oint_{r \to \infty} \partial_i \Phi dS^i = - \oint_{r \to \infty}
 \partial_i \psi dS^i = 2 \pi M_0 .
\end{equation}
We note that the lapse function, $\alpha$, obtained in this method is
always negative near the puncture. In the numerical simulation, we
modify the initial condition for $\alpha$ appropriately and ensure its
positivity.

For solving the momentum constraint, we decompose $\hat{A}_{ij}$ as
\begin{equation}
 \hat{A}_{ij} = \hat{\nabla}_i W_j + \hat{\nabla}_j W_i - \frac{2}{3}
 f_{ij} \hat{\nabla}_k W^k + K^{\rm P}_{ij} , \label{eq:ext_decomp}
\end{equation}
where $W_i$ is an auxiliary three-vector field, and $W^i = f^{ij}
W_j$. $K^{\rm P}_{ij}$ denotes a conformally weighted extrinsic
curvature associated with the linear momentum of the BH, written
by~\cite{brandtbrugmann1997}
\begin{equation}
 K^{\rm P}_{ij} = \frac{3}{2 r_{\rm BH}^2} [l_i P^{\rm BH}_j + l_j
 P^{\rm BH}_i - (f_{ij} - l_i l_j) l^k P^{\rm BH}_k] ,
\end{equation}
where $l^i = x^i_{\rm BH} / r_{\rm BH}$ is a unit radial vector, $l_i =
f_{ij} l^j$, and $P^{\rm BH}_i$ is the linear momentum of the BH, which
is determined by the condition in which the total linear momentum of the
system should vanish,
\begin{equation}
 P^{\rm BH}_i = - \int j_i \psi^6 d^3 x.
\end{equation}
$W_i$ obeys an elliptic equation
\begin{equation}
 \Delta W_i + \frac{1}{3} \hat{\nabla}_i \hat{\nabla}_j W^j = 8 \pi
 \psi^6 j_i ,
\end{equation}
which is derived by taking a derivative of Eq.~(\ref{eq:ext_decomp}) and
using the momentum constraint.

To summarize, we solve the elliptic equations for $\phi$, $\beta^i$,
$\eta$, and $W_i$ imposing outer boundary conditions derived from the
asymptotic flatness. In the present formalism, we do not have to impose
inner boundary conditions at the BH horizon unlike in the excision
method \cite{tbfs2007,tbfs2008}.

The basic equations for the hydrostatic equilibrium are derived from the
condition of irrotation, i.e., the zero relativistic vorticity
\begin{eqnarray}
 \omega_{\mu \nu} &=& \nabla_\mu (h u_\nu) - \nabla_\nu (h u_\mu)
 \nonumber \\
 &=& 0 ,
\end{eqnarray}
and from the helical symmetric relation for the specific momentum of the
fluid $\pounds_\xi (h u^\mu) = 0$. One result is the first integral of
the relativistic Euler equation,
\begin{equation}
 h \xi_\mu u^\mu = - C (= {\rm const}) .
\end{equation}
This equation determines $h$ (and subsequently $\rho$, $\varepsilon$,
and $P$ through an EOS) for an arbitrarily chosen constant $C$. The
irrotational flow condition implies the presence of a velocity potential
$\Psi$, which determines the four-velocity of the fluid by $h u_i = D_i
\Psi$, where $D_i$ is the covariant derivative associated with
$\gamma_{ij}$. The continuity equation $\nabla_\mu (\rho u^\mu) = 0$
then leads to an elliptic equation for the velocity potential $\Psi$.

A quasiequilibrium state is computed using an iteration method described
in detail in \cite{kst2009}. During the iteration, we fix the center of
mass of the binary with a 3PN-J condition described in
\cite{skyt2009,kst2009}; we determine the center of mass
phenomenologically so that the total angular momentum of the binary for
a given value of $\Omega m_0$ agrees with that derived by the third
post-Newtonian (3PN) approximation \cite{blanchet2002}. In this
condition, the initial orbital eccentricity is by a factor of $\sim 2$
smaller than that in other conditions tried to this time
\cite{shibatauryu2006,shibatauryu2007,shibatataniguchi2008}, and the
eccentricity at the onset of the merger becomes $\lesssim 1 \%$ for a
longterm simulation which tracks $\sim 5$ inspiral orbits.

\subsection{Piecewise polytropic equation of state}
\label{subsec:initial_pwp}

The temperature of NSs, except for newly born ones, are believed to be
much lower than the Fermi energy of the constituent particles
\cite{lattimerprakash}. This implies that we can safely neglect the
thermal effects and employ a cold EOS, for which the pressure, $P$, the
specific internal energy, $\varepsilon$, and other thermodynamical
quantities are written as a function of the rest-mass density $\rho$.
One of the simplest cold EOSs is a polytropic EOS,
\begin{equation}
 P = \kappa \rho^{1+1/n_{\rm p}} ,
\end{equation}
where $\kappa$ is the polytropic constant and $n_{\rm p}~(\geq 0)$ the
polytropic index: In the following, we often refer to the adiabatic
index defined by $\Gamma=1+1/n_{\rm p}$. The first law of the
thermodynamics,
\begin{equation}
 d \varepsilon = - P d \left( \frac{1}{\rho} \right) ,\label{first}
\end{equation}
determines the specific internal energy as $\varepsilon = P / [( \Gamma
-1 ) \rho]$ where we assume $\varepsilon=0$ at $\rho=0$. Then, the
specific enthalpy $h$ becomes
\begin{equation}
 h = 1 + \kappa \frac{\Gamma}{\Gamma-1} \rho^{\Gamma-1} .
\end{equation}

A piecewise polytropic EOS is a phenomenologically parametrized EOS,
which reproduces cold nuclear-theory-based EOSs at high density only
with a small number of polytropic constants and indices
\cite{rlof2009,rmsucf2009,ozelpsaltis2009}, i.e.,
\begin{equation}
 P (\rho) = \kappa_i \rho^{\Gamma_i}~~~{\rm for}~~\rho_{i-1} \leq \rho <
  \rho_i \; (1 \leq i \leq n),
 \label{eq:pwp}
\end{equation}
where $n$ is the number of the pieces used to parametrize an EOS and
$\rho_i$ denote boundary densities for which we provide an appropriate
value (see the method below). Here, $\rho_0=0$ and $\rho_n \rightarrow
\infty$. $\kappa_i$ is the polytropic constant and $\Gamma_i$ the
adiabatic index for each piece. We note that we could in principle match
to the known more realistic EOS at lower density. However, using a
single polytrope for the low-density EOS is justified to the extent that
the radius and deformability of the NS as well as resulting
gravitational waveforms in the merger phase are insensitive to the
low-density EOS.

At each boundary density, $\rho=\rho_i~(i=1,...,n-1)$, the pressure is
required to be continuous, i.e.,
$\kappa_i\rho_i^{\Gamma_i}=\kappa_{i+1}\rho_i^{\Gamma_{i+1}}$. Thus, if
we give $\kappa_1$, $\Gamma_i$, and $\rho_i~(i=1,...,n)$, the EOS is
totally determined. For the zero-temperature EOS, the first law of the
thermodynamics (\ref{first}) holds, and thus, $\varepsilon$ and $h$ are
also determined except for the choice of the integration constants,
which are fixed by the continuity condition of $\varepsilon$ (hence
equivalently $h$) at each $\rho_i$.

Recently, several authors have shown that the piecewise polytropic EOS
composed of one piece in the crust region and three pieces in the core
region approximately reproduces most of nuclear-theory-based EOSs at
high density \cite{rlof2009}. Here, three pieces in the core region are
required to reproduce a high-mass NS for which inner and outer cores
could have different stiffness due to the variation of properties of
high-density nuclear matter. In the present work, we pick up NSs of
relatively low mass 1.2--1.35$M_{\odot}$, taking into account that the
masses of the NSs in the observed binary are fairly small
\cite{stairs2004}. The highest density of such NSs is not high enough in
general that the EOS for the high-density part plays a critical role
(note that if the EOS is very soft, the EOS for the high-density region
is important, but we do not pursue this possibility in this paper). An
additional fact to be noted is that NSs in BH-NS binaries never achieve
the state of density higher than the initial value; their density should
decrease due to the tidal field of the companion BH during the
evolution. For these reasons, we employ a simple version of piecewise
polytropic EOS in this paper, in which only one piece is assigned for
the core region and one piece for the crust region as
in~\cite{rmsucf2009}. Following~\cite{rmsucf2009}, we employ the
parameters of the crust EOS for all the models as follows:
\begin{eqnarray}
&&\Gamma_1 = 1.35692395,\\
&&\kappa_1 / c^2 = 3.99873692 \times 10^{-8} {\rm g}^{1 - \Gamma_1} {\rm
 cm}^{3\Gamma_1 - 3} .
\end{eqnarray}
On the other hand, we vary the value of $\Gamma_2$ (the adiabatic index
for the core EOS). Authors in~\cite{rmsucf2009} propose that instead of
giving the density $\rho_1$, the pressure $p$ at the fiducial density
$\rho_{\rm fidu} = 10^{14.7}$ ${\rm g/cm}^3$ in the core region should
be provided because this parameter $p$ is closely correlated with the NS
radius and deformability \cite{lattimerprakash2001}. Thus, we have the
following relations:
\begin{eqnarray}
&& p = \kappa_2 \rho_{\rm fidu}^{\Gamma_2} , \\
&& \kappa_1 \rho_1^{\Gamma_1} = \kappa_2 \rho_1^{\Gamma_2} ( = P
 (\rho_1) ) .
\end{eqnarray}
These determine the values of $\kappa_2$ and $\rho_1$.

Table \ref{table:EOS} lists the parameters of the EOSs employed in this
paper, and several key quantities for each EOS. ``2H,'' ``H,'' ``HB,''
and ``B'' denote very stiff, stiff, moderately stiff, and soft EOSs,
respectively, for which $\Gamma_2=3.0$ universally, but the values of
$p$ are varied~\cite{rmsucf2009}. For ``HB,'' ``HBs,'' and ``HBss'' or
``B,'' ``Bs,'' and ``Bss'', we assign the same value of $p$ but
different values of $\Gamma_2$. The subscript ``s'' denotes that the
value of $\Gamma_2$ is smaller. For ``s'' and ``ss,'' $\Gamma_2=2.7$ and
2.4, respectively.

\begin{figure}[tbp]
 \includegraphics[width=90mm,clip]{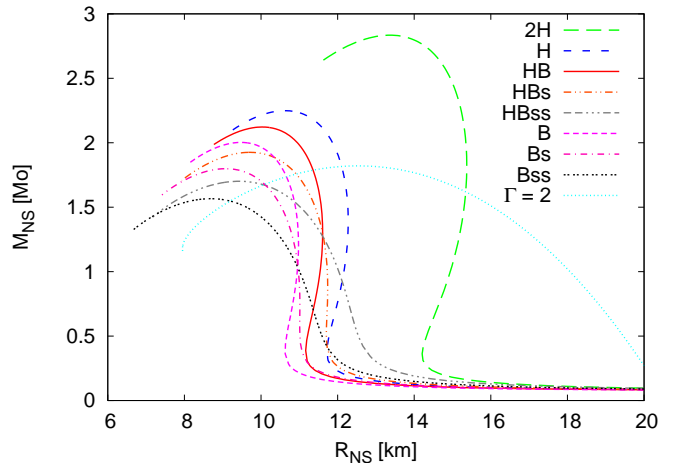} \caption{The relation
 between the mass and circumferential radius of the spherical NSs for
 piecewise polytropic EOSs adopted in this paper. For comparison, we
 also plot the curve for $\Gamma = 2$ polytropic EOS with $\kappa / c^2
 = 2 \times 10^{-16} {\rm g}^{-1} {\rm cm}^3$ (dotted curve).}
 \label{fig:MR}
\end{figure}

We calculate all the physical quantities for the spherical NS in
equilibrium both by solving the Tolman-Oppenheimer-Volkoff equation
directly and using the code to calculate initial conditions in the
isotropic gauge by LORENE, and check that numerical values agree with
each other within 0.03\%.  Figure~\ref{fig:MR} plots the relation
between the mass $M_{\rm NS}$ and circumferential radius $R_{\rm NS}$
for the spherical NSs with the adopted piecewise polytropic EOSs. For
comparison, we also plot the relation for $\Gamma = 2$ polytropic EOS
with $\kappa / c^2 = 2 \times 10^{-16} {\rm g}^{-1} {\rm cm}^3$ . Note
that in the polytropic EOS with a fixed adiabatic index, only the shape
of this relation has an invariant meaning and there is a freedom of the
absolute scaling, since all the dimensional quantities can be rescaled
through the polytropic length scale $R_{\rm poly} \equiv \kappa^{1/(2
\Gamma - 2)}$.

Figure~\ref{fig:MR} shows that for a given mass $\sim 1.35M_{\odot}$,
the radius depends strongly on the EOSs, whereas the radius for a given
piecewise polytropic EOS depends only weakly on the mass around the
canonical mass $\sim 1.35 M_\odot$. This weak dependence of the radius
on the mass is an often-seen feature for the nuclear-theory-based EOSs
\cite{lattimerprakash}. By contrast, the relation calculated with the
$\Gamma=2$ polytropic EOS does not show this feature. Figure
\ref{fig:MR} illustrates that the dependence of the radius $R_{\rm NS}$
on the mass $M_{\rm NS}$ becomes much stronger in this EOS than in the
piecewise polytropic EOSs. This illustrates that the $\Gamma=2$
polytropic EOS is not very realistic.

Comparison of the quantities among HB, HBs, and HBss EOS models in Table
\ref{table:EOS} reveals a complicated mass-radius relation: HB is not
always stiffer than HBss. Indeed, the radius with $M_{\rm NS}=1.2
M_\odot$ is largest for HBss and smallest for HB among three models,
whereas the radius with $M_{\rm NS}=1.35 M_\odot$ is largest for HB and
smallest for HBss. This complicated relation of the ``stiffness'' is due
to the choice for the combination ($\Gamma_2, p$)
(cf. Table~\ref{table:EOS}). For a density smaller than $\rho_{\rm
fidu}$, HBss EOS is stiffer than HB and HBs EOSs, whereas for a high
density $\rho > \rho_{\rm fidu}$, HB EOS is stiffer than the others. For
a given high-mass NS for which the central density is much larger than
$\rho_{\rm fidu}$, the radius with HB EOS should be larger than that
with other two EOSs. By contrast, for a given low-mass NS for which the
central density is not very high, the radius with HB EOS should be
smallest.

\begin{table*}[]
 \caption{Key ingredients of the adopted EOSs. $\Gamma_2$ is the
 adiabatic index in the core region and $p$ is the pressure at the
 fiducial density $\rho_{\rm fidu} = 10^{14.7}$ ${\rm g/cm}^3$, which
 determines the polytropic constant $\kappa_2$ of the core region and
 $\rho_{1}$: the critical rest-mass density separating the crust and
 core regions. $M_{\rm max}$ is the maximum mass of the NS for a given
 EOS. $R_{135}~(R_{12})$ and ${\cal C}_{135}~({\cal C}_{12})$ are the
 circumferential radius and the compactness of the NS with $M_{\rm NS} =
 1.35M_\odot~(1.2M_{\odot})$.}
 \begin{tabular}{ccc|cccccc} \hline
 Model & $\Gamma_2$ & $\log_{10} p$ (${\rm g/cm}^3$) & $\rho_{1}$
  ($10^{14}~{\rm g/cm}^3$) & $M_{\rm max} [M_\odot]$ & $R_{135}$ (km) &
  ${\cal C}_{135}$ & $R_{12}$ (km) & ${\cal C}_{12}$ \\
 \hline \hline
 2H & 3.0 & 13.95 & 0.7033 & 2.835 & 15.23 & 0.1309 &
 15.12 & 0.1172 \\ \hline
 H & 3.0 & 13.55 & 1.232 & 2.249 & 12.27 & 0.1624 & 12.25
 & 0.1447 \\ \hline
 HB & 3.0 & 13.45 & 1.417 & 2.122 & 11.61 & 0.1718 & 11.60
 & 0.1527 \\ \hline
 HBs & 2.7 & 13.45 & 1.069 & 1.926 & 11.57 & 0.1723 & 11.67
 & 0.1519 \\ \hline
 HBss & 2.4 & 13.45 & 0.6854 & 1.701 & 11.45 & 0.1741 & 11.74
 & 0.1509 \\ \hline
 B & 3.0 & 13.35 & 1.630 & 2.003 & 10.96 & 0.1819 & 10.98
 & 0.1614 \\ \hline
 Bs & 2.7 & 13.35 & 1.269 & 1.799 & 10.74 & 0.1856 & 10.88
 & 0.1629 \\ \hline
 Bss & 2.4 & 13.35 & 0.8547 & 1.566 & 10.27 & 0.1940 & 10.66
 & 0.1663 \\ \hline
 \end{tabular}
 \label{table:EOS}
\end{table*}

\subsection{Models} \label{sec:initial_model}

The previous works by three groups \cite{skyt2009,elsb2009,dfkpst2008}
have found that the NSs in BH-NS binaries with high mass ratio $Q
\gtrsim 4$ are barely subject to tidal disruption if the companion BH is
not spinning: At the merger, the BH swallows most of the NS matter at
one moment and the remnant disk mass is quite small or nearly equal to
zero. Namely, the NS behaves approximately as a point particle even at
the ISCO. Gravitational waves emitted in such a case have a similar
waveform to that from a BH-BH binary. Because the behavior of NSs with
high-mass BH companions does not show remarkable dependence on the EOS,
they are unsuitable for the purpose of this paper, i.e., to investigate
the effect of the EOS on gravitational waves and final outcomes. Thus,
we focus only on low mass-ratio binaries with $Q = 2$ and 3 in this
paper. Also, we choose relatively low-mass NSs, because two-piece EOSs
adopted in this paper may not be appropriate for modeling a high-mass NS
with high central density, due to the lack of model parameters in the
high-density region.

Table \ref{table:model} summarizes key quantities for the initial models
employed in the present numerical simulation. The labels for the models
denote the name of the EOS, the mass ratio, and the NS mass; e.g.,
2H-Q2M135 is modeled by 2H EOS, and its mass ratio and the NS mass are 2
and $1.35M_{\odot}$, respectively. The primary purpose of this paper is
to study the dependence of gravitational waveforms and the final outcome
on (i) the EOS of NSs, (ii) the mass ratio, and (iii) the NS mass. These
purposes are reflected in our choice of the initial models.

We prepare quasiequilibrium states basically with the same value of
$\Omega m_0$ for the same value of $Q$ irrespective of the EOS. The
value of $\Omega m_0$ is chosen to be small enough that the binaries
spend more than 5 inspiral orbits before the onset of the merger. For
$Q=2$ binaries, a smaller value of initial angular velocity is required
only for 2H EOS, because the NS with this EOS has a much larger radius
than with other EOSs and is sensitive to the BH tidal force even for a
larger orbital separation; to track $\agt 5$ inspiral orbits before the
tidal disruption, we have to choose the value of $\Omega m_0$ by $\sim
10\%$ as small as that for other EOSs. For the case of $Q=3$, we also
choose smaller values of $\Omega m_0$ for $M_{\rm NS} = 1.2 M_\odot$
cases.

\begin{table*}[]
 \caption{Key parameters and quantities for the initial conditions
 adopted in the numerical simulations. The adopted EOS, mass ratio
 ($Q$), NS mass in isolation ($M_{\rm NS}$), angular velocity ($\Omega$)
 in units of $c^3/Gm_0$, baryon rest mass ($M_*$), compactness of the NS
 in isolation (${\cal C}$), maximum rest-mass density ($\rho_{\rm
 max}$), ADM mass of the system ($M_0$), and total angular momentum of
 the system ($J_0$), respectively.}
 \begin{tabular}{c|cccc|ccccc} \hline
 Model & EOS & $Q$ & $M_{\rm NS} [M_\odot$] & $G \Omega m_0/c^3$ & $M_*
  [M_\odot]$~~ & ~~${\cal C}$~~ & ~~~$\rho_{\rm max} ({\rm g/cm}^3)$~~ &
  ~~$M_0 [M_\odot]$~~ & $J_0 [G M^2_\odot / c]$\\ \hline
 \hline
  2H-Q2M135 & 2H & 2 & 1.35 & 0.0250 & 1.455 & 0.1309 & 3.740 $\times
  10^{14}$ & 4.015 & 14.39 \\
  H-Q2M135 & H & 2 & 1.35 & 0.0280 & 1.484 & 0.1624 & 7.018 $\times
  10^{14}$ & 4.013 & 14.02 \\
  HB-Q2M135 & HB & 2 & 1.35 & 0.0280 & 1.493 & 0.1718 & 8.262 $\times
  10^{14}$ & 4.013 & 14.02 \\
  HBs-Q2M135 & HBs & 2 & 1.35 & 0.0280 & 1.489 & 0.1723 & 9.154 $\times
  10^{14}$ & 4.013 & 14.02 \\
  HBss-Q2M135 & HBss & 2 & 1.35 & 0.0280 & 1.485 & 0.1741 & 1.082
  $\times 10^{15}$ & 4.013 & 14.02 \\
  B-Q2M135 & B & 2 & 1.35 & 0.0280 & 1.503 & 0.1819 & 9.761 $\times
  10^{14}$ & 4.013 & 14.02 \\
  Bs-Q2M135 & Bs & 2 & 1.35 & 0.0280 & 1.501 & 0.1856 & 1.137 $\times
  10^{15}$ & 4.013 & 14.02 \\
  Bss-Q2M135 & Bss & 2 & 1.35 & 0.0280 & 1.501 & 0.1940 & 1.490 $\times
  10^{15}$ & 4.013 & 14.02 \\
  \hline
  2H-Q3M135 & 2H & 3 & 1.35 & 0.0280 & 1.455 & 0.1309 & 3.737 $\times
  10^{14}$ & 5.359 & 21.05 \\
  H-Q3M135 & H & 3 & 1.35 & 0.0300 & 1.484 & 0.1624 & 7.011 $\times
  10^{14}$ & 5.358 & 20.74 \\
  HB-Q3M135 & HB & 3 & 1.35 & 0.0300 & 1.493 & 0.1718 & 8.254 $\times
  10^{14}$ & 5.358 & 20.74 \\
  B-Q3M135 & B & 3 & 1.35 & 0.0300 & 1.503 & 0.1819 & 9.751 $\times
  10^{14}$ & 5.357 & 20.74 \\
  \hline
  2H-Q2M12 & 2H & 2 & 1.20 & 0.0220 & 1.282 & 0.1172 & 3.466 $\times
  10^{14}$ & 3.571 & 11.71 \\
  H-Q2M12 & H & 2 & 1.20 & 0.0280 & 1.303 & 0.1447 & 6.421 $\times
  10^{14}$ & 3.567 & 11.08 \\
  HB-Q2M12 & HB & 2 & 1.20 & 0.0280 & 1.310 & 0.1527 & 7.522 $\times
  10^{14}$ & 3.567 & 11.08 \\
  B-Q2M12 & B & 2 & 1.20 & 0.0280 & 1.317 & 0.1614 & 8.832 $\times
  10^{14}$ & 3.567 & 11.08 \\
  \hline
  HB-Q3M12 & HB & 3 & 1.20 & 0.0280 & 1.310 & 0.1527 & 7.517 $\times
  10^{14}$ & 4.763 & 1.663 \\
  B-Q3M12 & B & 3 & 1.20 & 0.0280 & 1.317 & 0.1614 & 8.826 $\times
  10^{14}$ & 4.763 & 1.663 \\
  \hline
 \end{tabular}
 \label{table:model}
\end{table*}

\section{Methods of simulations} \label{sec:simulation}

Numerical simulation is performed using an adaptive-mesh refinement
(AMR) code {\small SACRA} \cite{yst2008}. The formulation, the gauge
conditions, the numerical scheme, and the methods of diagnostics are
essentially the same as those described in~\cite{yst2008,skyt2009}
except for the EOS. Thus, we here only briefly review them. We also
describe the present setup of the computational domain for the AMR
algorithm and grid resolution in Sec.~\ref{subsec:simulation_grids}.

\subsection{Formulation and numerical methods} \label{subsec:simulation_method}

In {\small SACRA}, we solve Einstein's evolution equation in the BSSN
formalism~\cite{shibatanakamura1995,baumgarteshapiro1998} with the
moving-puncture method~\cite{brandtbrugmann1997,clmz2006,bcckm2006}. We
evolve a conformal factor $W \equiv \gamma^{-1/6}$, the conformal
three-metric $\tilde{\gamma}_{ij} =\gamma^{-1/3} \gamma_{ij}$, the trace
of the extrinsic curvature $K$, the conformal trace-free part of the
extrinsic curvature $\tilde{A}_{ij} =\gamma^{-1/3} (K_{ij} - K
\gamma_{ij} / 3)$, and an auxiliary variable $\tilde{\Gamma}^i \equiv -
\partial_j \tilde{\gamma}^{ij}$. The spatial derivatives in the
evolution equations are evaluated by a fourth-order centered finite
difference except for the advection terms which is evaluated by a
fourth-order noncentered finite difference. A fourth-order Runge-Kutta
method is employed for the time evolution.

Following \cite{bghhst2008}, we employ a moving-puncture gauge in the
form
\begin{eqnarray}
 ( \partial_t - \beta^j \partial_j ) \alpha & = & - 2 \alpha K , \\
 ( \partial_t - \beta^j \partial_j ) \beta^i & = & (3/4) B^i , \\
 ( \partial_t - \beta^j \partial_j ) B^i & = & ( \partial_t - \beta^j
 \partial_j ) \tilde{\Gamma}^i - \eta_s B^i ,
\end{eqnarray}
where $B^i$ is an auxiliary variable and $\eta_s$ is an arbitrary
constant. In this work, we typically set $\eta_s \approx M_\odot /
M_{\rm BH}$.

For the hydrodynamics, we evolve $\rho_* \equiv \rho \alpha u^t W^{-3}$,
$\hat{u}_i \equiv h u_i$, and $e_* \equiv h \alpha u^t - P / (\rho
\alpha u^t)$. To handle the advection terms, we adopt a high-resolution
central scheme by Kurganov and Tadmor \cite{kurganovtadmor2000} with a
third-order piecewise parabolic interpolation for the cell
reconstruction.

With regards to the EOS, we decompose the pressure and the specific
internal energy into cold and thermal parts as follows (e.g.,
\cite{stu2005})
\begin{equation}
 P = P_{\rm cold} + P_{\rm th} \; , \; \varepsilon = \varepsilon_{\rm
 cold} + \varepsilon_{\rm th} .
\end{equation}
Here, the thermal part is nonzero only in the presence of shock heating,
and thus, this part plays a role for the evolution only in the merger
phase. Once the primitive variables $\rho$ and $\varepsilon$ are
recovered from the conserved variables $\rho_*$, $\hat u_i$, and $e_*$,
we calculate zero-temperature parts $P_{\rm cold}$ and $\varepsilon_{\rm
cold}$ from $\rho$ using the piecewise polytropic EOS
(\ref{eq:pwp}). Then, the thermal part of the specific internal energy
is calculated by $\varepsilon_{\rm th} = \varepsilon -\varepsilon_{\rm
cold}$, and finally the thermal part of the pressure $P_{\rm th}$ is
determined. In this paper, we adopt a simple $\Gamma$-law, ideal-gas EOS
for the thermal part as (e.g., \cite{stu2005})
\begin{equation}
 P_{\rm th} = ( \Gamma_{\rm th} - 1 ) \rho \varepsilon_{\rm th} ,
\end{equation}
where $\Gamma_{\rm th}$ is an adiabatic index for the thermal part. We
choose $\Gamma_{\rm th}$ equal to the adiabatic index in the crust
region, $\Gamma_1$, for simplicity.

Because the vacuum is not allowed in any conservative hydrodynamic
scheme, an artificial atmosphere of small density is distributed outside
the NS in the same manner as done in our previous work
\cite{skyt2009}. The rest-mass density of the atmosphere is set to be
$\rho_{\rm atm}=10^{-9}\rho_{\rm max} \approx 10^6~{\rm g/cm^3}$ for the
inner computational domain. For the outer domain with $r \geq r_{\rm c}
\approx 20R_{\rm NS}$, a smaller density is assigned according to the
rule $\rho=\rho_{\rm atm} e^{1-r/r_{\rm c}}$. The total rest mass of the
atmosphere is always less than $10^{-5} M_\odot$, and hence, we can
safely neglect spurious effects by accretion of the atmosphere onto the
remnant accretion disk as long as the disk mass is much larger than
$10^{-5} M_\odot$.

\subsection{Diagnostics} \label{subsec:simulation_diagnostics}

Gravitational waves are extracted calculating the outgoing part of the
complex Weyl scalar $\Psi_4$, which we evaluate at a finite coordinate
radii $r=300$--$400 M_\odot$. Gravitational waveforms are obtained by
integrating $\Psi_4$ twice in time as
\begin{equation}
 h_+(t) - i h_\times(t) =-\int^t dt' \int^{t'} dt'' \Psi_4 (t'') ,
 \label{eq:h}
\end{equation}
and then by subtracting the quadratic function $a_2 t^2 + a_1 t + a_0$
from the obtained waveform using the least-square fitting for
determining the constants $a_0$, $a_1$, and $a_2$. The purpose of this
subtraction is to eliminate unphysical components in numerically
calculated Weyl scalar, $\Psi_4$ \cite{reisswigpollney2010}, as
described in \cite{skyt2009} (see also \footnote{In the previous work,
we subtract quadratic functions by the least-square fitting also from
$\Psi_4$ itself and $\int \Psi_4 dt$. We have found that we do not have
to perform this procedure.}). We also calculate the amount of radiated
energy $\Delta E$ and angular momentum $\Delta J$ by integrating the
emission rate calculated from the Weyl scalar $\Psi_4$ as
\begin{eqnarray}
 \frac{dE}{dt} &=& \frac{r^2}{16 \pi} \oint_S \left| \int \Psi_4 dt
					      \right|^2 dA , \\
 \frac{d J_z}{dt} &=& - \frac{r^2}{16 \pi} {\rm Re} \biggl[ \oint_S
  \left( \int \bar{\Psi}_4 dt \right) \nonumber \\
 && \hskip 2cm \times
  \left( \int \int \partial_\varphi \Psi_4 dt dt^\prime \right) dA
  \biggl] ,
\end{eqnarray}
where $S$ denotes a coordinate sphere of $r = {\rm const}$, $dA = r^2 d(
\cos \theta ) d\varphi$ is the surface element of $S$, and
$\bar{\Psi}_4$ is the complex conjugate of $\Psi_4$. We decompose
$\Psi_4$ into $s = -2$ spin-weighted spherical harmonics of $2 \le l \le
4$. Among them, $(l,|m|) = (2,2)$ modes are always dominant but higher
$l$ modes such as $(l,|m|)=(3,3)$, $(4,4)$ and $(2,1)$ modes contribute
to the totally radiated energy and angular momentum by larger than 1\%.

We compare numerical waveforms with those derived by the Taylor-T4
formula in the post-Newtonian approximation \cite{bbkmpsct2007} for two
point masses in quasicircular orbits. Assuming that both the BH and NS
have no spin angular momentum, we calculate the evolution of the orbital
angular velocity $\Omega (t)$ through $X (t) = [ m_0 \Omega (t) ]^{2/3}$
and the orbital phase $\Theta (t)$ up to 3.5PN order by solving the
ordinary differential equations \cite{bcp2007}
\begin{widetext}
 \begin{eqnarray}
 \frac{dX}{dt} &=& \frac{64 \nu X^5}{5 m_0} \biggl[ 1 - \frac{743 + 924
  \nu}{336} X + 4 \pi X^{3/2} + \left( \frac{34103}{18144} +
				 \frac{13661}{2016} \nu + \frac{59}{18}
				 \nu^2 \right) X^2 \nonumber \\
 && - \left( \frac{4159}{672} + \frac{15876}{672} \nu \right) \pi
  X^{5/2} + \biggl\{ \frac{16447322263}{139708800} - \frac{1712}{105}
  \gamma_E + \frac{16}{3} \pi^2 - \left( \frac{56198689}{217728} -
				   \frac{451}{48} \pi^2 \right) \nu
  \nonumber \\
 && + \frac{541}{896} \nu^2 - \frac{5605}{2592} \nu^3 - \frac{856}{105}
  \ln (16X) \biggl\} X^3 - \left( \frac{4415}{4032} -
			    \frac{358675}{6048} \nu - \frac{91495}{1512}
			    \nu^2 \right) \pi X^{7/2} \biggl] ,
 \end{eqnarray}
\end{widetext}
\begin{equation}
 \frac{d \Theta}{dt} = \frac{X^{3/2}}{m_0} ,
\end{equation}
where $\nu = Q / (1 + Q)^2$ and $\gamma_E$ is the Euler constant. After
$X$ and $\Theta$ are obtained, we calculate complex gravitational-wave
amplitude $h^{22}$ of $(l,m) = (2,2)$ mode, assuming that the binary is
orbiting on the equatorial $(\theta = \pi/2)$ plane, up to 3PN order
using the formula \cite{kidder2008}
\begin{widetext}
 \begin{eqnarray}
 h^{22} &=& - 8 \sqrt{\frac{\pi}{5}} \frac{\nu m_0}{D} e^{-2 i \Theta} X
  \biggl[ 1 - \left( \frac{107}{42} - \frac{55}{42} \nu \right) X + 2
  \pi X^{3/2} - \left( \frac{2173}{1512} + \frac{1069}{216} \nu -
		 \frac{2047}{1512} \nu^2 \right) X^2 \nonumber \\
 && - \left\{ \left( \frac{107}{21} - \frac{34}{21} \nu \right) \pi + 24
       i \nu \right\} X^{5/2} + \biggl\{ \frac{27027409}{646800} -
 \frac{856}{105} \gamma_E + \frac{2}{3} \pi^2 - \frac{1712}{105} \ln 2 -
 \frac{428}{105} \ln X \nonumber \\
 && - \left( \frac{278185}{33264} - \frac{41}{96} \pi^2 \right) \nu -
  \frac{20261}{2772} \nu^2 + \frac{114635}{99792} \nu^3 +
  \frac{428}{105} i \pi \biggr\} X^3 \biggr] ,
 \end{eqnarray}
\end{widetext}
where $D$ is a distance between the binary and an observer. We also
compute a gravitational-wave spectrum from the waveform obtained in this
way.

We determine the properties of the BHs formed after the merger such as
masses and spins using the quantities associated with the apparent
horizon. The apparent horizon is determined in the same manner as
described in \cite{yst2008}.

The BH mass may be estimated by two methods. In the first method, we
measure the circumferential radius $C_e$ of the apparent horizon along
the equatorial plane and calculate $C_e / 4\pi$, which gives the BH mass
in the stationary vacuum BH spacetime. By this method, we estimate the
mass of the remnant BH after the spacetime settles to an approximately
steady state (assuming that deviation from Kerr spacetime due to the
presence of surrounding materials is negligible). In the second method,
we measure the irreducible mass of the BH, $M_{\rm irr}$, which is
determined from the area of the apparent horizon $A_{\rm AH}$ as
\begin{equation}
 M_{\rm irr} = \sqrt{\frac{A_{\rm AH}}{16 \pi}} .
\end{equation}
For the Kerr spacetime, $M_{\rm irr}$ is written by the mass and
dimensionless spin parameter $a \equiv J_{\rm BH} / M_{\rm BH}^2$ of a
BH (where $J_{\rm BH}$ is the spin angular momentum of the BH) as
\begin{equation}
 M_{\rm irr} = M_{\rm BH} \sqrt{{1 + \sqrt{1 - a^2} \over 2}} .
\end{equation}
Thus, if either the BH spin or mass is known, the BH mass or spin is
determined (again assuming that deviation from Kerr spacetime due to the
presence of surrounding materials is negligible). For the Kerr
spacetime, this relation may be written as
\begin{equation}
 M_{\rm irr} = \frac{C_e}{4 \sqrt{2} \pi} \sqrt{1 + \sqrt{1 - a^2}} .
 \label{eq:irrmass}
\end{equation}
Thus, we may say that the spin is estimated by calculating $M_{\rm irr}$
and $C_e$.

The dimensionless spin parameter of the BH is estimated also using the
quantities defined on the apparent horizon. For a Kerr BH with spin
parameter $a$, the ratio of the circumferential radius along the
meridional plane $C_p$ to the one along the equatorial plane $C_e$ is
written as
\begin{equation}
 \frac{C_p}{C_e} = \frac{\sqrt{2 \hat{r}_+}}{\pi} E \left( \frac{a^2}{2
						     \hat{r}_+} \right)
 ,
\end{equation}
where $\hat{r}_+ = 1 + \sqrt{1 - a^2}$ is the normalized radius of the
horizon and $E(z)$ is an elliptic integral
\begin{equation}
 E(z) = \int_0^{\pi/2} \sqrt{1 - z \sin^2 \theta} d\theta .
\end{equation}
Assuming that this relation holds for a BH surrounded by materials
again, we estimate the spin parameter of the remnant BH.

Comparison of the spin obtained from $C_p/C_e$ with that derived from
Eq.~(\ref{eq:irrmass}) provides a consistency check. It is found that
these two values agree with each other within the error $\Delta a
=0.003$ irrespective of the model of BH-NS binaries. For this reason, in
the following, we only present the spin determined from $C_p/C_e$.

In addition to the quantities for the remnant BHs, we calculate the
total rest mass of materials located outside the apparent horizon by
integrating the rest-mass density with respect to the proper volume
element,
\begin{equation}
 M_{r > r_{\rm AH}} \equiv \int_{r > r_{\rm AH}} \rho_* d^3 x ,
 \label{eq:diskmass}
\end{equation}
where $r_{\rm AH} = r_{\rm AH} (\theta, \varphi)$ denotes the radius of
the apparent horizon as a function of the angular coordinates $(\theta,
\varphi)$. $M_{r > r_{\rm AH}}$ is regarded as the mass of the remnant
disk when the system settles to a quasistationary state after the
merger.

\begin{table}
 \caption{Setup of the grid structure for the computation with our AMR
 algorithm. $\Delta x = h_6 = L / (2^6 N)$ is the grid spacing at the
 finest-resolution domain with $L$ being the location of the outer
 boundaries for each axis. $R_{\rm diam}/\Delta x$ denotes the grid
 number assigned inside the semimajor diameter of the NS. $\lambda_0$ is
 the gravitational wavelength of the initial configuration.}
 \begin{tabular}{cccc} \hline
  Model & $\Delta x / M_0$ & $R_{\rm diam} / \Delta x$ & $L /
  \lambda_0$ \\ \hline \hline
  2H-Q2M135 & 0.0471 & 90.8 & 1.189 \\
  H-Q2M135 & 0.0377 & 86.2 & 1.065 \\
  HB-Q2M135 & 0.0347 & 87.0 & 0.982 \\
  HBs-Q2M135 & 0.0353 & 85.2 & 0.998 \\
  HBss-Q2M135 & 0.0353 & 84.0 & 0.998 \\
  B-Q2M135 & 0.0330 & 85.1 & 0.932 \\
  Bs-Q2M135 & 0.0324 & 84.4 & 0.915 \\
  Bss-Q2M135 & 0.0270 & 95.4 & 0.825 \\ \hline
  2H-Q3M135 & 0.0353 & 89.0 & 0.998 \\
  H-Q3M135 & 0.0282 & 84.7 & 0.856 \\
  HB-Q3M135 & 0.0269 & 82.7 & 0.816 \\
  B-Q3M135 & 0.0247 & 83.8 & 0.749 \\ \hline
  2H-Q2M12 & 0.0565 & 86.9 & 1.255 \\
  H-Q2M12 & 0.0453 & 83.1 & 1.281 \\
  HB-Q2M12 & 0.0420 & 83.6 & 1.188 \\
  B-Q2M12 & 0.0392 & 83.4 & 1.109 \\ \hline
  HB-Q3M12 & 0.0306 & 84.6 & 0.866 \\
  B-Q3M12 & 0.0278 & 86.9 & 0.786 \\ \hline
 \end{tabular}
 \label{table:grid}
\end{table}

\subsection{Setup of AMR grids} \label{subsec:simulation_grids}

Numerical simulation is performed using an AMR algorithm described
in~\cite{yst2008}, to which the reader may refer for details. In the
present work, we prepare seven refinement levels to ensure that the
computational domain extends to the local wave zone for initial
quasiequilibrium states and that both compact objects are resolved with
a sufficient grid resolution (e.g., Table \ref{table:grid}). Each
refinement domain consists of the uniform, vertex-centered grids with
$(2N+1, 2N+1, N+1)$ grid points for $(x,y,z)$ with the equatorial plane
symmetry at $z=0$ imposed. In the present work, we typically choose $N =
50$, with the exception that $N = 54$ for model Bss-Q2M135, in which the
NS is quite compact and needs to be resolved with a better grid
resolution. For several models arbitrarily chosen, we performed
numerical simulations with lower grid resolutions, $N=36$ and 42, to
check the convergence of the numerical results (see the Appendix). The
edge length of the largest domain is denoted by $2L$ and the grid
spacing for each domain is then $h_l = L / (2^l N)$, where $l=0$--6. In
all the simulations, two sets of four finer domains comoving with
compact objects cover the region in the vicinity of two objects, and the
other three coarser domains cover both objects by a wider domain with
their origins being fixed at the approximate center of mass of the
binary. Namely, we prepare 11 refinement domains in total for all the
simulations.

Table \ref{table:grid} summarizes the parameters of the grid structure
for the simulations in this paper. As mentioned above, the value of $L$
is chosen to be $\approx \lambda_0$, where $\lambda_0 \equiv \pi /
\Omega_0$ is the gravitational wavelength at $t=0$ and $\Omega_0$ is the
orbital angular velocity of the initial configuration. Because the
gravitational wavelength decreases during the evolution of the binaries,
the outer boundary of the computational domains is guaranteed to be
located in the wave zone throughout the simulation. Each of the two
finest domains covers the semimajor axis of the NS with 42--48 grid
points and the BH radius (the coordinate radius of the apparent horizon)
with typically $\approx 20$ grid points, respectively. For $N=54$ run,
the total memory required for the simulations is about 11.6 G bytes. We
perform numerical simulations with personal computers of 12 G bytes
memory and of core-i7X processors with clock speed 3.2 or 3.33 GHz. We
only use two processors to perform one job with an OPEN-MP
library. Typical computational time required to perform one simulation
(for $\sim 40$ ms in physical time of coalescence) is 7--10 weeks.

\section{Numerical results} \label{sec:result}

\subsection{Orbital evolution and general merger process}
\label{subsec:result_orbit}

\begin{figure}
 \includegraphics[width=80mm,clip]{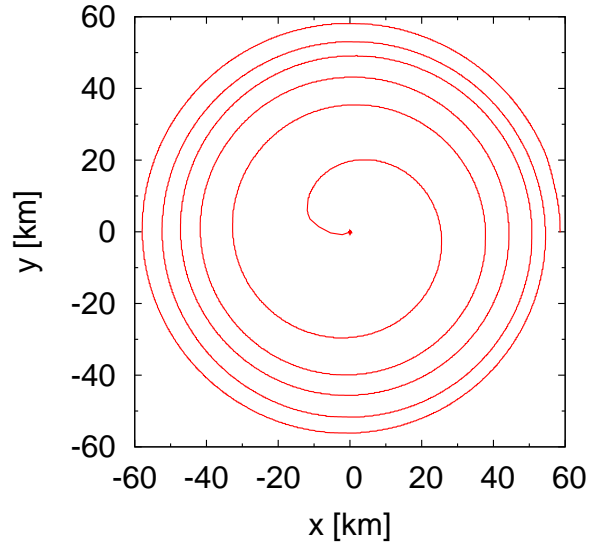} \caption{Evolution of the
 coordinate separation of the binary $x^i_{\rm sep}$ for model
 HB-Q2M135.} \label{fig:orb}
\end{figure}

To obtain a realistic numerical result for gravitational waveforms and
the final outcome formed after the merger, it is necessary to exclude
spurious effects associated with a noncircularity in the orbital motion
as much as possible. To assess the circularity of the orbital motion, we
plot the evolution of the coordinate separation $x^i_{\rm sep} =
x^i_{\rm NS} - x^i_{\rm BH}$ for model HB-Q2M135 in
Fig.~\ref{fig:orb}. Here, the position of the maximum rest-mass density
is identified as the coordinate of the NS, $x^i_{\rm NS}$, and the
location of the puncture, $x^i_{\rm P}$, is the coordinate of the BH,
$x^i_{\rm BH}$. This figure suggests that the orbital eccentricity
appears to be low throughout the whole evolution. Because $\gtrsim 5$
orbits are tracked, the eccentricity, which is likely to be nonzero
initially, should be suppressed by gravitational radiation reaction. We
note that for all the models, similar trajectories are found.

\begin{figure}
 \includegraphics[width=90mm,clip]{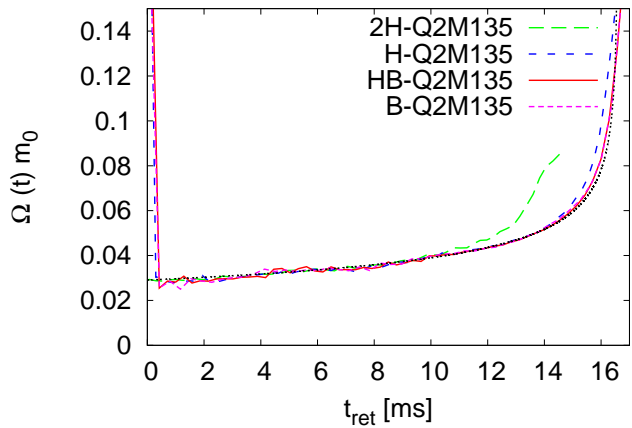}
\vspace{-6mm}
 \caption{Time evolution of the orbital angular velocity $\Omega(t) m_0$
 for models 2H-Q2M135, H-Q2M135, HB-Q2M135, and B-Q2M135 as a function
 of a retarded time defined by Eq.~(\ref{eq:retardedtime}) with an
 appropriate time shift. The dotted curve denotes the evolution of the
 orbital angular velocity calculated by the Taylor-T4 formula.}
 \label{fig:omega}
\end{figure}

\begin{figure*}
 \begin{tabular}{ccc}
 \includegraphics[width=55mm,clip]{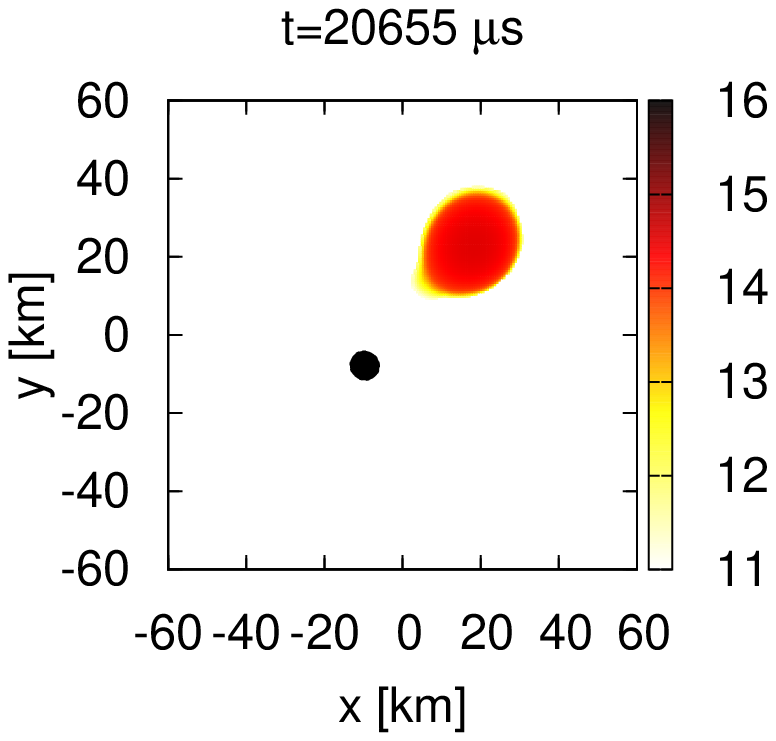} &
 \includegraphics[width=55mm,clip]{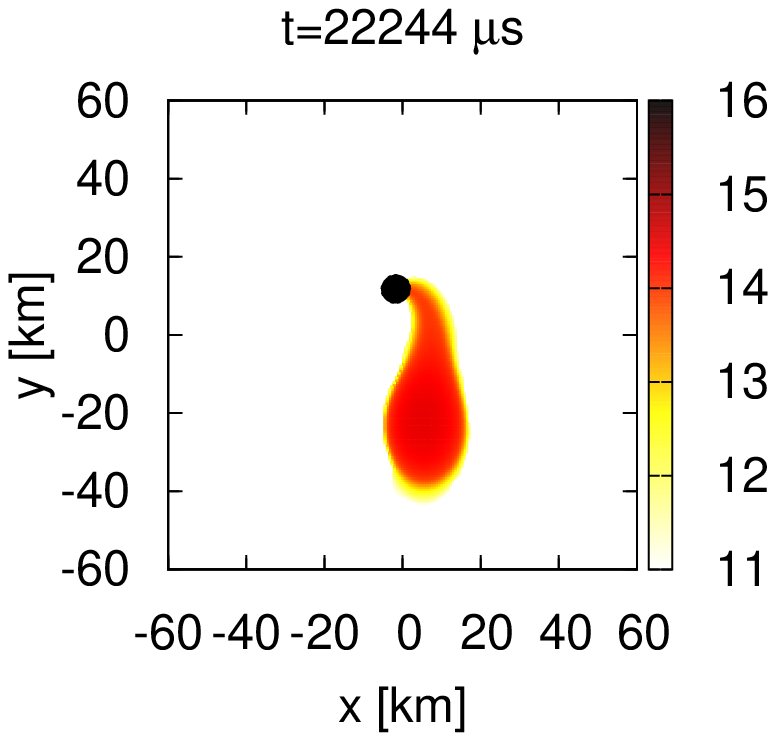} &
 \includegraphics[width=55mm,clip]{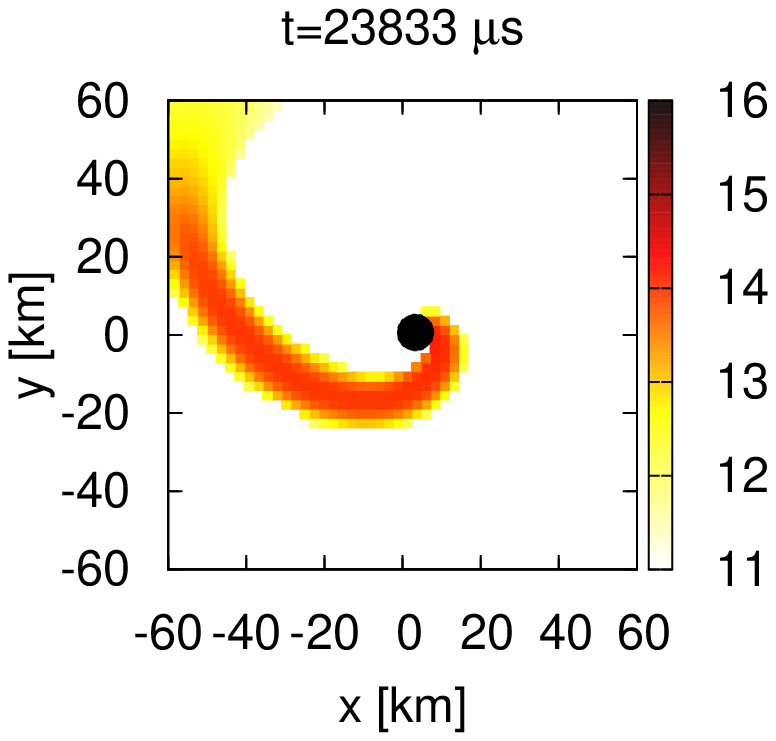} \\
 \includegraphics[width=55mm,clip]{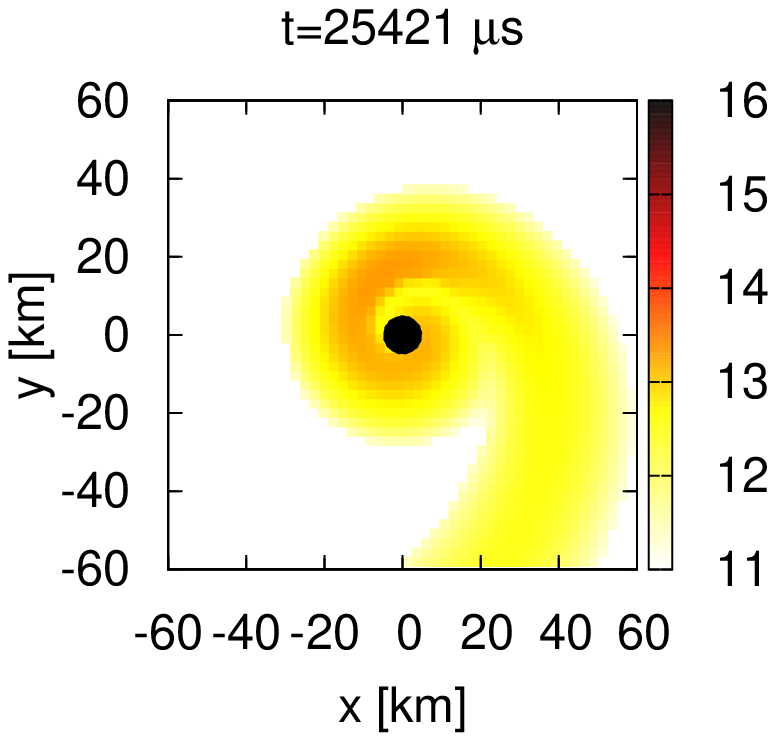} &
 \includegraphics[width=55mm,clip]{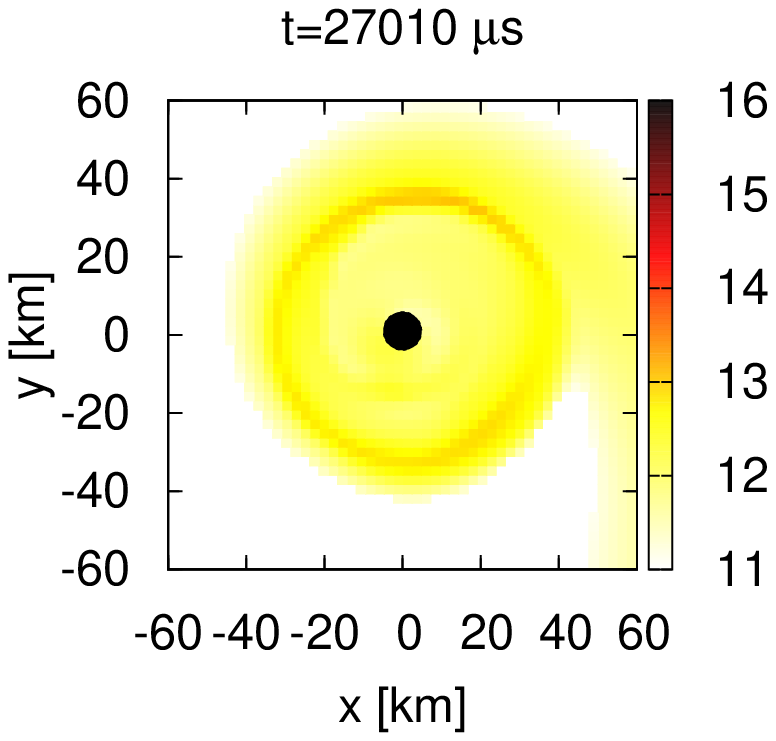} &
 \includegraphics[width=55mm,clip]{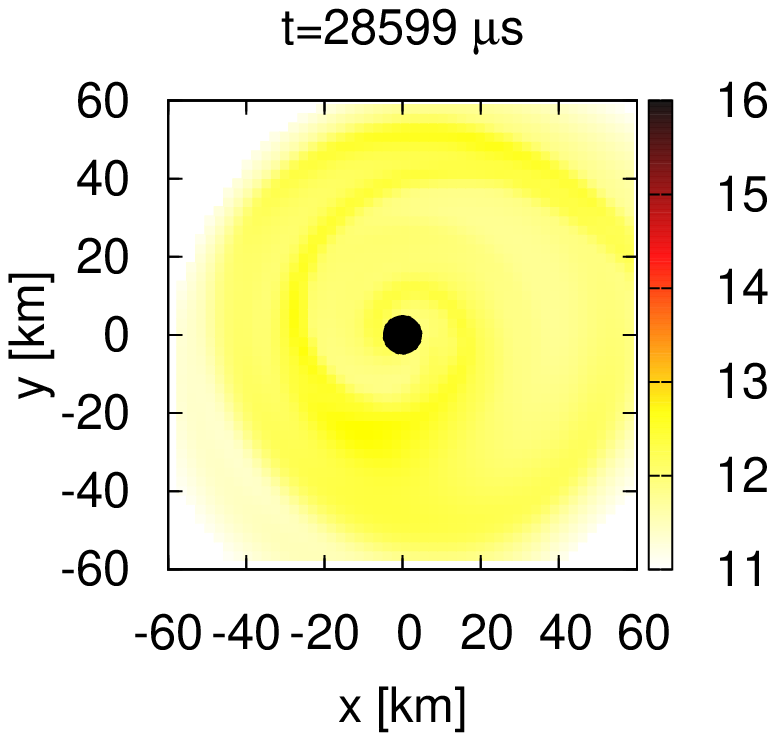}
 \end{tabular}
\vspace{-5mm}
 \caption{Evolution of the rest-mass density profile in units of ${\rm
 g/cm}^3$ and the location of the apparent horizon on the equatorial
 plane for model 2H-Q2M135. The filled circles denote the regions inside
 the apparent horizons. The color panels on the right-hand side of each
 figure show ${\rm log}_{10}(\rho)$.} \label{fig:snapshot1}
\end{figure*}

\begin{figure*}
 \begin{tabular}{ccc}
 \includegraphics[width=55mm,clip]{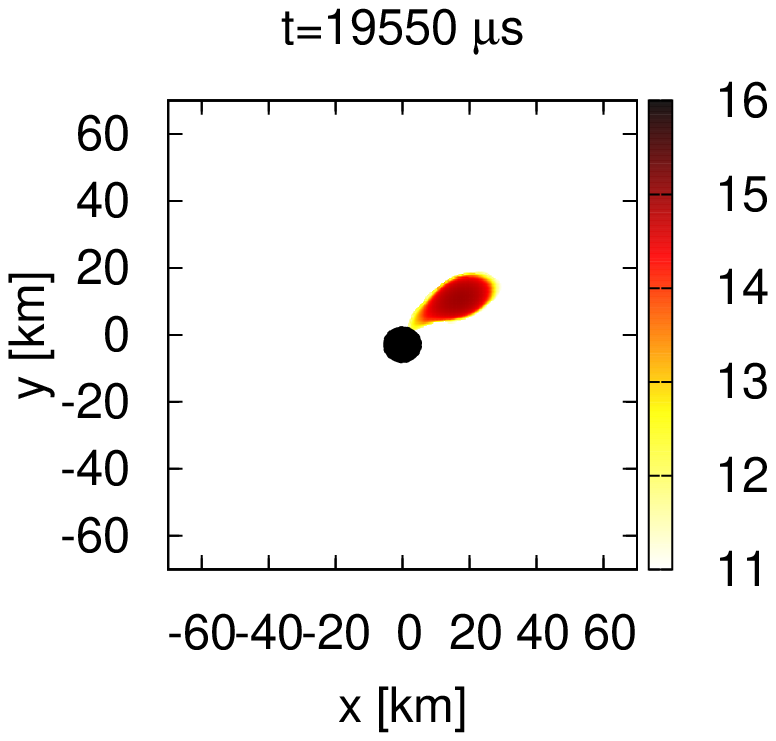} &
 \includegraphics[width=55mm,clip]{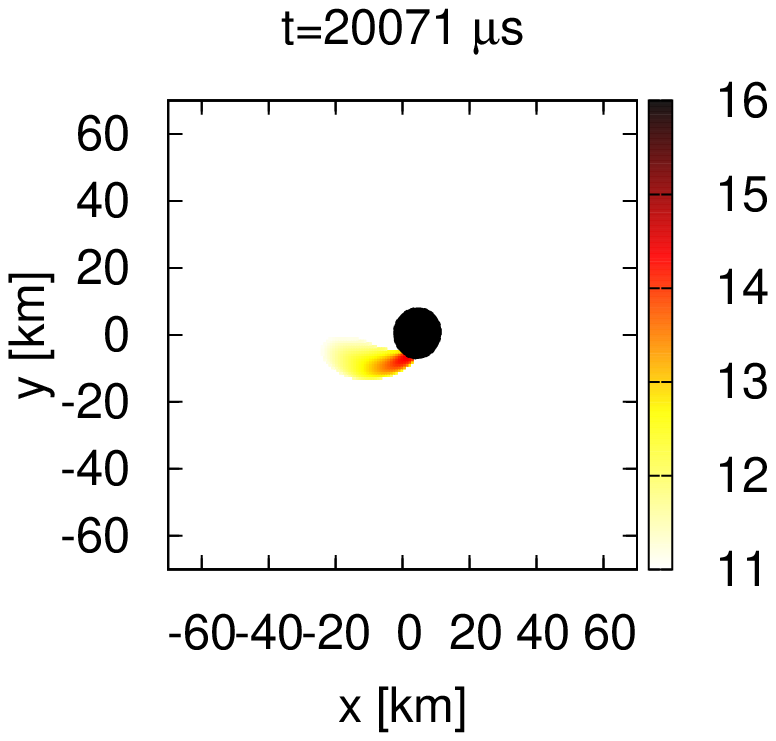} \\
 \includegraphics[width=55mm,clip]{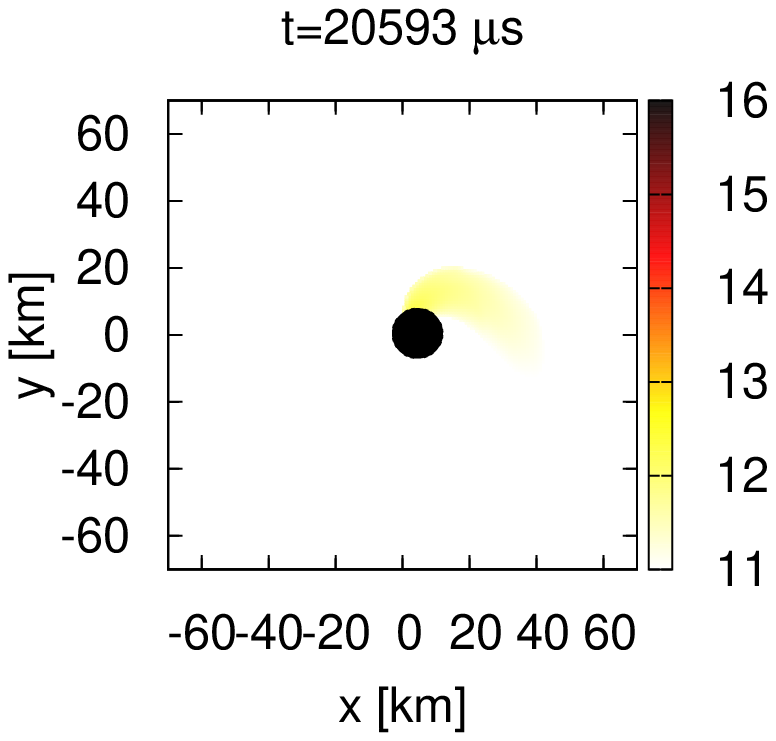} &
 \includegraphics[width=55mm,clip]{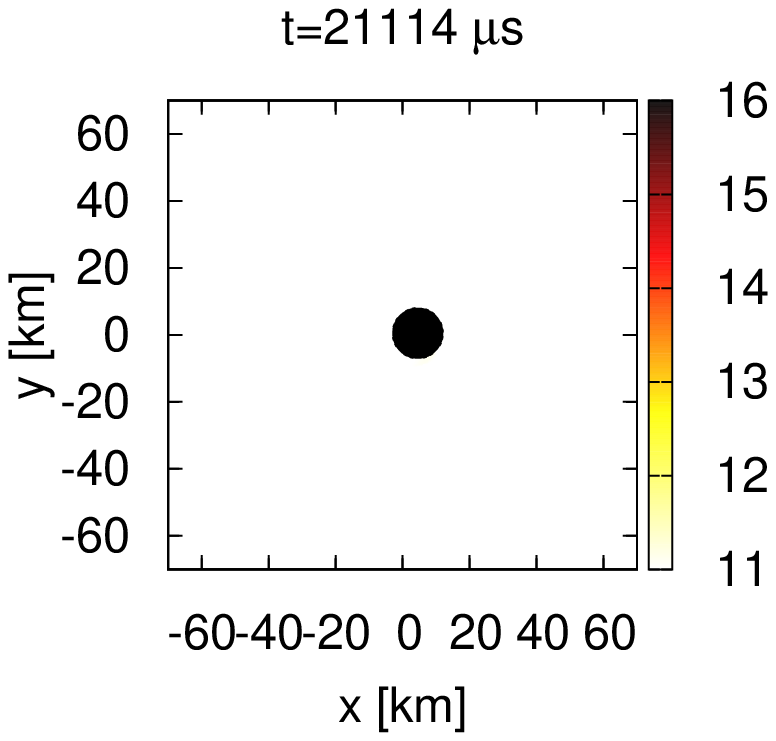}
 \end{tabular}
\vspace{-5mm}
 \caption{The same as Fig.~\ref{fig:snapshot1} but for model B-Q3M135.}
 \label{fig:snapshot2}
\end{figure*}

The coordinate separation shown above is a gauge-dependent quantity. To
show a stronger evidence that the eccentricity is suppressed to a small
level, it is better to plot a gauge-independent
quantity. Figure~\ref{fig:omega} plots the evolution of the orbital
angular velocity defined from the $(l,m) = (2,2)$ mode of $\Psi_4$ by
\begin{equation}
 \Omega (t) = \frac{1}{2} \frac{| \Psi_4 (l=m=2)|}{| \int \Psi_4 (l=m=2)
 dt |} ,
 \label{eq:omega}
\end{equation}
for models 2H-Q2M135, H-Q2M135, HB-Q2M135, and B-Q2M135. Here, the
horizontal axis is chosen to be an approximate retarded time defined by
\begin{equation}
t_{\rm ret} = t - D - 2M_0 \ln (D/M_0).\label{eq:retardedtime}
\end{equation}
We here do not plot the curve after the onset of tidal disruption. For
comparison, the angular velocity derived from the Taylor-T4 formula is
also plotted. To align the curve in the inspiral phase for $\Omega(t)
m_0 \leq 0.05$, we appropriately shift the time for each model. For
$t_{\rm ret} \alt 0$ ms, an unphysical (a junk wave) component contained
in the initial data dominates the waveform, and hence, $\Omega (t)$
derived from Eq.~(\ref{eq:omega}) does not give the angular velocity.

Figure~\ref{fig:omega} shows that the angular velocity obtained in
numerical simulations agrees with that by the Taylor-T4 formula within a
small modulation of $\Delta \Omega / \Omega \lesssim 5\%$ irrespective
of the models. With the fact that the orbital eccentricity is
approximately estimated as $e \approx 2 \Delta \Omega / 3 \Omega$ for $e
\ll 1$, we conclude that the orbital eccentricity is suppressed within
$\sim 3\%$. Figure~\ref{fig:omega} also shows that the deviation from
the Taylor-T4 result becomes remarkable in an earlier time for models
with stiffer EOSs such as 2H and H EOSs. This is due to the fact that
the tidal elongation and disruption of the NS occur at slightly earlier
stages of the inspiral orbits for models with the stiffer EOSs. This
illustrates the fact that the stiffness of the EOS is reflected clearly
in the gravitational-wave frequency (and gravitational-wave phase) as a
function of time.

Figures~\ref{fig:snapshot1} and \ref{fig:snapshot2} plot the snapshots
of the rest-mass density profiles and the location of the apparent
horizon on the equatorial plane at selected time slices for models
2H-Q2M12 and B-Q3M135. Figure \ref{fig:snapshot1} illustrates the
process in which the NS is tidally disrupted to form a disk surrounding
the companion BH. In this case, the NS is disrupted far outside the ISCO
and then forms a one-armed spiral arm with large angular momentum. As a
consequence of the angular momentum transport in the arm, a large amount
of materials spread outward and then form a disk around the BH. We will
report more details about the remnant disk in
Sec.~\ref{subsec:result_disk}. Figure~\ref{fig:snapshot2} illustrates
the case in which the NS is not tidally disrupted before it is swallowed
by the BH. In this case, mass of the disk formed after the onset of the
merger is negligibly small.

\subsection{Gravitational waveforms} \label{subsec:result_waveform}

\begin{figure*}[tbp]
 \begin{tabular}{cc}
  \includegraphics[width=85mm,height=50mm]{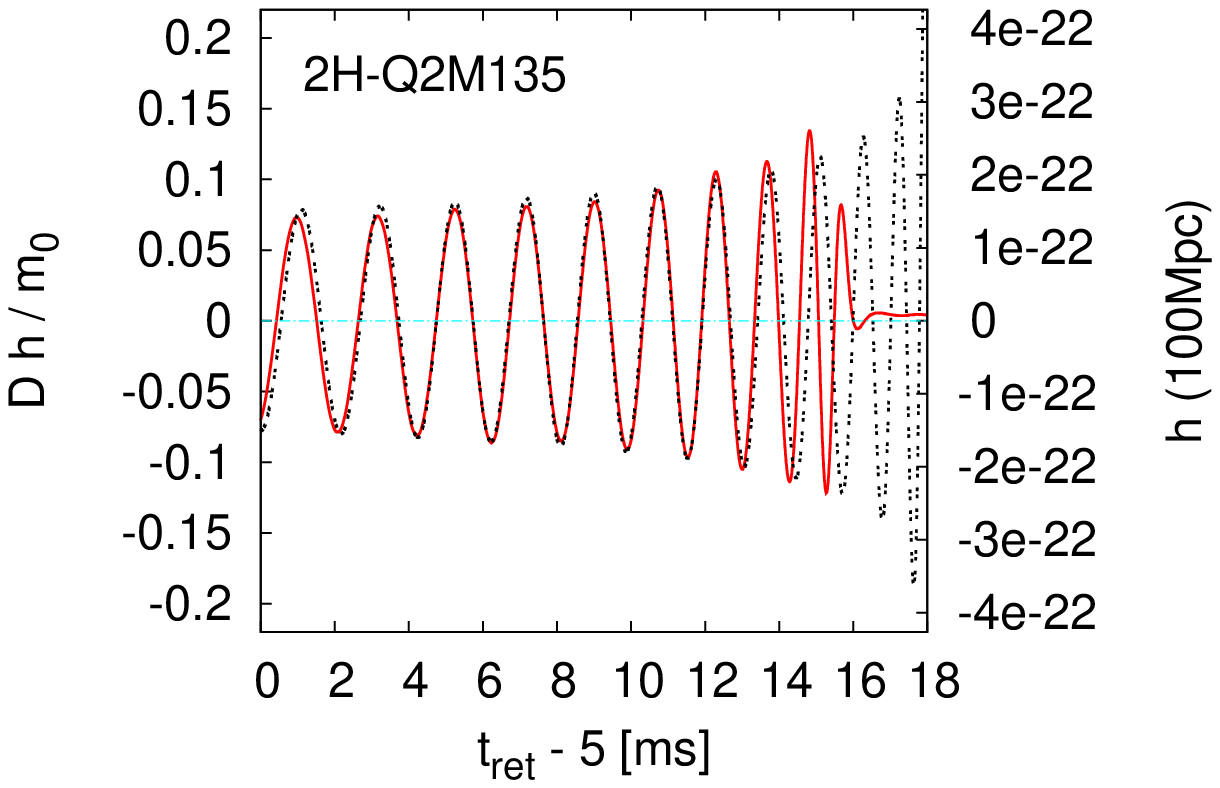} &
  \includegraphics[width=85mm,height=50mm]{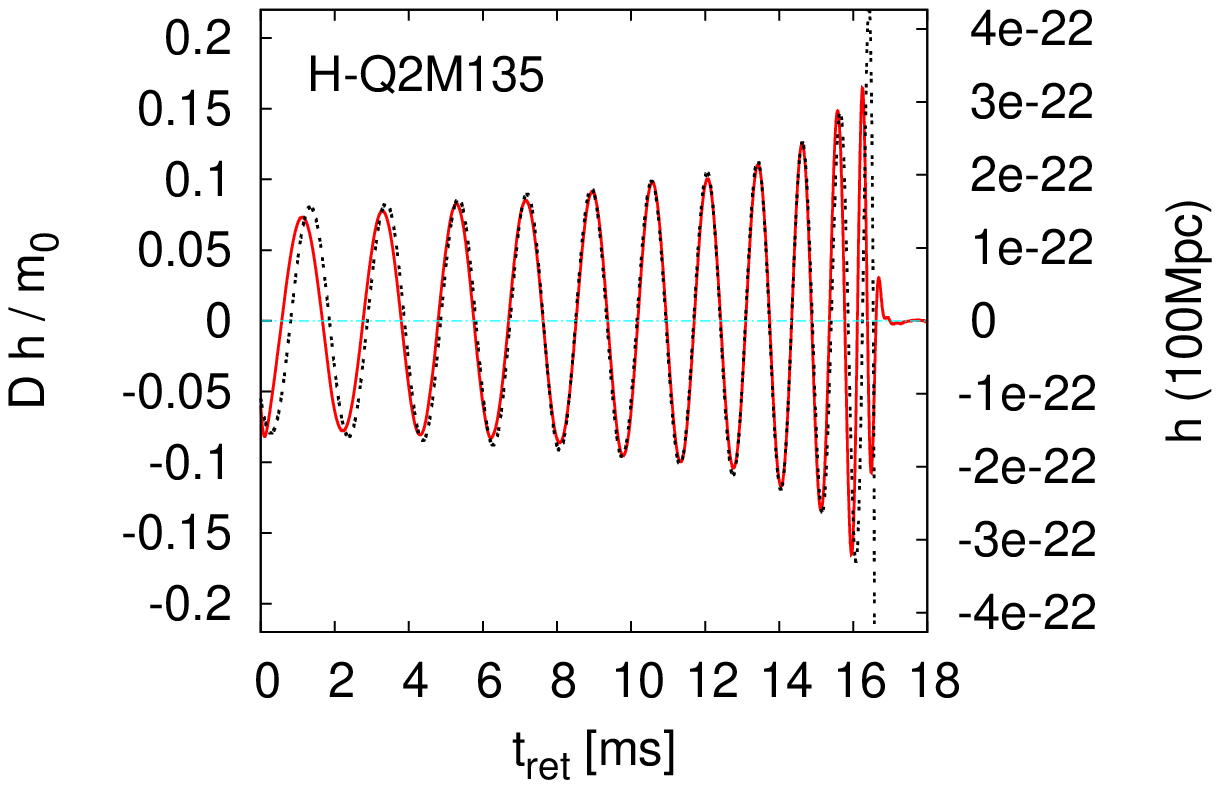} \\
  \includegraphics[width=85mm,height=50mm]{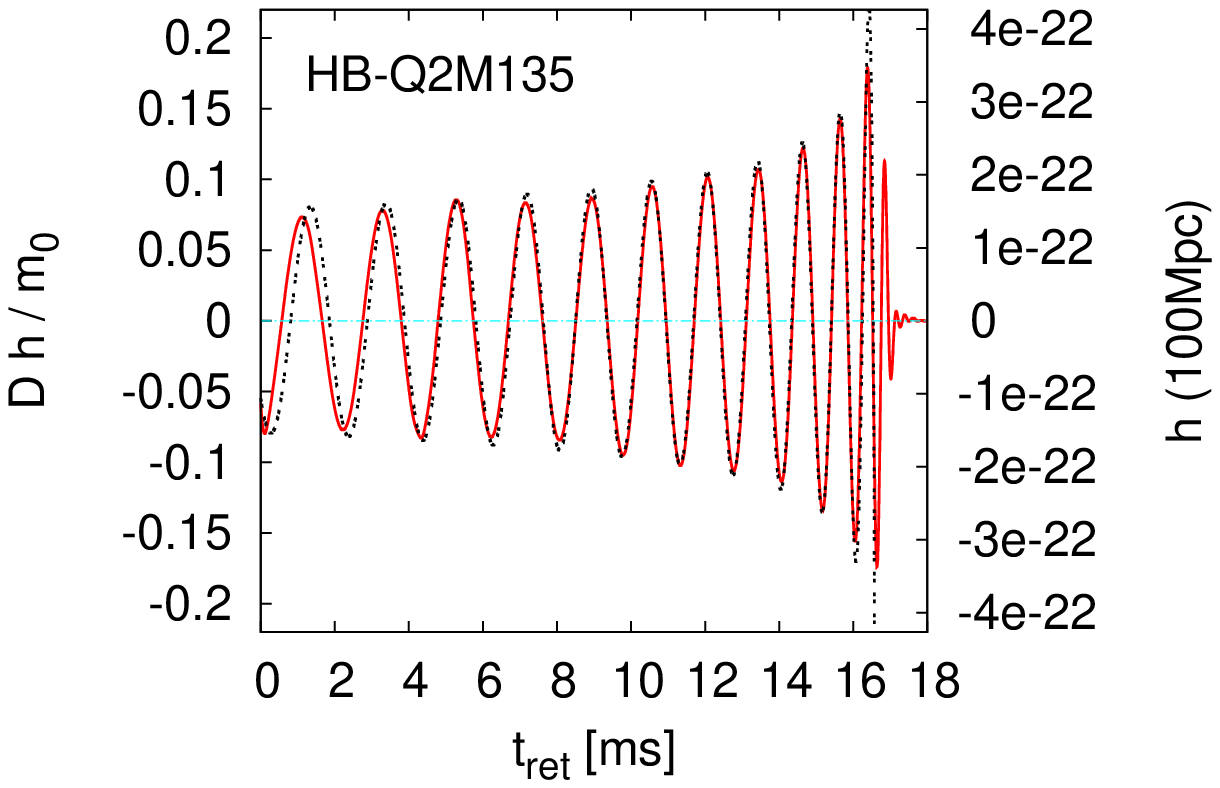} &
  \includegraphics[width=85mm,height=50mm]{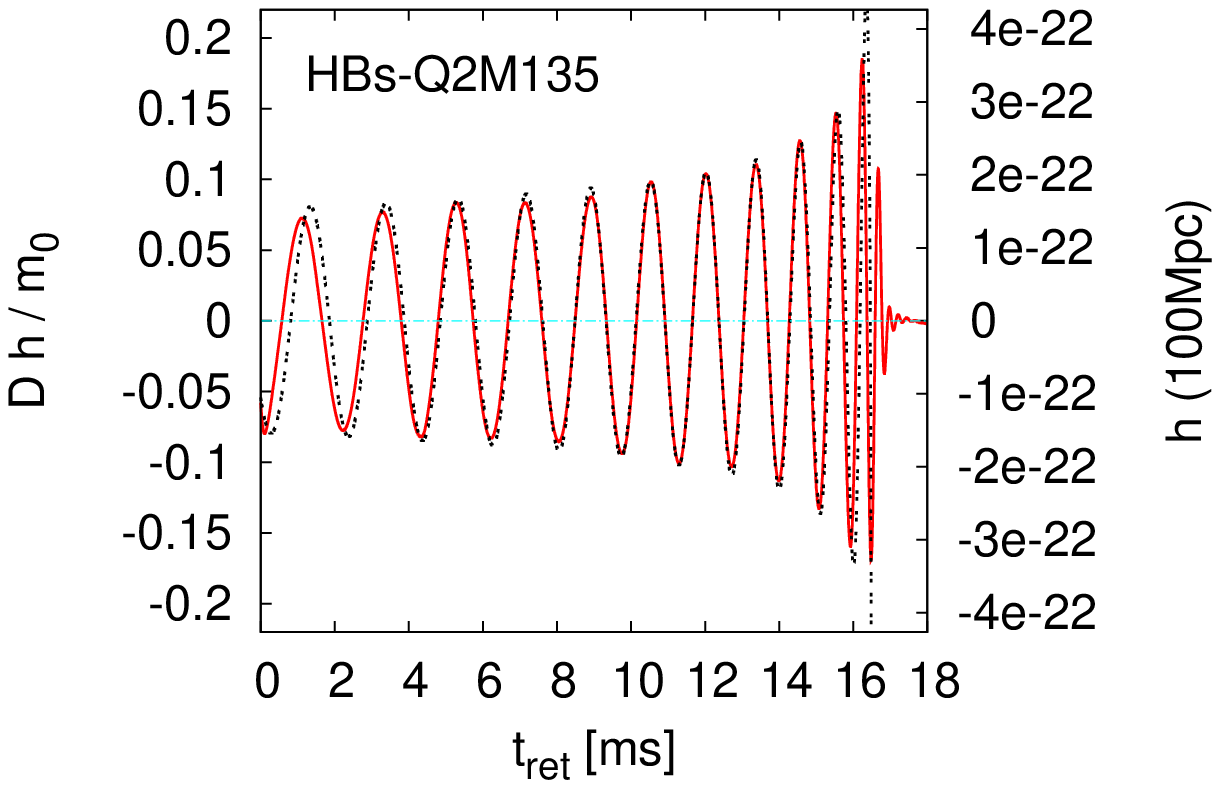} \\
  \includegraphics[width=85mm,height=50mm]{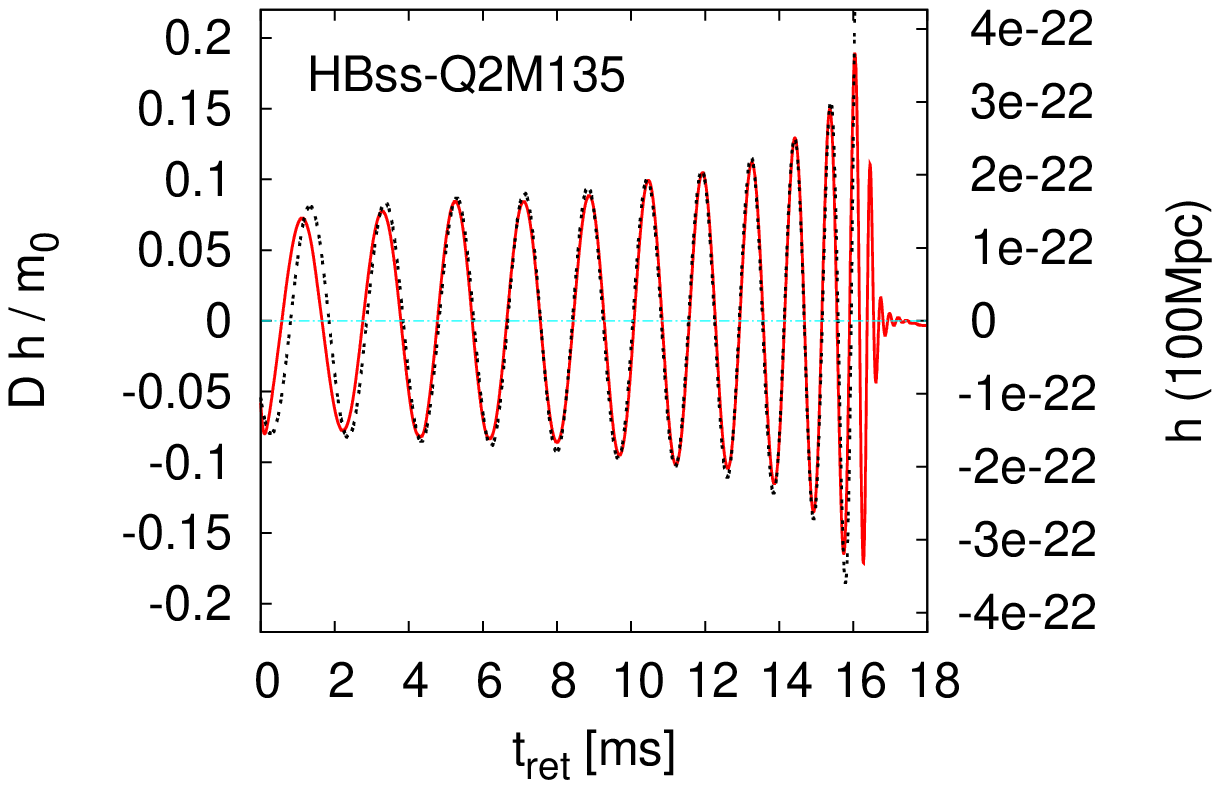} &
  \includegraphics[width=85mm,height=50mm]{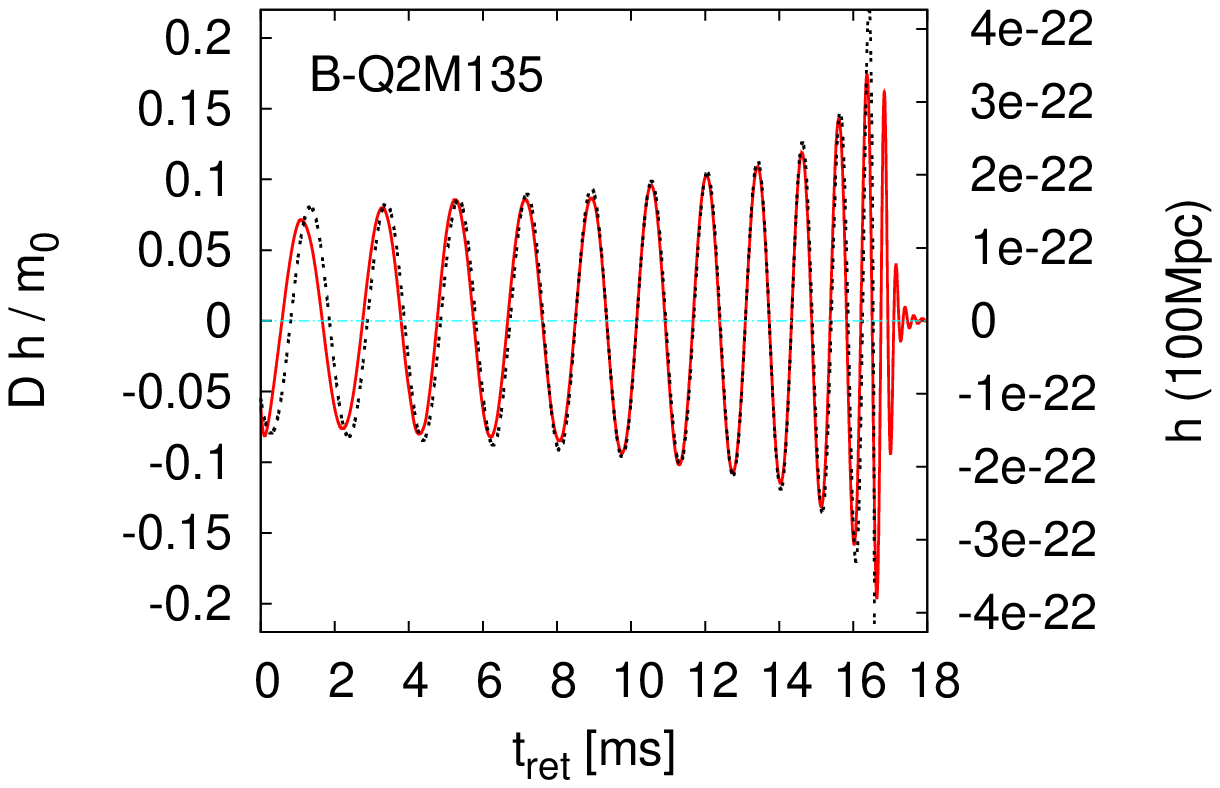} \\
  \includegraphics[width=85mm,height=50mm]{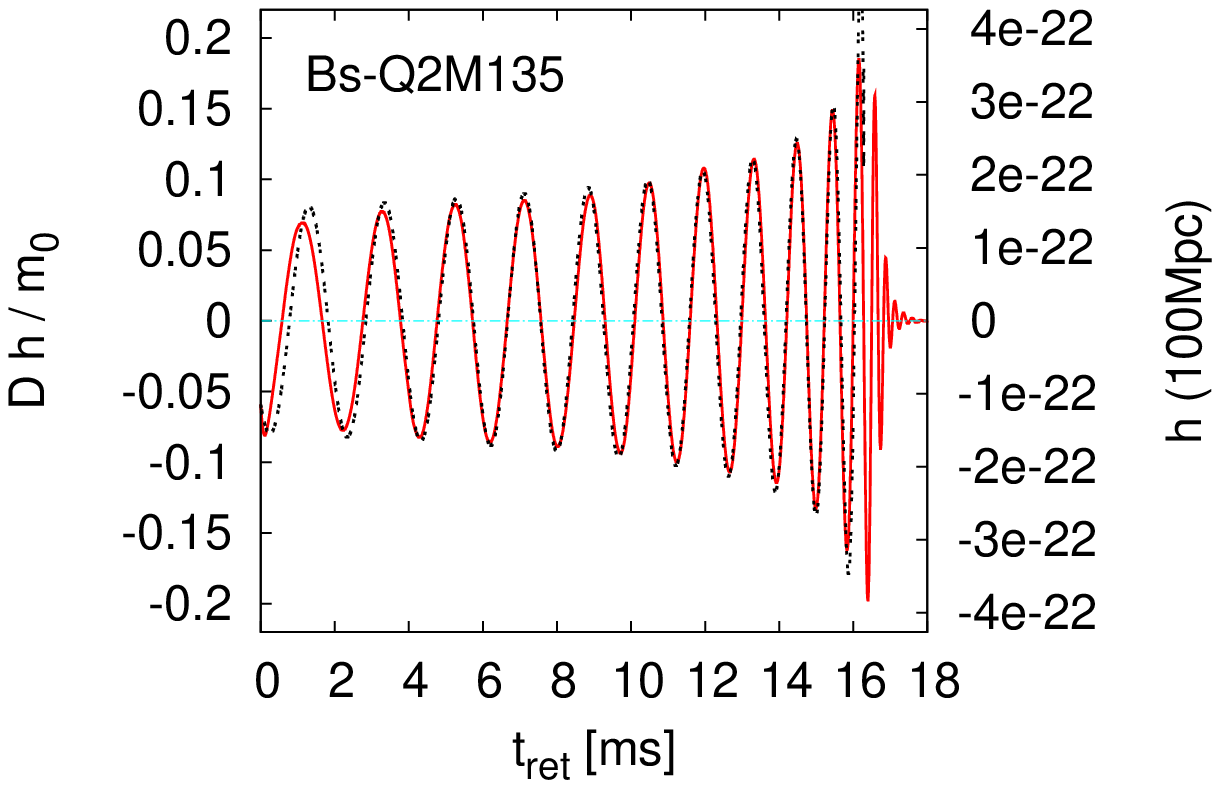} &
  \includegraphics[width=85mm,height=50mm]{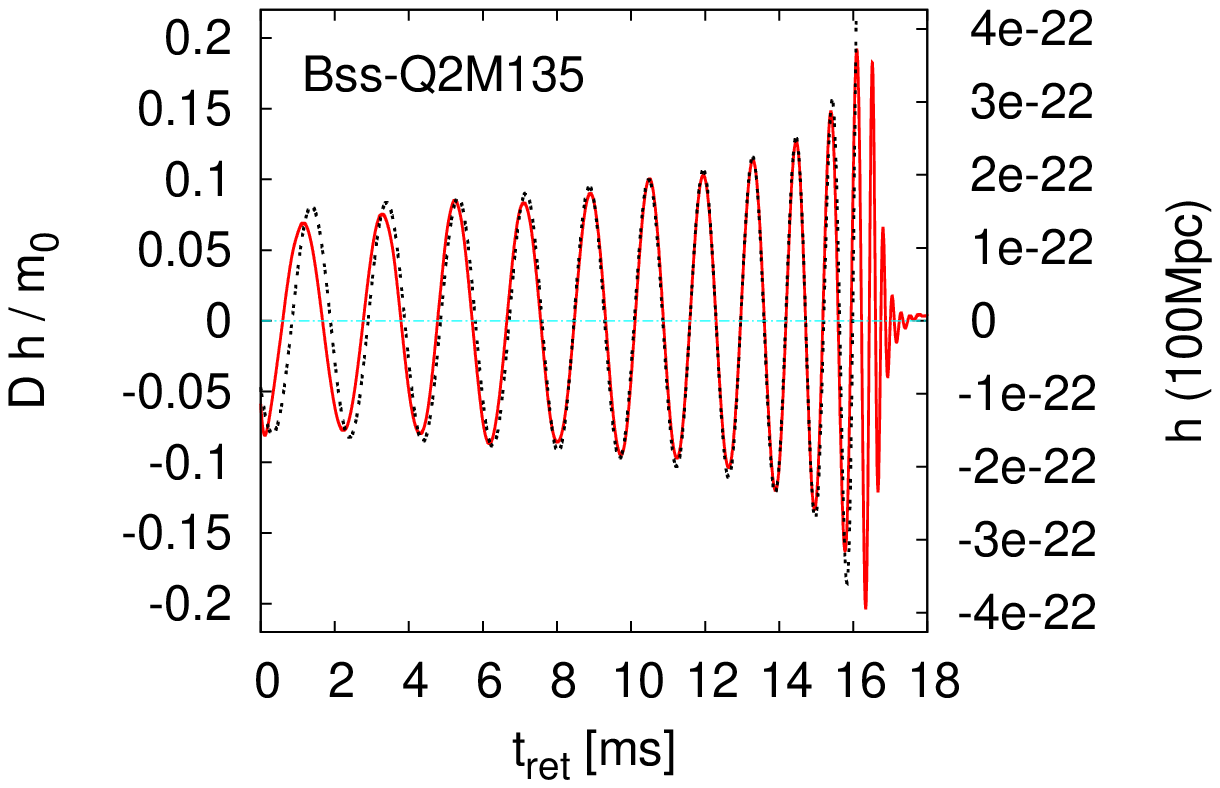}
 \end{tabular}
 \vspace{-5mm} \caption{$(l,m)=(2,2)$, plus-mode gravitational
 waveforms for models 2H-Q2M135, H-Q2M135, HB-Q2M135, HBs-Q2M135,
 HBss-Q2M135, B-Q2M135, Bs-Q2M135, and Bss-Q2M135. All the waveforms are
 shown for an observer located along the $z$ axis (axis perpendicular to
 the orbital plane) and plotted as a function of a retarded time. For
 model 2H-Q2M135, the waveform is plotted as a function of $t_{\rm
 ret}-5$ ms to align it with other waveforms (note that the initial
 value of $\Omega$ only for this model is smaller than those for other
 models). The left axis denotes the amplitude normalized by the distance
 from the binary $D$ and the total mass $m_0$. The right axis denotes
 the physical amplitude of gravitational waves observed at a
 hypothetical distance 100 Mpc. The dotted curves denote the waveform
 calculated by the Taylor-T4 formula.} \label{fig:GW1}
\end{figure*}

\begin{figure*}[tbp]
 \begin{tabular}{cc}
 \includegraphics[width=85mm,height=50mm]{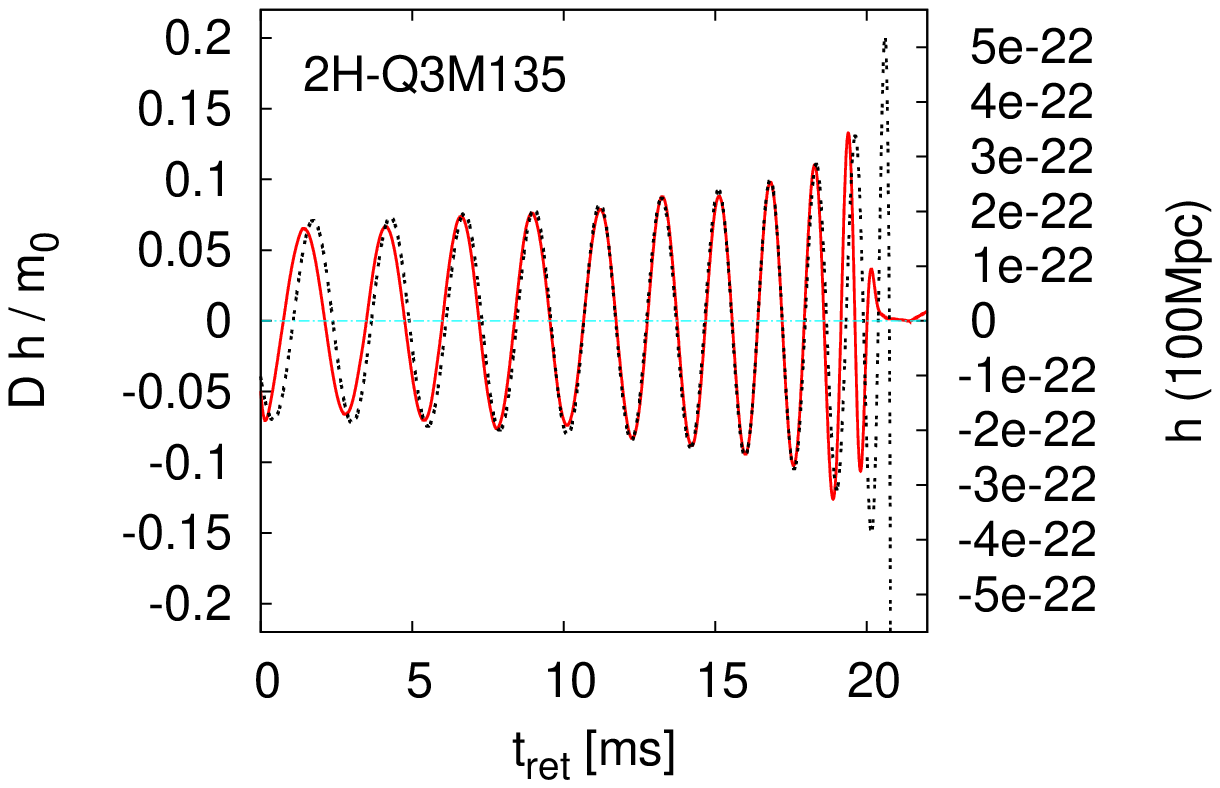} &
 \includegraphics[width=85mm,height=50mm]{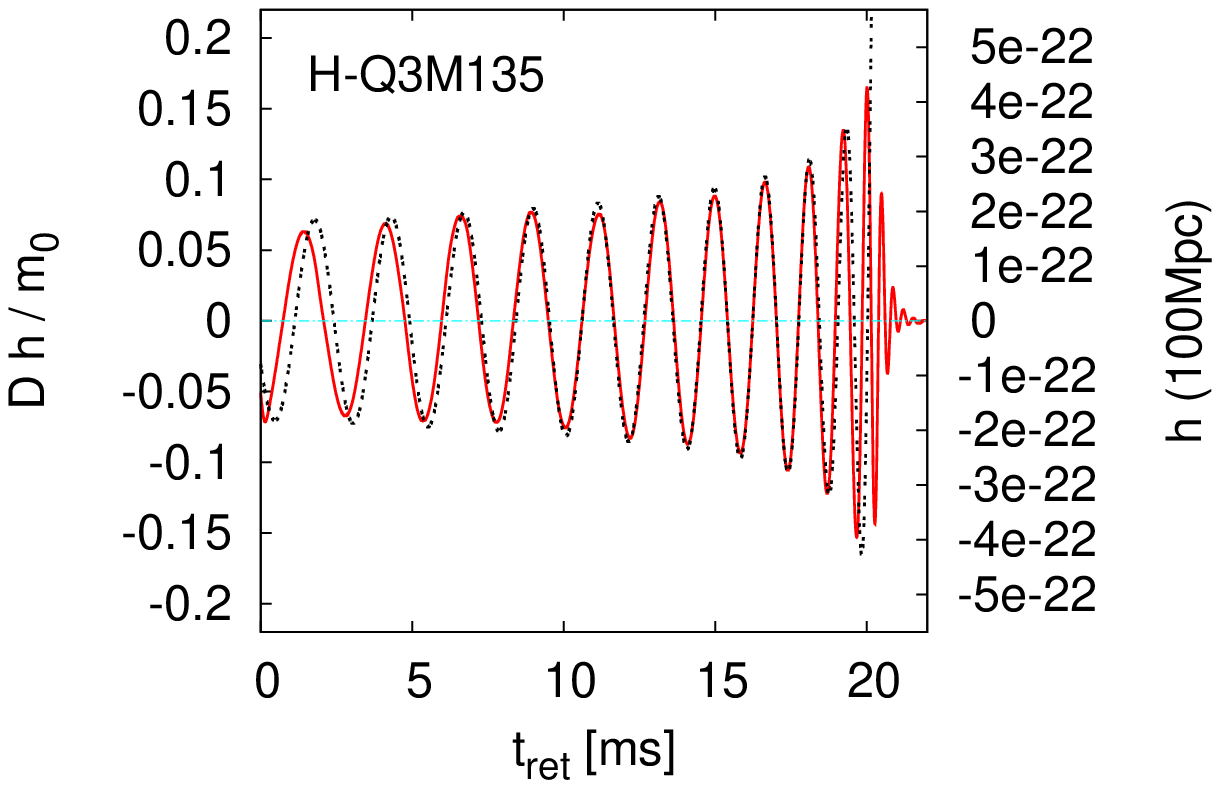} \\
 \includegraphics[width=85mm,height=50mm]{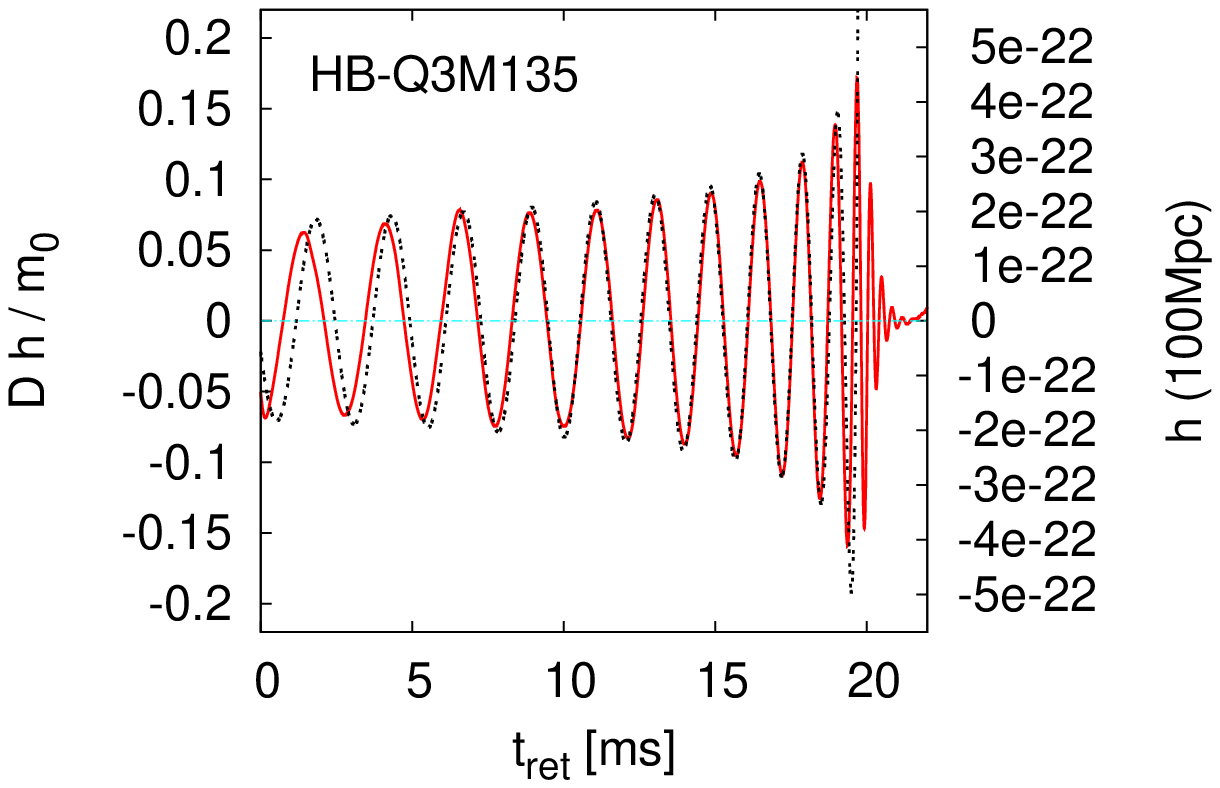} &
 \includegraphics[width=85mm,height=50mm]{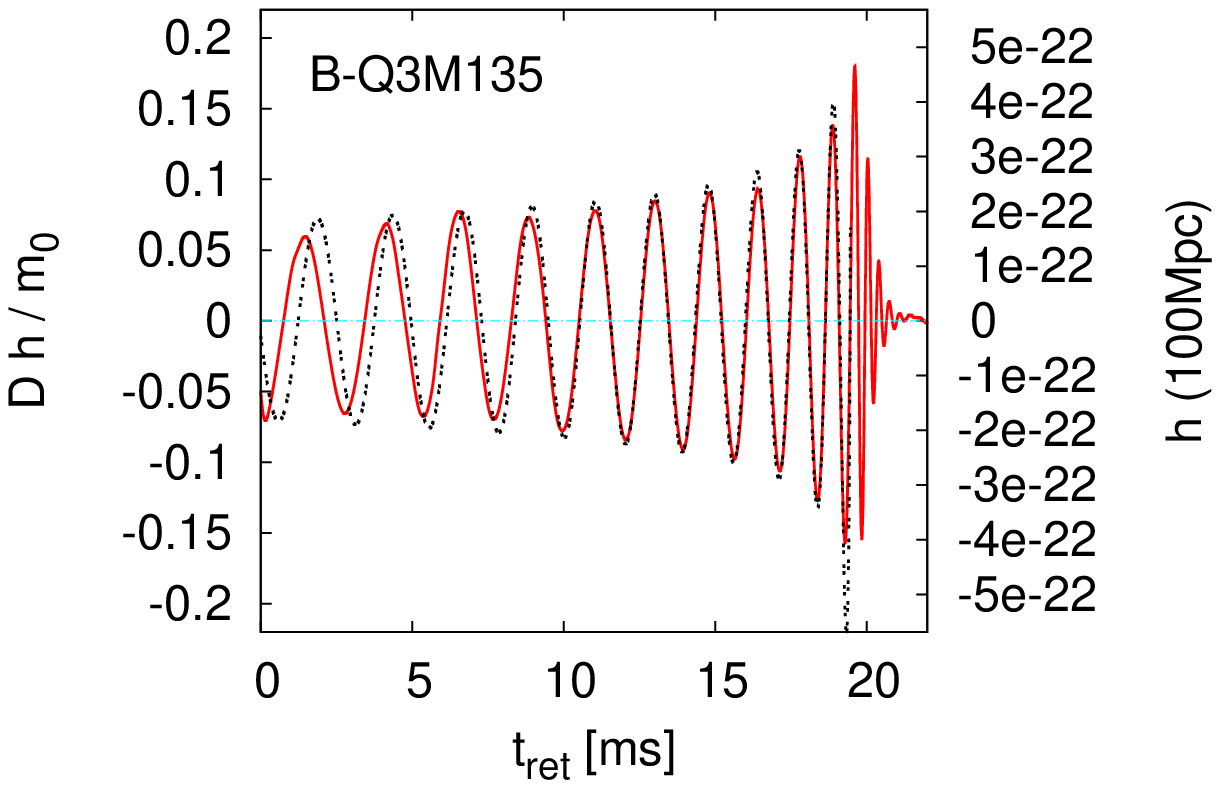} \\
 \includegraphics[width=85mm,height=50mm]{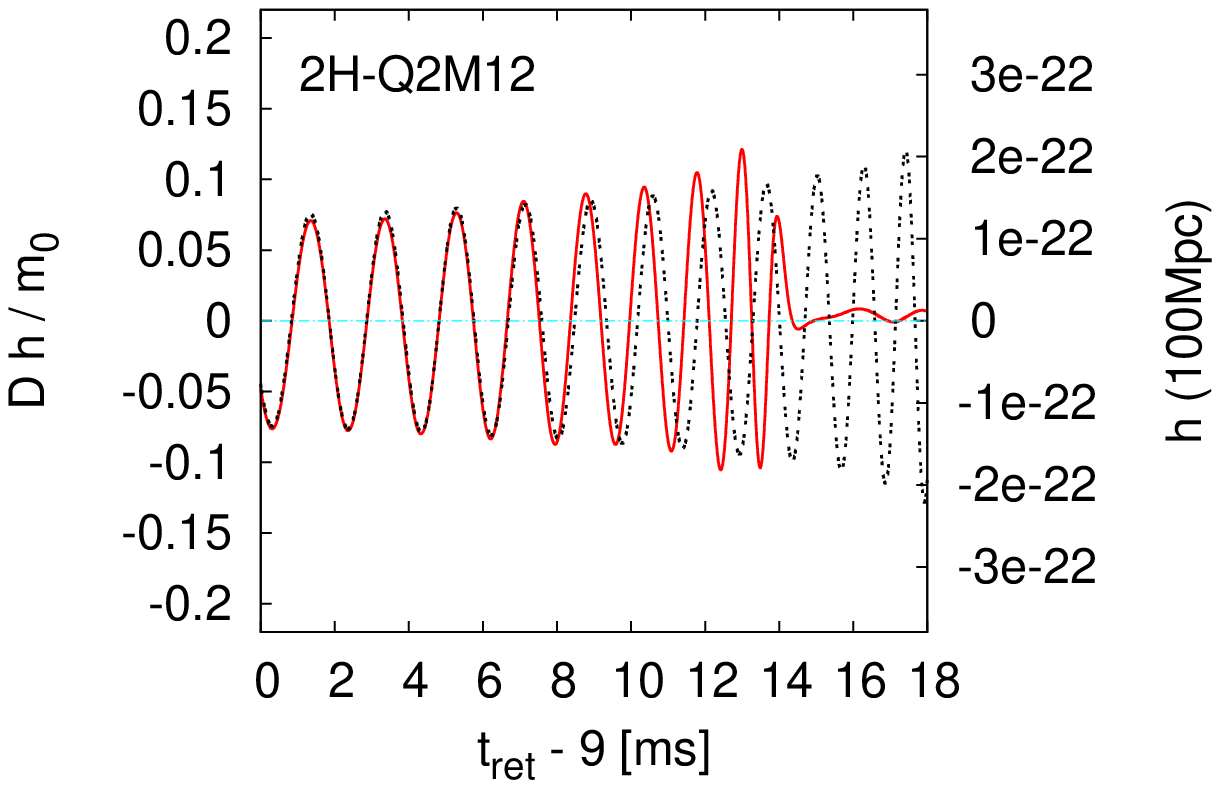} &
 \includegraphics[width=85mm,height=50mm]{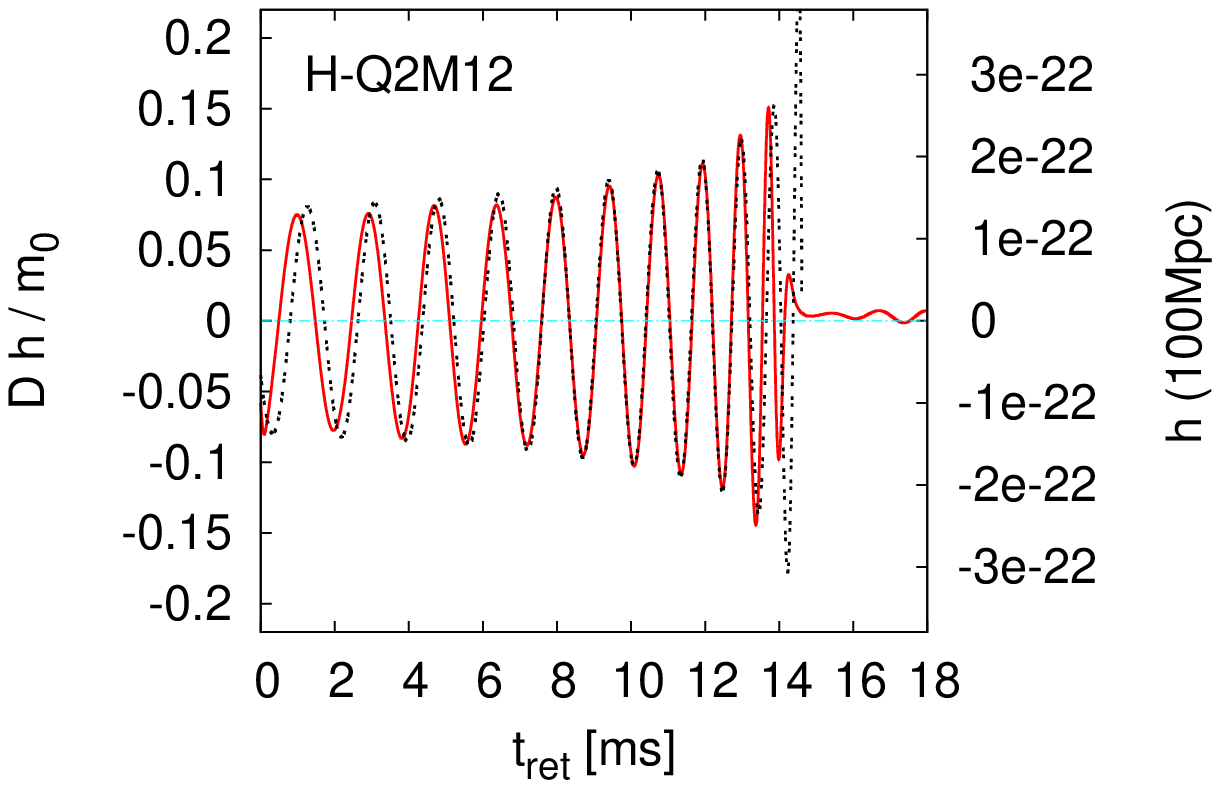} \\
 \includegraphics[width=85mm,height=50mm]{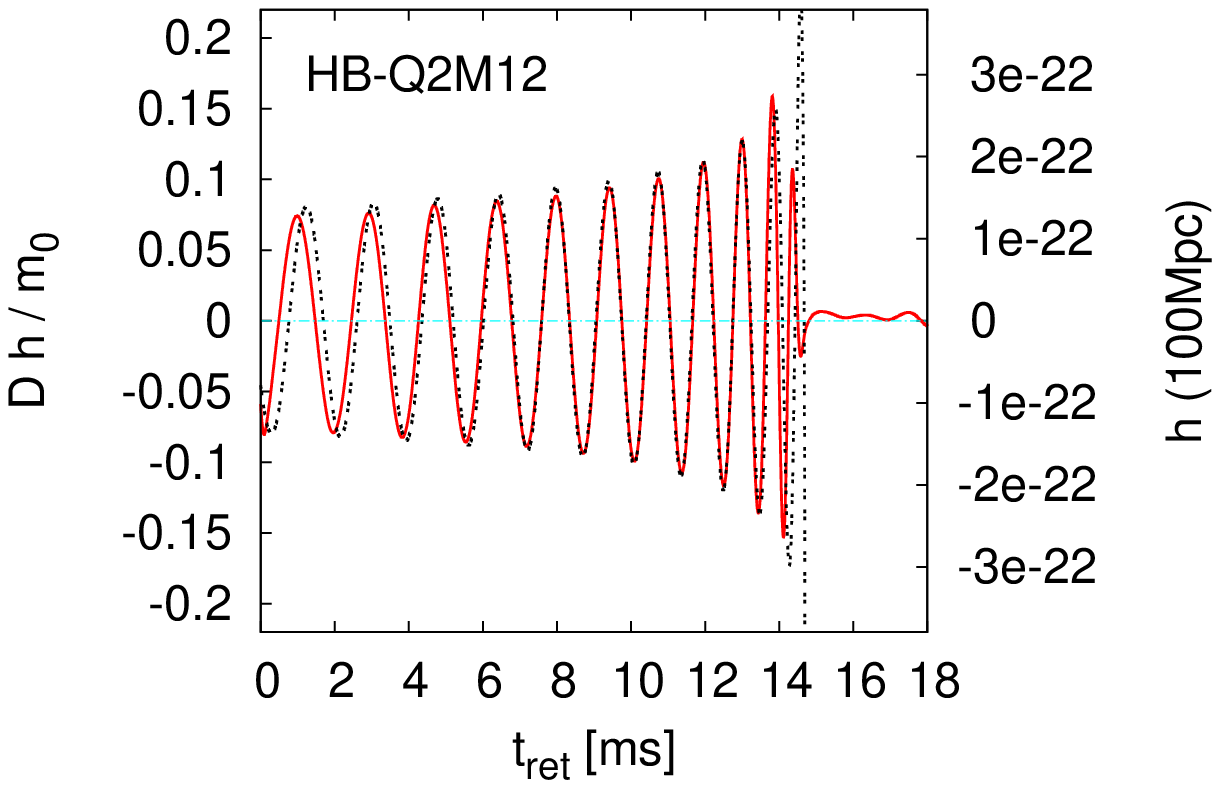} &
 \includegraphics[width=85mm,height=50mm]{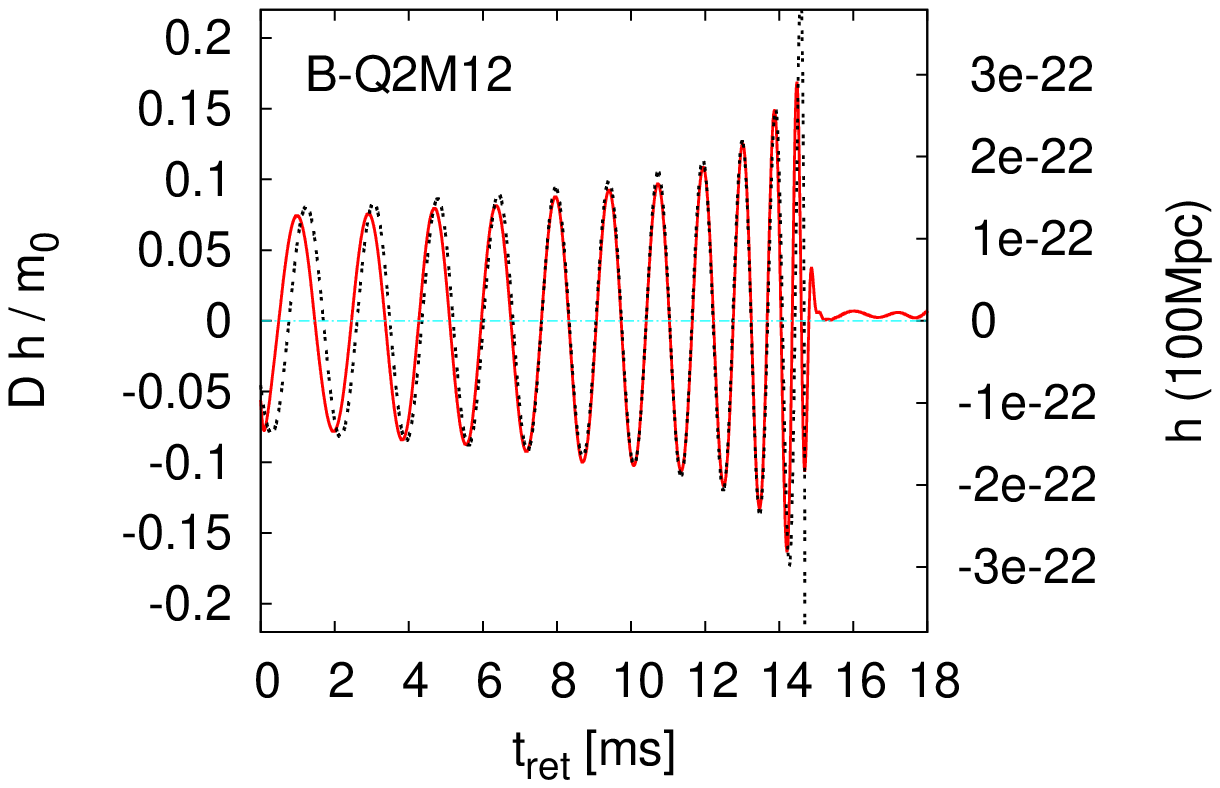}
 \end{tabular} \caption{The same as Fig.~\ref{fig:GW1} but for models
 2H-Q3M135, H-Q3M135, HB-Q3M135, B-Q3M135, 2H-Q2M12, H-Q2M12, HB-Q2M12,
 and B-Q2M12. Again, the waveform for model 2H-Q2M12 is plotted as a
 function of $t_{\rm ret}-9$ ms.} \label{fig:GW2}
\end{figure*}

Figures~\ref{fig:GW1} and \ref{fig:GW2} plot the $(l,m) = (2,2)$,
plus-mode gravitational waveforms obtained numerically (hereafter
referred to as $h_+$). All the waveforms are shown for an observer
located along the $z$ axis (axis perpendicular to the orbital plane) and
plotted as a function of a retarded time $t_{\rm ret}$. We plot the
amplitude in a normalized form, $D h_+ / m_0$, and the physical
amplitude observed by an observer located at a hypothetical distance $D
= 100$ Mpc.

To validate the numerical waveforms, we compare them with the Taylor-T4
waveform, which is accurate up to 3.5PN order in phase and 3PN order in
amplitude, with an appropriate time shift; the time shift is carried out
to align the curve of $\Omega(t)$ as performed in
Sec.~\ref{subsec:result_orbit}. Figures~\ref{fig:GW1} and \ref{fig:GW2}
show that these two waveforms agree with each other irrespective of
models during the inspiral phase, except for 2--3 initial cycles. The
reasons for this initial disagreement are that an approaching velocity
associated with gravitational radiation reaction is not taken into
account in the initial data and also the initial condition does not
exactly model a quasicircular state, because we do not fully solve
Einstein's equation for deriving it.

The numerical waveforms in the merger phase also (but due to a physical
reason) deviate from the Taylor-T4 ones both in phase and amplitude, in
particular for models with stiff EOSs, e.g., 2H-Q2M135 and 2H-Q2M12.
For such models, ringdown waveforms associated with the BH quasinormal
mode are not seen in the merger and ringdown phases, and instead, the
gravitational-wave amplitude damps suddenly in the middle of the
inspiral phase. The reason for this quick damping is that the NS is
tidally disrupted by the companion BH at an orbit in the inspiral phase
within one orbital period, and then, the disrupted material forms a
relatively low-density and nearly axisymmetric matter distribution
around the BH, suppressing time variation of a mass quadrupole moment.
Because the gravitational-wave emission stops in the middle of the
inspiral motion, the maximum amplitude of gravitational waves is smaller
for such a binary than for a binary with no tidal disruption, as shown
in Fig.~\ref{fig:GW1}. All these facts illustrate that the finite size
effect of the NS significantly modifies gravitational waves derived in
the point-particle approximation (in the Taylor-T4 formula). On the
other hand, ringdown gravitational waves are clearly seen for models
with soft EOSs (for which tidal disruption does not occur) such as model
B-Q3M135, in which the numerical and the Taylor-T4 waveforms are in more
excellent agreement even in the late inspiral phase.

Table \ref{table:GW} presents total radiated energy $\Delta E$ and
angular momentum $\Delta J$ carried away by gravitational waves. The
contribution from all the $l=2$--4 modes is taken into account for
$\Delta E$ and $\Delta J$. We estimate systematic errors in the
presented values to be less than 10\%, which are associated mainly with
the finite grid resolution and partly with the finite extraction radii
(cf. the Appendix). We note that the $(l,|m|) = (2,2)$ modes always
contribute by $\gtrsim 90 \%$ to both for $\Delta E$ and $\Delta J$. The
fraction of these modes is larger for binaries composed of less-compact
NSs, because only binaries which escape the tidal disruption in the late
inspiral phase can efficiently emit higher $l$-mode gravitational
waves. Among other modes, $(3,3)$ and $(4,4)$ modes constitute most of
the remaining part of $\Delta J$, whereas the order of magnitude of the
$(2,1)$ mode is as large as that of the $(4,4)$ mode for $\Delta E$.

The numerical results shown in Table~\ref{table:GW} illustrate a
quantitative dependence of gravitational-wave emission on the
compactness of the NS: For a given mass ratio, gravitational-wave
emission continues for a longer duration and consequently total radiated
energy and angular momentum are larger for binaries composed of more
compact NSs. Comparison among the models with $Q=2$ and $M_{\rm NS} =
1.35 M_\odot$ and with the same initial value of $\Omega m_0$ shows that
both $\Delta E / M_0$ and $\Delta J / J_0$ are monotonically increasing
functions of the NS compactness ${\cal C}$. This point is also
recognized from Figs.~\ref{fig:GW1} and \ref{fig:GW2}, e.g., from the
comparison among gravitational waves for models H-Q2M135, HB-Q2M135, and
B-Q2M135 (note that for model 2H-Q2M135 the simulation is started from a
lower value of $\Omega m_0$ and it is not suitable for this comparison).
Table \ref{table:GW} also shows that $\Delta J / \Delta E$ decreases as
the EOS softens. This is due to the fact that $\Delta J / \Delta E
\approx m / \Omega$ for a given angular harmonic of $m$, and for a soft
EOS, more radiation is emitted at large angular velocity, $\Omega$.

\begin{table}
 \caption{Total radiated energy $\Delta E$ and angular momentum $\Delta
 J$ carried away by gravitational waves. $\Delta E$ and $\Delta J$ are
 normalized with respect to the initial ADM mass $M_0$ and angular
 momentum $J_0$, respectively. We also show the ratio between $\Delta J$
 and $\Delta E$.}
 \begin{tabular}{cccc}
  \hline Model & $ \Delta E / M_0 (\%)$ & $ \Delta J / J_0 (\%)$ &
  $(\Delta J / J_0) / (\Delta E / M_0)$ \\
  \hline \hline
  2H-Q2M135 & 0.55 & 14 & 26 \\
  H-Q2M135 & 1.1 & 20 & 18 \\
  HB-Q2M135 & 1.4 & 22 & 16 \\
  HBs-Q2M135 & 1.4 & 22 & 16 \\
  HBss-Q2M135 & 1.5 & 23 & 15 \\
  B-Q2M135 & 1.7 & 24 & 14 \\
  Bs-Q2M135 & 1.9 & 25 & 13 \\
  Bss-Q2M135 & 2.2 & 27 & 12 \\ \hline
  2H-Q3M135 & 0.64 & 15 & 23 \\
  H-Q3M135 & 1.4 & 22 & 16 \\
  HB-Q3M135 & 1.6 & 23 & 14 \\
  B-Q3M135 & 1.8 & 24 & 13 \\ \hline
  2H-Q2M12 & 0.41 & 12 & 30 \\
  H-Q2M12 & 0.74 & 16 & 21 \\
  HB-Q2M12 & 0.89 & 18 & 20 \\
  B-Q2M12 & 1.1 & 20 & 18 \\ \hline
  HB-Q3M12 & 1.2 & 21 & 18 \\
  B-Q3M12 & 1.4 & 23 & 16 \\ \hline
 \end{tabular}
 \label{table:GW}
\end{table}

\subsection{Gravitational-wave spectrum} \label{subsec:result_spectrum}

\begin{figure*}[tbp]
 \includegraphics[width=120mm,clip]{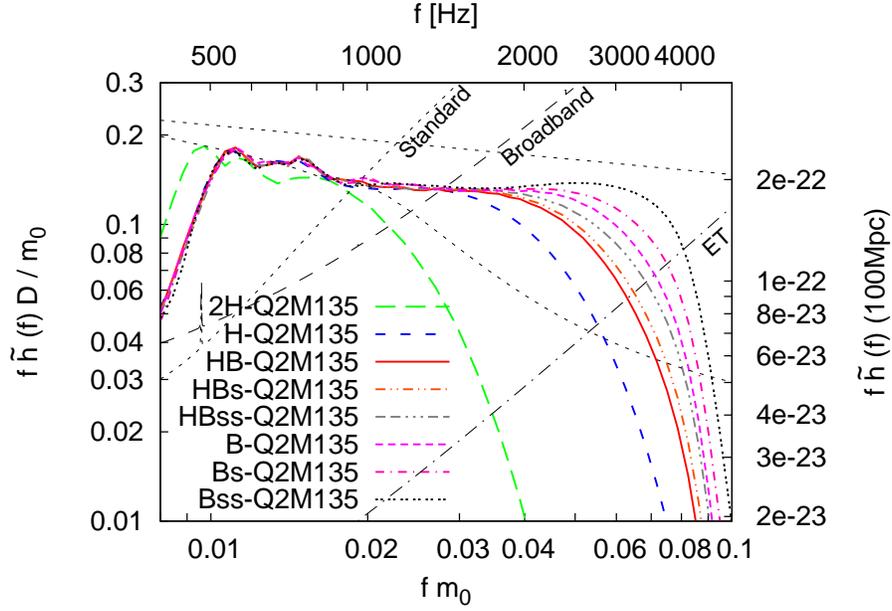} \caption{Spectra of
 gravitational waves from BH-NS binaries for $Q=2$ and $M_{\rm NS} =
 1.35 M_\odot$ with all the EOSs chosen in this paper. The bottom axis
 denotes the normalized dimensionless frequency $f m_0(=G f m_0 /c^3)$
 and the left axis the normalized amplitude $f \tilde{h} (f) D /
 m_0$. The top axis denotes the physical frequency $f$ in Hz and the
 right axis the effective amplitude $f \tilde{h} (f)$ observed at a
 distance of 100 Mpc from the binaries. The short-dashed slope line
 plotted in the upper left region denotes a planned noise curve of the
 Advanced-LIGO \cite{ligo2009} optimized for $1.4 M_\odot$ NS-NS
 inspiral detection (``Standard''), the long-dashed slope line denotes a
 noise curve optimized for the burst detection (``Broadband''), and the
 dot-dashed slope line plotted in the lower right region denotes a
 planned noise curve of the Einstein Telescope (``ET'') \cite{et}. The
 upper transverse dashed line is the spectrum derived by the quadrupole
 formula and the lower one is the spectrum derived by the Taylor-T4
 formula, respectively.}  \label{fig:spec1}
\end{figure*}

\begin{figure}[tbp]
 \begin{tabular}{c}
  \includegraphics[width=90mm,clip]{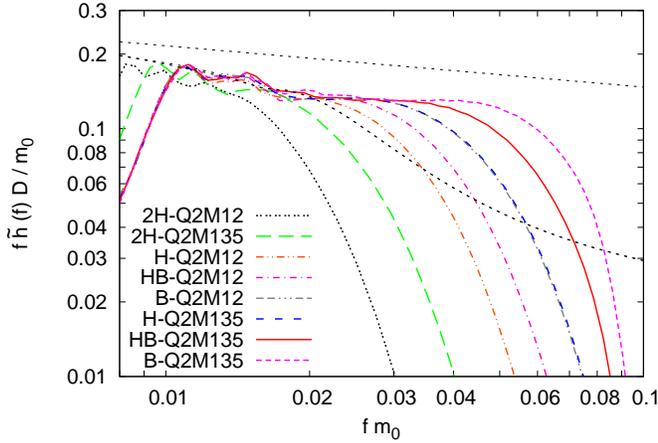}
 \end{tabular}
 \caption{The same as Fig.~\ref{fig:spec1} but for $Q=2$ and for $M_{\rm
 NS}=1.35 M_\odot$ and $1.2 M_\odot$. Only the normalized amplitude $f
 \tilde{h} (f) D / m_0$ as a function of the dimensionless frequency $f
 m_0$ is shown.} \label{fig:spec2}
\end{figure}

\begin{figure*}[tbp]
 \begin{tabular}{cc}
 \includegraphics[width=90mm,clip]{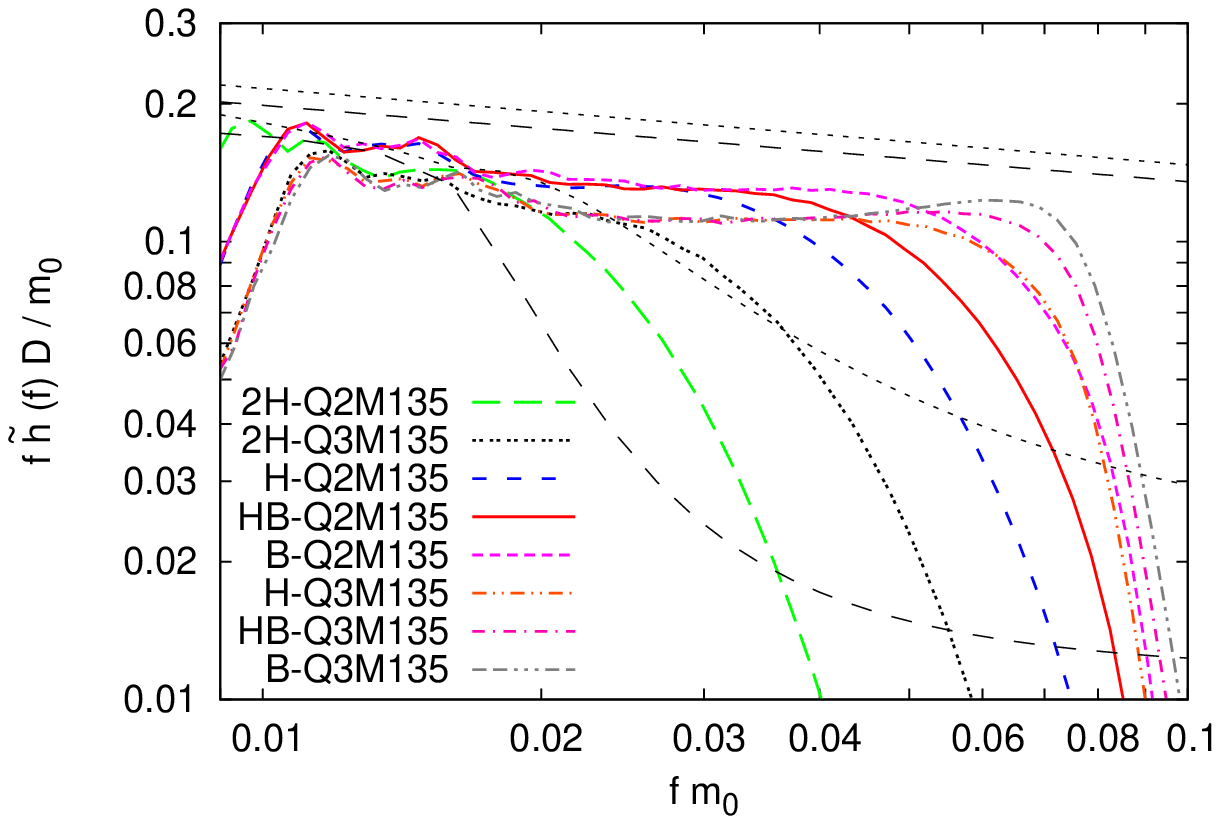} &
 \includegraphics[width=90mm,clip]{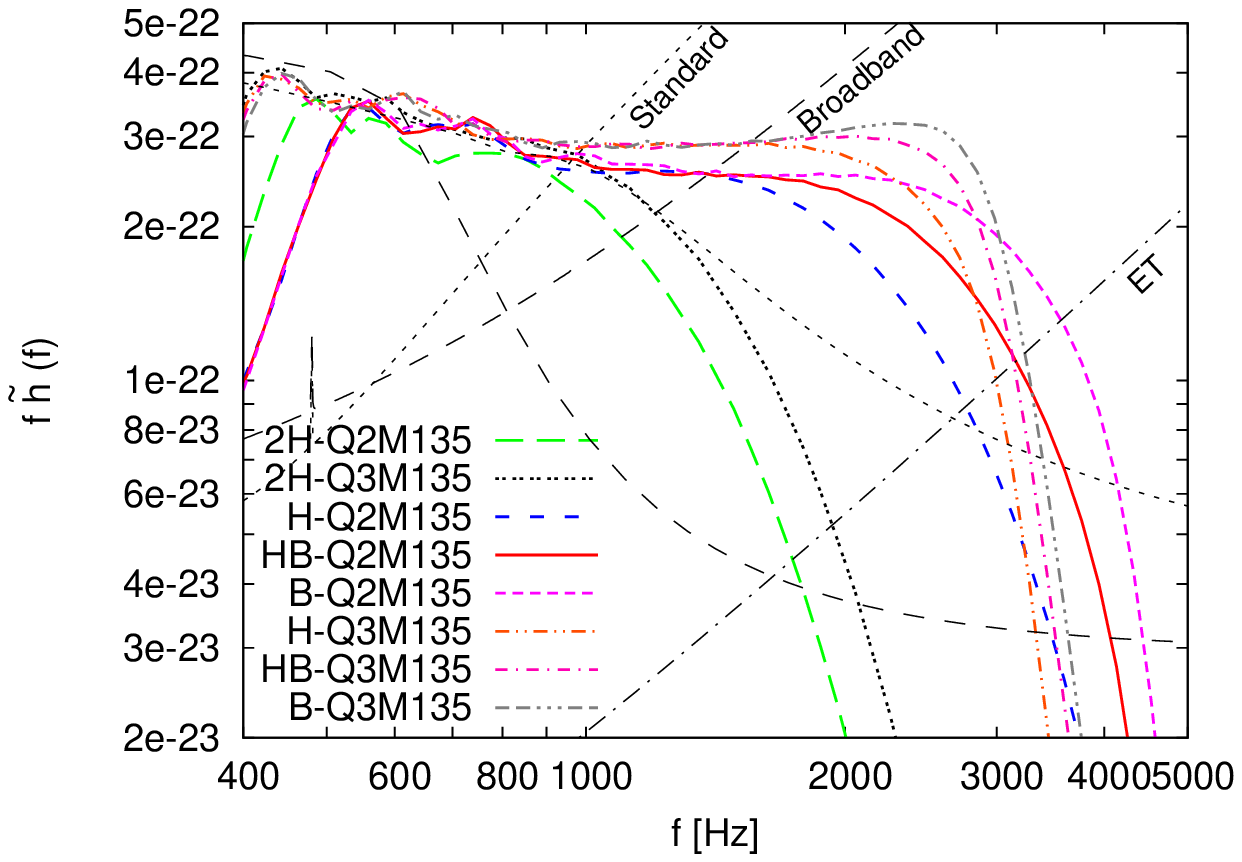}
 \end{tabular}
 \caption{The same as Fig.~\ref{fig:spec1} but for $M_{\rm NS} = 1.35
 M_\odot$ and for $Q = 2$ and 3. The left panel shows the normalized
 amplitude $f \tilde{h} (f) D / m_0$ as a function of the dimensionless
 frequency $f m_0$. The right panel shows the spectra observed at a
 distance of 100 Mpc. The spectra derived from the quadrupole formula
 and the Taylor-T4 formula are plotted by the short-dashed ($Q=2$) and
 long-dashed lines ($Q=3$).} \label{fig:spec3}
\end{figure*}

Characteristic features of a gravitational waveform, such as
characteristic frequencies and their dependence on the EOS, are well
reflected in the Fourier spectrum.
Figures~\ref{fig:spec1}--\ref{fig:spec3} display gravitational-wave
spectra for all the models with the mass ratio $Q=2$ and the models with
the mass ratio $Q=3$ and the NS mass $M_{\rm NS} = 1.35 M_\odot$. As
before \cite{skyt2009}, we define the Fourier spectrum as a sum of each
Fourier component of two independent polarizations of the $(l,m) =
(2,2)$ mode as
\begin{eqnarray}
 \tilde{h} (f) &=& \sqrt{\frac{| \tilde{h}_+ (f) |^2 + |
 \tilde{h}_\times (f) |^2}{2}} , \\
 \tilde{h}_A (f) &=& \int e^{2 \pi i f t} h_A (t) dt ,
\end{eqnarray}
where $A$ denotes two polarization modes, $+$ or $\times$. In
calculating $\tilde{h} (f)$ from a numerically obtained Weyl scalar,
$\Psi_4$, we always omit the unphysical radiation component extracted at
$t_{\rm ret} \lesssim 0$ ms using a step function of retarded time as
the window function so that the spurious radiation component does not
introduce unphysical oscillations in the gravitational-wave
spectrum. The spectrum amplitude for a low-frequency region of $f
\approx \Omega (t_{\rm ret} = 0) / \pi$ changes slightly if we include
the spurious radiation component. However, we believe that our use of
the window function is physically reasonable \footnote{In the previous
work \cite{skyt2009}, we matched the Taylor-T4 waveform with numerical
waveforms in the inspiral phase to compensate lack of numerical
waveforms at low frequencies, and then, performed the Fourier
transformation. In the present work, we do not perform such a procedure,
because that is not necessary to study the dependence of the Fourier
spectrum on EOSs near $f \sim f_{\rm cut}$.}. We always show the
spectrum based on gravitational waves observed along the $z$ axis (axis
perpendicular to the orbital plane), which is the most optimistic
direction for the gravitational-wave detection. (To obtain an averaged
amplitude, we only need to multiply a factor of 0.4; e.g., see
\cite{bcp2007}.) Because the Fourier components of any dimensionless
quantity have the dimension of time, we define a dimensionless effective
amplitude $f \tilde{h} (f)$. In the figure, we plot this quantity
observed at a hypothetical distance 100 Mpc as a function of $f$ (Hz) or
a normalized amplitude $f \tilde{h} (f) D / m_0$ as a function of
dimensionless frequency $f m_0$.

Figure \ref{fig:spec1} plots gravitational-wave spectra for $Q=2$ and
$M_{\rm NS} = 1.35 M_\odot$ with all the EOSs employed in this
paper. For all these models, the total mass is universally $m_0 = 4.05
M_\odot$, and thus, a nondimensional quantity, $f m_0(=G f m_0/c^3)$, is
plotted at the bottom and $f$ in units of Hz is plotted at the top.
Also, a normalized amplitude, $f \tilde{h} (f) D / m_0$, is plotted at
the left side and $f \tilde{h} (f)$ observed at a distance of 100 Mpc
from the binary is at the right side. For comparison, we also plot the
spectra derived from the quadrupole formula (e.g.,
\cite{cutlerflanagan1994}) and the Taylor-T4 formula (dashed curves).

General qualitative features of the gravitational-wave spectrum by BH-NS
binaries are summarized as follows. In the early stage of the inspiral
phase, during which the orbital frequency is $\alt 1$ kHz and the PN
point-particle approximation works well, the gravitational-wave spectrum
is approximately reproduced by the Taylor-T4 formula. For this phase,
the spectrum amplitude of $f \tilde{h}(f)$ decreases as $f^{-n_i}$ where
$n_i=1/6$ for $f \ll 1$ kHz and the value of $n_i$ increases with $f$
for $f \alt 1$ kHz. As the orbital separation decreases, both the
nonlinear effect of general relativity and the finite size effect of the
NS come into play, and as a result, the PN point-particle approximation
breaks down. If the tidal disruption sets in for a relatively large
separation (e.g. for 2H EOS), the amplitude of the gravitational-wave
spectra damps for a low frequency in the middle of the inspiral phase
(before the ISCO is reached). By contrast, if the tidal disruption does
not occur or occurs at a close orbit near the ISCO, the spectrum
amplitude for a high frequency region ($f \agt 1$ kHz) is larger than
that predicted by the Taylor-T4 formula (i.e., the value of $n_i$
decreases). In this case, an inspiral-like motion continues even inside
the ISCO for a dynamical time scale and gravitational waves with a high
amplitude are emitted. As a result, $f \tilde{h}(f)$ becomes a slowly
varying function of $f$ for 1 kHz $\alt f \alt f_{\rm cut}$, where
$f_{\rm cut} \sim 2$--3 kHz is the so-called cutoff frequency which
depends on the binary parameters as well as the EOS of the NSs. (A more
strict definition of $f_{\rm cut}$ will be given below.) A steep damping
of the spectra for $f \agt f_{\rm cut}$ is universally observed, and for
softer EOSs with a smaller radius of NSs, the frequency of $f_{\rm cut}$
is higher. This cutoff frequency is determined by the frequency of
gravitational waves emitted when the NS is tidally disrupted for the
stiff EOSs or by the frequency of a quasinormal mode of the formed BH
for the soft EOSs. Therefore, the cutoff frequency provides potential
information for the EOS through the tidal-disruption event of the NSs,
in particular for the stiff EOSs.

Hereafter, we pay special attention to the cutoff frequency determined
by the tidal disruption. It is natural to expect that the NS compactness
${\cal C}$ primarily determines the cutoff frequency in the combination,
$f_{\rm cut} m_0$, because the orbital angular velocity at the onset of
mass shedding, $R_{\rm shed}$, is written as a function of $Q$ and
${\cal C}$ as \cite{tbfs2007,tbfs2008}
\begin{equation}
 \Omega m_0 \propto \frac{{\cal C}^{3/2} (1+Q)^{3/2}}{\sqrt{Q}} .
 \label{eq:fcut}
\end{equation}
In fact, we found a qualitative correlation between ${\cal C}$ and
$f_{\rm cut} m_0$ in the previous work \cite{skyt2009}. To reconfirm
this, we first plot gravitational-wave spectra [$f \tilde{h} (f) D /
m_0$ as a function of $f m_0$] for $Q=2$ with the different NS mass
$M_{\rm NS} = 1.35 M_\odot$ and $1.2 M_\odot$ in Fig.~\ref{fig:spec2}.
This indeed shows $f_{\rm cut} m_0$ increases monotonically with ${\cal
C}$ irrespective of the NS mass for the given mass ratio.

\begin{figure}[tbp]
 \includegraphics[width=90mm,clip]{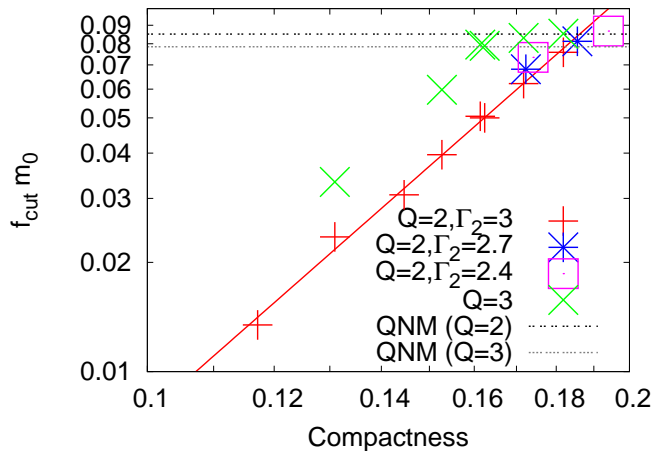} \caption{$f_{\rm cut} m_0$
 as a function of ${\cal C}$ in logarithmic scales. The solid line is
 obtained by a linear fitting of the data for $Q=2$ and $\Gamma_2 =
 3$. The short-dashed and long-dashed lines show approximate frequencies
 of quasinormal mode of the remnant BH for $Q=2$ and $Q=3$,
 respectively.}  \label{fig:fcut}
\end{figure}

Figure \ref{fig:spec3} shows the gravitational-wave spectrum for $M_{\rm
NS}=1.35M_{\odot}$ and for $Q=2$ and 3. The left panel plots $f
\tilde{h} (f) D / m_0$ as a function of $f m_0$ and the right panel $f
\tilde h(f)$ as a function of $f$ for $D=100$ Mpc. This shows that
dependence of $f_{\rm cut} m_0$ on ${\cal C}$ for $Q=3$ is weaker than
for $Q=2$. The reason for this is that the tidal effect is weaker for
$Q=3$, as discussed in Sec.~\ref{subsec:result_disk}. (As later shown in
Fig.~\ref{fig:fcut}, $f_{\rm cut}$ for models H-Q3M135, HB-Q3M135, and
B-Q3M135 are not determined by the orbital frequency at tidal disruption
but by the quasinormal-mode frequency of the remnant BH, which sets an
approximate upper limit on the frequency of gravitational waves emitted
in the merger.) Hence, the information of the EOS is not encoded in
gravitational waves for $Q=3$ as strongly as for $Q=2$. The right panel
shows that $f_{\rm cut}$ is between $\sim 1$ and 3 kHz depending weakly
on the value of $Q$.

To analyze the cutoff frequency quantitatively and to strictly study its
dependence on EOSs, we perform a systematic fitting procedure. As
in~\cite{skyt2009}, we fit all the spectra by a function with seven free
parameters
\begin{eqnarray}
 \tilde{h}_{\rm fit} (f) &=& \tilde{h}_{\rm 3PN} (f) e^{-(f / f_{\rm
 ins})^{\sigma_{\rm ins}}} \nonumber \\
 &+& \frac{A m_0}{Df} e^{-(f / f_{\rm cut})^{\sigma_{\rm cut}}} [1 -
 e^{-(f / f_{\rm ins2})^{\sigma_{\rm ins2}}}] , \label{eq:fit}
\end{eqnarray}
where $\tilde{h}_{\rm 3PN} (f)$ is the Fourier spectrum calculated by
the Taylor-T4 formula and $f_{\rm ins}$, $f_{\rm ins2}$, $f_{\rm cut}$,
$\sigma_{\rm ins}$, $\sigma_{\rm ins2}$, $\sigma_{\rm cut}$, and $A$ are
free parameters. The first and second terms of Eq.~(\ref{eq:fit}) denote
the spectrum models for the inspiral and merger phases, respectively. We
determine these free parameters by searching the minimum for a weighted
norm defined by
\begin{equation}
 \sum_i \left\{ [ f_i \tilde{h} (f_i) - f_i \tilde{h}_{\rm fit} (f_i) ]
	 f_i^{1/3} \right\}^2 ,
 \label{eq:norm}
\end{equation}
where $i$ denotes the data point for the spectrum. In the previous work
\cite{skyt2009}, we fix $\sigma_{\rm ins} = 3.5$ and $\sigma_{\rm ins2}
= 5$ to save the computational costs. Here, these are chosen to be free
parameters to reproduce a more consistent spectrum with the original
one.

Among these seven free parameters, we focus on $f_{\rm cut}$ because it
depends most strongly on the compactness ${\cal C}$ and the EOS of the
NS. Figure~\ref{fig:fcut} plots $f_{\rm cut} m_0$, obtained in this
fitting procedure, as a function of ${\cal C}$. Also the typical
quasinormal-mode frequencies, $f_{\rm QNM}$, of the remnant BH
calculated in Sec.~\ref{subsec:result_disk} are plotted by the two
horizontal lines, which show that the values of $f_{\rm cut} m_0$ for
models H-Q3M135, HB-Q3M135, and B-Q3M135 agree approximately with
$f_{\rm QNM}$ and indicates that $f_{\rm cut}$ for these models are
irrelevant to the tidal disruption. For $Q=3$, $f_{\rm cut}m_0$ depends
clearly on the EOS only for ${\cal C} \alt 0.16$. This agrees with the
result with $\Gamma=2$ polytropic EOS~\cite{skyt2009}. By contrast,
$f_{\rm cut} m_0$ for $Q=2$ depends strongly on the NS compactness
${\cal C}$ irrespective of $M_{\rm NS}$ not only for the piecewise
polytropic EOS but also for $\Gamma=2$ polytrope \cite{skyt2009}. The
solid line in Fig.~\ref{fig:fcut} is the linear fitting of $\ln(f_{\rm
cut} m_0)$ as a function of $\ln({\cal C})$ for $Q=2$ and for the
piecewise polytrope with $\Gamma_2 = 3$, and denoted by a
well-approximated relation
\begin{equation}
 \ln ( f_{\rm cut} m_0 ) = (3.87 \pm 0.12) \ln {\cal C} + (4.03 \pm
  0.22). \label{eq:cal4}
\end{equation}
Thus, $f_{\rm cut} m_0$ is approximately proportional to ${\cal
C}^{3.9}$ (for $Q=3$ and $\Gamma_2 = 3$, $f_{\rm cut} m_0$ also appears
to be proportional to ${\cal C}^4$, although the number of data points
is small and thus this is not conclusive). This is a note-worthy point
because the power of ${\cal C}$ is much larger than 1.5, which is
expected from the relation for the mass-shedding limit,
Eq.~(\ref{eq:fcut}). Qualitatively, this increase in the power is
natural because the duration of a NS for the survival against tidal
disruption after the onset of mass shedding is in general longer for a
more compact NS due to a stronger central condensation of the
mass. Equation (\ref{eq:cal4}) implies that the ratio $f_{\rm
cut}/f_{\rm shed}~(>1)$, where $f_{\rm shed}$ is the frequency of
gravitational waves at the onset of mass shedding, is larger for the
larger values of ${\cal C}$. This is the preferable feature, for an
observer of gravitational waves from BH-NS binaries who tries to
constrain the EOS of the NSs, because the dependence of $f_{\rm cut}
m_0$ on the EOS is enhanced.

Comparison of the values of $f_{\rm cut} m_0$ for models HB-Q2M135
($\Gamma_2 = 3.0$ and ${\cal C} = 0.1718$), HBs-Q2M135 ($\Gamma_2 = 2.7$
and ${\cal C} = 0.1723$), and HBss-Q2M135 ($\Gamma_2 = 2.4$ and ${\cal
C} = 0.1741$), for which the value of ${\cal C}$ is approximately
identical, shows that $f_{\rm cut} m_0$ depends also on the adiabatic
index of EOS in the central region, $\Gamma_2$. The reason for this is
that the NSs with smaller values of $\Gamma_2$ (but with the same value
of ${\cal C}$) have more centrally condensed density profile as can be
seen from the value of $\rho_{\rm max}$ in Table~\ref{table:model}, and
hence, are less subject to tidal disruption ($f_{\rm cut}m_0$ becomes
larger). Quantitatively, the value of $f_{\rm cut}m_0$ increases by
$\sim 20$\%, when the value of $\Gamma_2$ is varied from 3 to 2.4. This
result suggests that it may be possible to constrain not only the
compactness of a NS but also its density profile and detailed function
of $P(\rho)$ for the EOS, if gravitational waves emitted during the
merger of low-mass BH-NS binaries are detected.

\subsection{Properties of the disk} \label{subsec:result_disk}

\begin{figure*}[tbp]
 \begin{tabular}{cc}
 \includegraphics[width=90mm,clip]{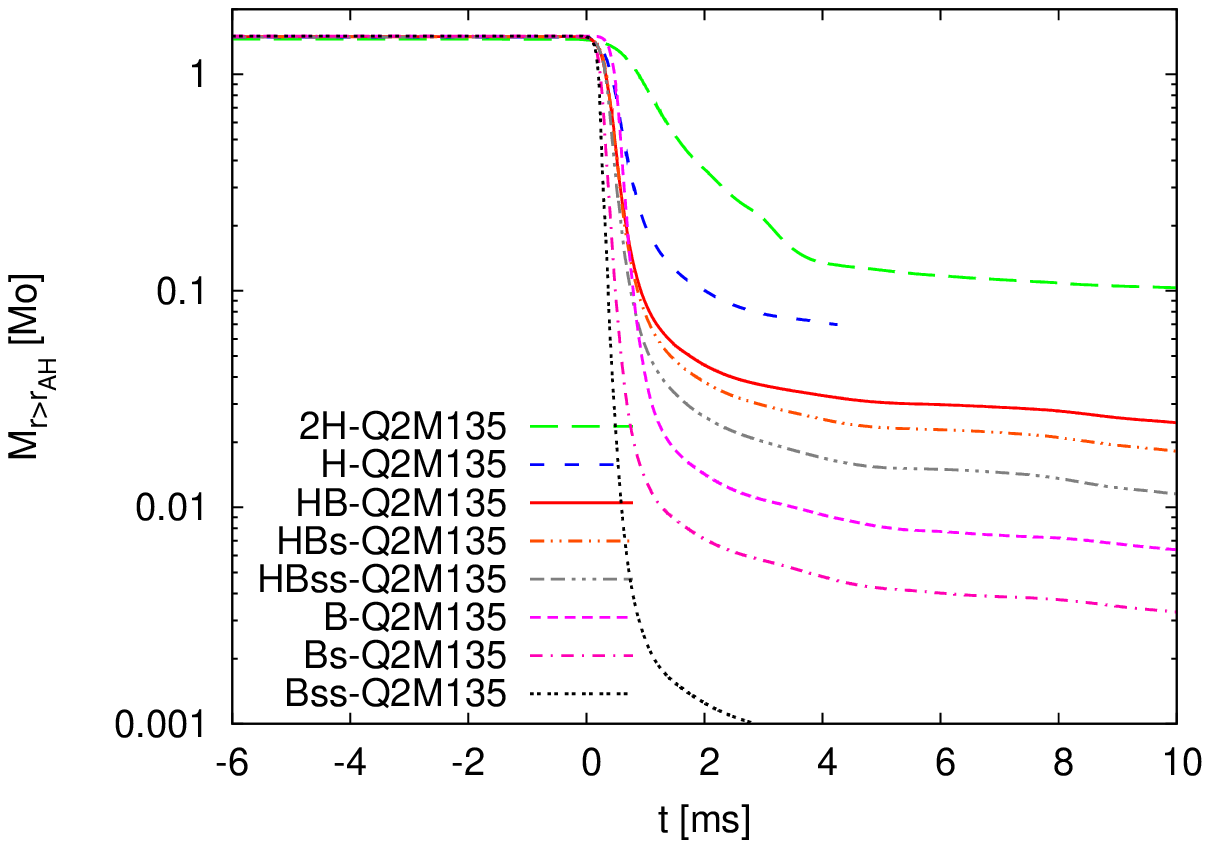} &
 \includegraphics[width=90mm,clip]{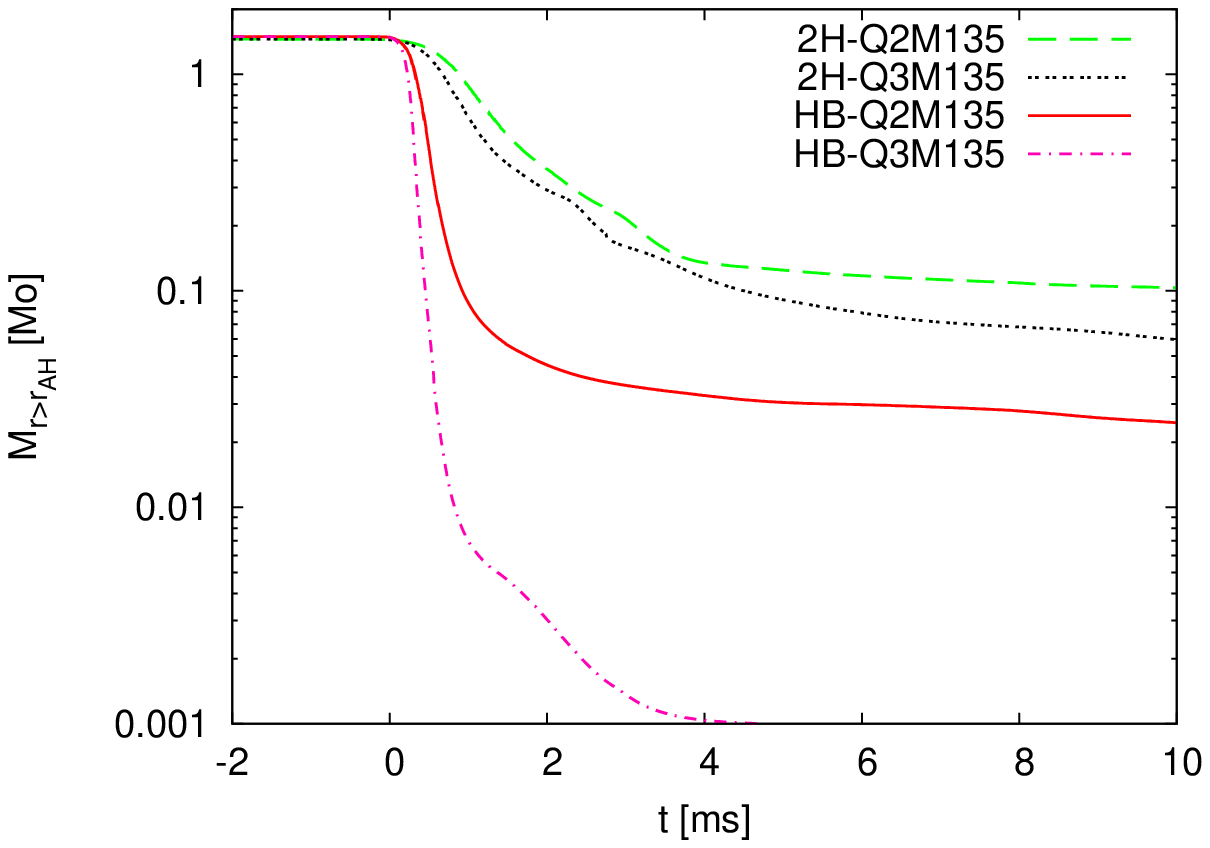} \\
 \includegraphics[width=90mm,clip]{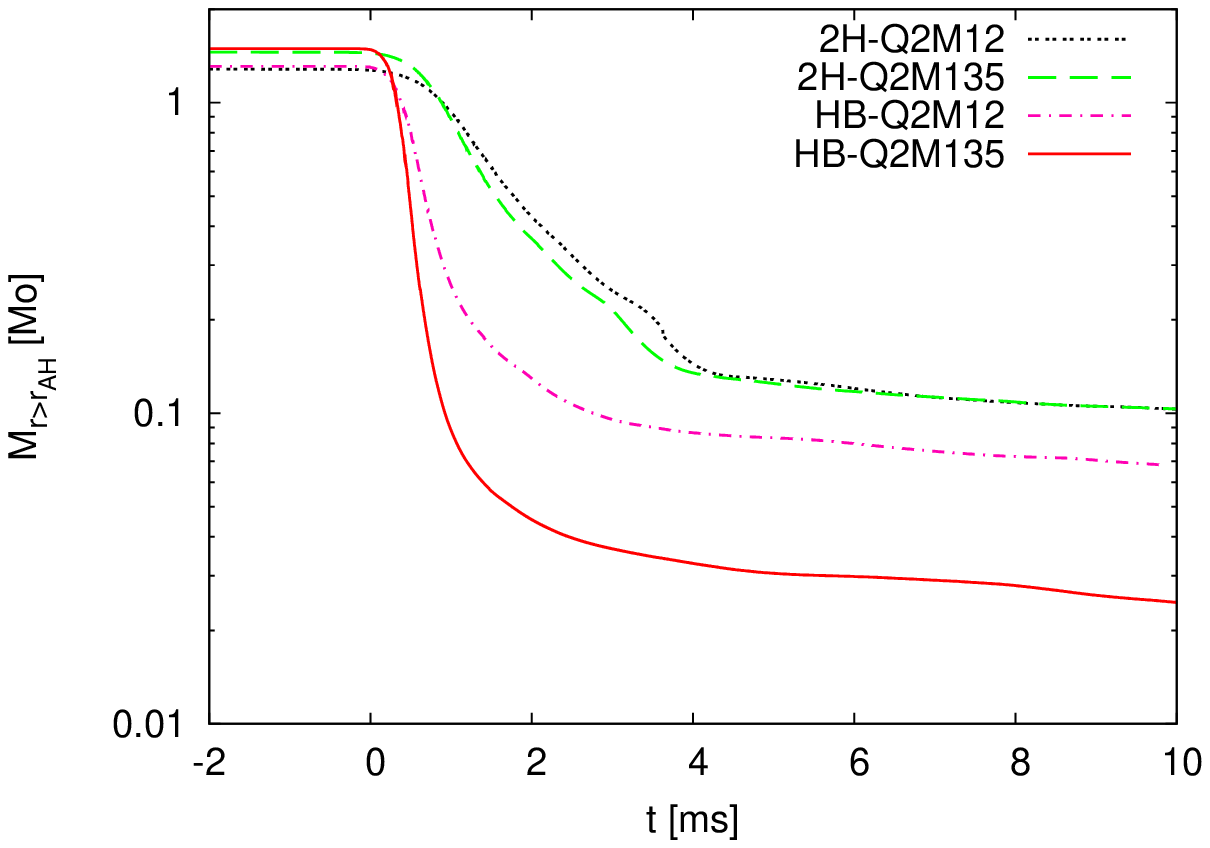} &
 \includegraphics[width=90mm,clip]{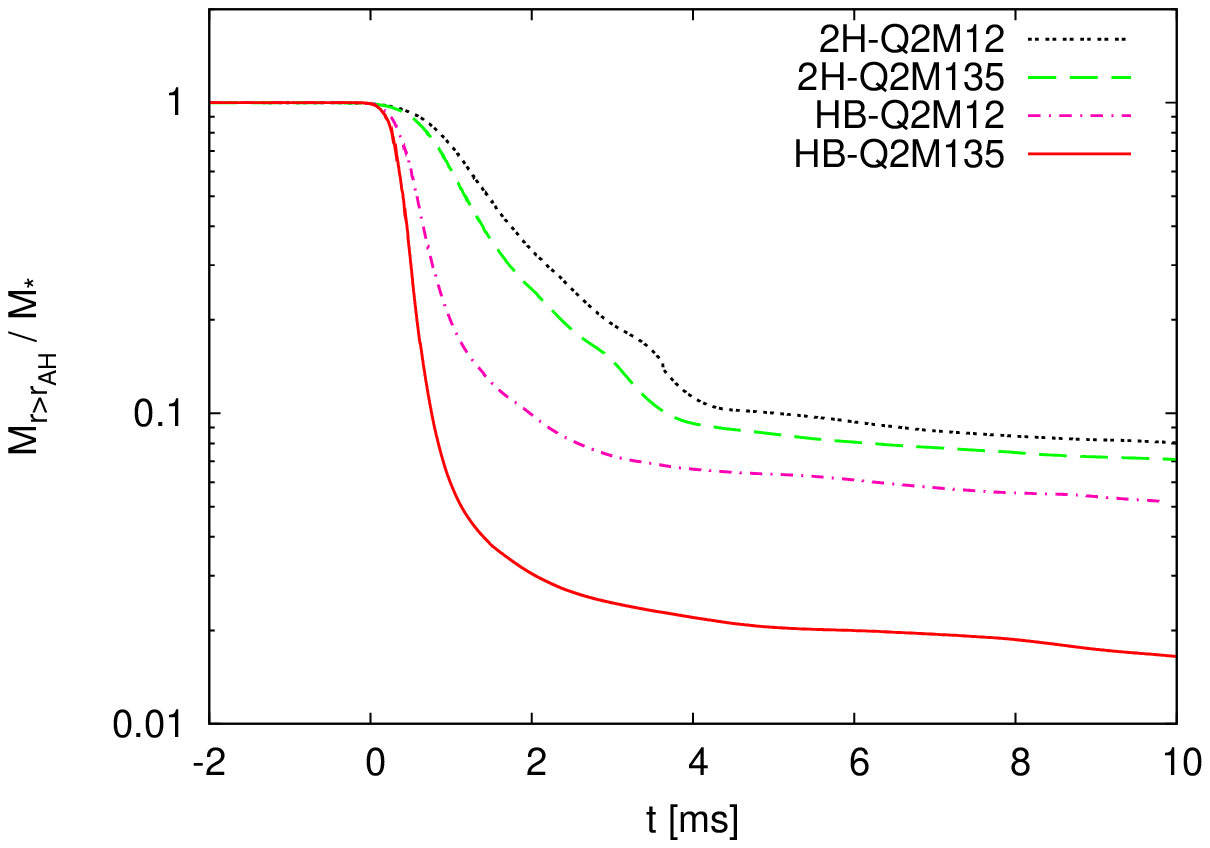} \end{tabular}
 \caption{Evolution of the rest mass of the material located outside the
 apparent horizon, $M_{r > r_{\rm AH}}$, with an appropriate time shift;
 in these plots, the time at the onset of the merger is taken as the
 time origin. The top-left panel shows the results for models with $Q=2$
 and $M_{\rm NS} = 1.35 M_\odot$ for all the EOSs employed in this paper
 (we note that the simulation for model H-Q2M135 unfortunately
 terminated in the middle of the accretion process due to the electrical
 outage at our institute). The top-right panel shows the results for
 selected models with $M_{\rm NS} = 1.35 M_\odot$ but with different
 values of $Q$. The bottom-left panel shows the results for selected
 models with $Q=2$ but with the different NS mass $M_{\rm NS}$. The
 bottom-right panel is the same as the bottom-left panel except for the
 normalization of the mass, with respect to the initial rest mass
 $M_*$.} \label{fig:disk}
\end{figure*}

\begin{figure}[tbp]
 \includegraphics[width=90mm,clip]{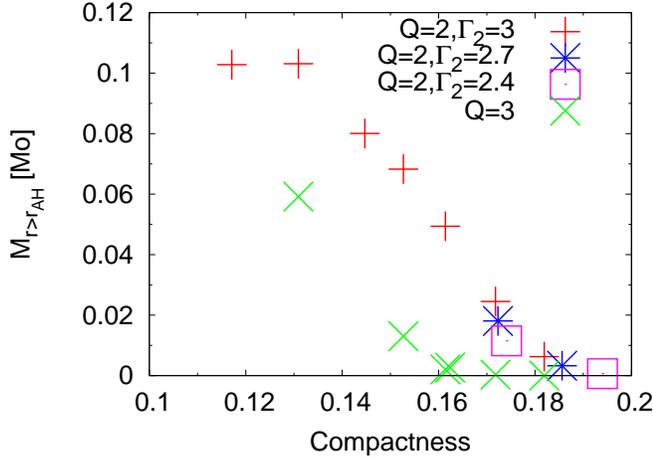} \caption{Disk mass $M_{r >
 r_{\rm AH}}$ at $t-t_{\rm merger} \approx 10$ ms as a function of the
 NS compactness ${\cal C}$. Note that the disk mass for model Bss-Q2M135
 is estimated at the end of the simulations, 4.83 ms, because it became
 already very small at that time and we stopped the simulation.}
 \label{fig:mtoc}
\end{figure}

\begin{figure}[tbp]
 \includegraphics[width=90mm,clip]{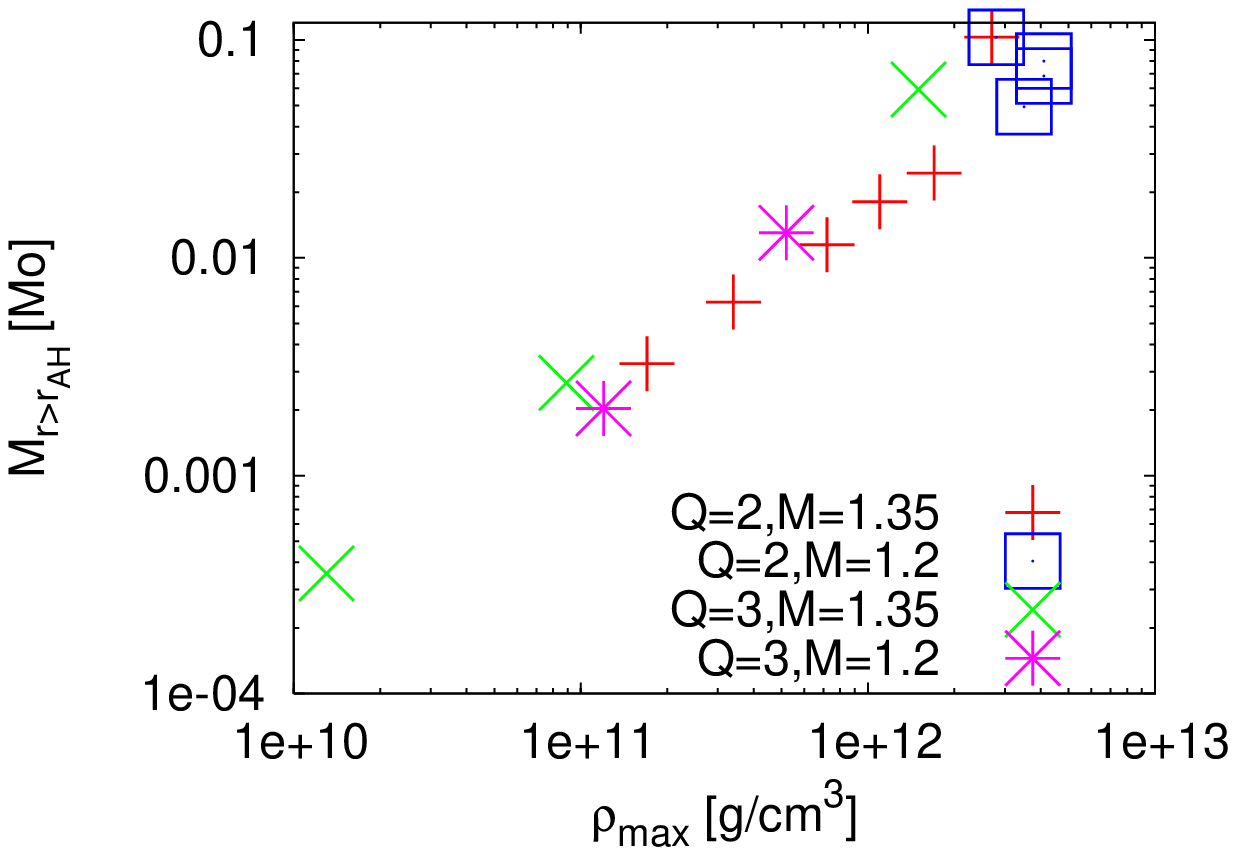} \caption{Relation between
 disk mass $M_{r > r_{\rm AH}}$ and the maximum density, $\rho_{\rm
 max}$, estimated at $t-t_{\rm merger} \approx 10$ ms. The maximum
 density oscillates with time even in the quasistationary phase, and we
 here plot a value averaged in one oscillation period.}
 \label{fig:rhotom}
\end{figure}

If a NS is tidally disrupted before it is swallowed by the companion BH,
a disk may be formed around the BH. Figure \ref{fig:disk} plots the time
evolution of the rest mass of the material located outside the apparent
horizon $M_{r>r_{\rm AH}}$ defined by Eq.~(\ref{eq:diskmass}). This
shows that most of the material is swallowed by the BH soon after the
onset of the merger (or tidal disruption) within $\sim 1$ ms, but
1\%--10\% of total rest mass survives around the BH to be a disk, if the
tidal disruption occurs (see Table~\ref{table:remnant} which lists the
numerical results of $M_{r > r_{\rm AH}}$ at the end of the simulations
for all the models).

To clarify that the disk will survive for a time duration longer than
the dynamical time scale of the system, we estimate an accretion time
scale. Figure \ref{fig:disk} shows that for $t-t_{\rm merger} \agt 5$
ms, $M_{r > r_{\rm AH}}$ for each model behaves approximately as $C \exp
( - t / t_{\rm d} )$ where $C$ is a constant and $t_{\rm d}$ is the
accretion time scale which we determine by a least-square fitting of
$M_{r > r_{\rm AH}}(t)$ at $t-t_{\rm merger} \approx 10$ ms. The fourth
column of Table~\ref{table:remnant} lists the numerical results. It is
found that the accretion time scale is always longer than the dynamical
time scale of the remnant disk $\sim 10$ ms, and hence, we conclude that
the BH-NS merger always forms a long-lived accretion disk, if the disk
is formed \footnote{Note that in the presence of magnetic fields,
angular momentum transport by them works efficiently, and thus, the
accretion time scale may be shorter than that presented here in
reality.}.

Figure~\ref{fig:mtoc} plots the values of $M_{r > r_{\rm AH}}$ estimated
at $t-t_{\rm merger} \approx 10$ ms as a function of the NS compactness
${\cal C}$ and clarifies the dependence of the disk mass on the EOS. The
disk mass for model Bss-Q2M135 is estimated at the end of the
simulations, 4.83 ms, because it already became very small at that time
and we stopped the simulation. (We also note that the result for model
H-Q2M135 is not included in Fig.~\ref{fig:mtoc}, because the simulation
for this model unfortunately terminated just after the disk formation
due to the electrical outage at our institute.) This figure summarizes
the key features as follows: (i) for a given mass ratio and for a given
adiabatic index of the core, $\Gamma_2$, the disk mass decreases
monotonically with the increase of ${\cal C}$ for $M_{r>r_{\rm AH}}
\lesssim 0.1 M_\odot$; (ii) for a given mass ratio and for a given NS
compactness, the disk mass increases slightly with the increase of
$\Gamma_2$; and (iii) the disk mass is highly sensitive to the mass
ratio of the binary, $Q$, for a given mass and EOS of the NS. In the
following, we observe these features from Fig.~\ref{fig:disk} in detail.

The top left panel of Fig.~\ref{fig:disk} plots the disk-mass evolution
for binaries with $Q=2$, $M_{\rm NS} = 1.35 M_\odot$ and for all the
EOSs employed in this paper. For this sample, ${\cal C} \propto R_{\rm
NS}^{-1}$ since $M_{\rm NS}$ is identical, and we find that the disk
mass increases monotonically with ${\cal C}^{-1}$ (see
Table~\ref{table:EOS} for ${\cal C}$ of each model); the disk mass is
larger for a model for which the tidal disruption occurs at a more
distant orbit (i.e., for smaller value of $f_{\rm cut}$,
cf. Fig.~\ref{fig:fcut}). This is quite reasonable because the earlier
onset of tidal disruption helps more materials to remain outside the
ISCO of the BH.

Comparison of the results for models HB-Q2M135 ($\Gamma_2 = 3.0$ and
${\cal C} = 0.1718$), HBs-Q2M135 ($\Gamma_2 = 2.7$ and ${\cal C} =
0.1723$), and HBss-Q2M135 ($\Gamma_2 = 2.4$ and ${\cal C} = 0.1741$)
indicates that the disk mass depends not only on the compactness of the
NS but also on the adiabatic index of the core, $\Gamma_2$; a higher
value of $\Gamma_2$ is preferable for forming a massive disk. This
dependence on $\Gamma_2$ is consistent with the result reported in
\cite{ism2005}; the NS with a larger value of the adiabatic index is
more subject to tidal disruption (tidal disruption occurs for more
distant orbital separation). The physical interpretation for this result
is that the degree of central mass concentration for NSs of larger
values of the adiabatic index is weaker, helping earlier tidal
disruption (in other words, we may say that the tidal Love number or
deformability is larger for the larger value of $\Gamma_2$).

The top right panel of Fig.~\ref{fig:disk} plots the disk-mass evolution
for the NS with the same mass ($M_{\rm NS}=1.35M_{\odot}$) but with
different mass ratio $Q=2$ and $Q=3$ and with HB and 2H EOSs. This,
together with Fig.~\ref{fig:mtoc}, shows that the disk mass depends
strongly on the mass ratio, in particular for the soft EOS. The reason
for this is simply that the NS is less subject to tidal disruption for a
larger BH mass (i.e., for weaker tidal force near the ISCO). The present
result suggests that the disk mass is much smaller than $0.01M_{\odot}$
for BH-NS binaries with the typical NS mass of $M_{\rm NS} = 1.2$--$1.35
M_\odot$ and ${\cal C} \gtrsim 0.16$, if the BH is nonspinning and
$M_{\rm BH} \gtrsim 4 M_\odot$. Only for the case ${\cal C} \lesssim
0.16$, the disk mass may be larger than $0.01M_{\odot}$ even with a
high-mass BH companion. This conclusion is in agreement with the
previous studies \cite{skyt2009,elsb2009,dfkpst2008}.

The two bottom panels of Fig.~\ref{fig:disk} compare the disk-mass
evolution for models 2H-Q2M12 and 2H-Q2M135 and for models HB-Q2M12 and
HB-Q2M135. In the left panel we plot the disk mass in units of $M_\odot$
while the bottom right panel plots the disk mass in units of $M_*$. We
note that the NS radius depends weakly on the mass for $1.2M_{\odot}
\leq M_{\rm NS} \leq 1.35M_{\odot}$ for both EOSs, and also the mass
ratio $Q$ is identical for these models. Nevertheless, the disk mass
depends strongly on the NS mass except for models with stiff 2H EOS as
seen in Table \ref{table:remnant}; it decreases with the increase of
$M_{\rm NS}$. Thus, not the NS radius $R_{\rm NS}$ but ${\cal C}$ is the
key parameter for determining the disk mass.

Before closing this section, we summarize several key properties of the
remnant disk. Figure~\ref{fig:rhotom} plots the relation between $M_{r >
r_{\rm AH}}$ and the maximal rest-mass density $\rho_{\rm max}$ of the
remnant disk estimated at $t-t_{\rm merger} \approx 10$ ms. This clearly
shows a strong correlation between two quantities. The value of $M_{r >
r_{\rm AH}}$ increases approximately linearly with $\rho_{\rm max}$ for
$M_{r > r_{\rm AH}}\alt 0.1M_{\odot}$, and for $M_{r > r_{\rm AH}} \geq
0.01 M_{\odot}$, $\rho_{\rm max}$ is larger than $4 \times 10^{11}~{\rm
g/cm^3}$. Because the density is high and the temperature should be also
high enough ($\sim 10$ MeV if viscous effects or magnetohydrodynamic
effects are taken into account \cite{srj2006,lrp2005,sst2007}),
neutrinos will be copiously produced in such a disk in reality. Because
of the high density and temperature, the cross section to the nucleon
will be large enough ($\sim 10^{-41}~{\rm cm}^2$) to trap neutrinos
inside the disk of nucleon number density $n_n=\rho/m_n \agt
10^{35}~{\rm cm}^{-3}$ where $m_n$ is nucleon mass $1.66 \times
10^{-24}$ g \cite{npk2001,dpn2002,kohrimineshige2002}. Therefore, a
neutrino-dominated accretion disk will be always produced, if BH-NS
binaries result in a system composed of the BH and surrounding disk of
mass larger than $0.01M_{\odot}$.

\subsection{Properties of the remnant BH} \label{subsec:result_BH}

\begin{table*}
 \caption{Several key quantities for the merger remnants. All the
 quantities are estimated when we stopped the simulation at $t=t_{\rm
 end}$. $t_{\rm merger}$ denotes the time of the merger and the time
 duration for following the disk evolution, $t_{\rm end} - t_{\rm
 merger}$, is shown in the second column. $M_{r>r_{\rm AH}}$ is the rest
 mass of the disk surrounding the BH; because the accretion is still
 ongoing at the end of simulations due to the hydrodynamic angular
 momentum transport process, the values listed give only an approximate
 mass of the long-lived accretion disk (especially for model H-Q2M135;
 see Sec.~\ref{subsec:result_disk}), which survives for a time scale
 longer than the dynamical time scale $\sim 10$ ms. $t_{\rm d}$ is the
 approximate accretion time scale estimated around $\sim 10$ ms after
 the merger, which we show only for the case $M_{r > r_{\rm AH}} \gtrsim
 0.001 M_\odot$. $C_e$ and $C_p$ are the circumferential radii of the
 apparent horizon along the equatorial plane and meridional plane,
 respectively, and $C_e/4\pi$ is the approximate mass of the remnant BH.
 $M_{\rm irr}$ is the irreducible mass of the remnant BH. $a$ is the
 nondimensional spin parameter of the remnant BH estimated from
 $C_p/C_e$.}
 \begin{tabular}{cc|cccccc} \hline
  Model & $t_{\rm end} - t_{\rm merger}$ (ms) & $M_{r>r_{\rm AH}}
  [M_\odot]$ & $t_{\rm d}$ (ms) & $C_e / 4 \pi M_0$ & $M_{\rm irr} /
  M_0$ & $C_p / C_e$ & ~~~~$a$~~~~ \\ \hline
  \hline
  2H-Q2M135 & 12.6 & 0.097 & 30 & 0.947 & 0.891 & 0.913 & 0.64 \\
  H-Q2M135 & 4.22 & 0.070 & $\cdots$ & 0.968 & 0.905 & 0.905 & 0.66 \\
  HB-Q2M135 & 11.9 & 0.022 & 18 & 0.978 & 0.912 & 0.902 & 0.67 \\
  HBs-Q2M135 & 13.7 & 0.015 & 15 & 0.980 & 0.914 & 0.902 & 0.67 \\
  HBss-Q2M135 & 13.5 & 0.0093 & 12 & 0.981 & 0.916 & 0.903 & 0.67 \\
  B-Q2M135 & 20.0 & 0.0045 & 20 & 0.980 & 0.917 & 0.905 & 0.66 \\
  Bs-Q2M135 & 12.5 & 0.0029 & 19 & 0.979 & 0.917 & 0.906 & 0.66 \\
  Bss-Q2M135 & 4.83 & $6 \times 10^{-4}$ & $\cdots$ & 0.977 & 0.917 &
  0.910 & 0.65 \\ \hline
  2H-Q3M135 & 20.4 & 0.044 & 19 & 0.961 & 0.925 & 0.944 & 0.52 \\
  H-Q3M135 & 21.4 & 0.0015 & 11 & 0.984 & 0.942 & 0.937 & 0.55 \\
  HB-Q3M135 & 16.5 & $2 \times 10^{-4}$ & $\cdots$ & 0.983 & 0.942 &
  0.937 & 0.55 \\
  B-Q3M135 & 15.1 & $< 10^{-5}$ & $\cdots$ & 0.982 & 0.941 & 0.939 &
  0.55 \\
  \hline
  2H-Q2M12 & 12.6 & 0.097 & 31 & 0.939 & 0.886 & 0.918 & 0.62 \\
  H-Q2M12 & 11.9 & 0.077 & 30 & 0.959 & 0.899 & 0.907 & 0.66 \\
  HB-Q2M12 & 9.72 & 0.068 & 30 & 0.964 & 0.902 & 0.906 & 0.66 \\
  B-Q2M12 & 15.4 & 0.043 & 24 & 0.972 & 0.908 & 0.903 & 0.67 \\ \hline
  HB-Q3M12 & 12.2 & 0.011 & 15 & 0.979 & 0.937 & 0.937 & 0.55 \\
  B-Q3M12 & 10.6 & 0.0019 & 17 & 0.982 & 0.940 & 0.936 & 0.56 \\ \hline
 \end{tabular}
 \label{table:remnant}
\end{table*}

Table \ref{table:remnant} shows several quantities associated with the
remnant BH such as the mass and spin, in addition to the disk mass.
Unlike the disk mass, the mass and spin of the remnant BH depend weakly
on the EOS of the NS. For given values of $Q$ and $M_{\rm NS}$, the BH
mass tends to be slightly smaller for stiffer EOS, primarily because the
fraction of the NS mass swallowed by the BH is smaller (the disk mass is
larger). The spin does not show such a clear dependence. The reason is
that the spin angular momentum of the remnant BH is affected by two
competing processes; one is the orbital angular momentum dissipation due
to gravitational radiation reaction and the other is the distribution of
the angular momentum to the disk surrounding the BH. The former
dissipation effect is important for the case in which the NS is compact
and the tidal disruption does not occur as stated in
Sec.~\ref{subsec:result_waveform}. By contrast, the latter effect is
more important for the case in which the NS is less compact and the
tidal disruption occurs in the relatively early stage of the inspiral
phase. Although the relation $\Delta J > J_{r > r_{\rm AH}}$ (where
$J_{r > r_{\rm AH}}$ denotes the angular momentum of disk) always holds
for all the models, we may also have the relation $\Delta E \agt M_{r >
r_{\rm AH}}$. As a result, a nondimensional spin parameter, which may be
approximately estimated by
\begin{eqnarray}
 &&{{\rm Spin~angular~momentum} \over {\rm (Mass)}^2} \nonumber \\
 &&\approx {(J_0-\Delta J -J_{r > r_{\rm AH}}) \over (M_0-\Delta E-M_{r
  > r_{\rm AH}})^2} ,
\end{eqnarray}
does not depend simply on the EOS.

The spin of the remnant BH is primarily determined by the mass ratio,
$Q$; $a = 0.66 \pm 0.03$ for $Q=2$ and $a = 0.54 \pm 0.02$ for $Q=3$
(here $\pm$ signs do not imply the error bars but signify differences
due to the EOS). Thus, the spin parameter is modified by the EOS only in
$\pm 5\%$.

From the typical value of the spin parameter $a$ and mass of the remnant
BH $M_{\rm BH,f}$, we estimate typical quasinormal-mode frequencies
$f_{\rm QNM}$ of the remnant BH by the latest fitting formula
\cite{bcs2009}
\begin{equation}
 f_{\rm QNM} M_{\rm BH,f} \approx \frac{1}{2\pi} [ 1.5251 - 1.1568 ( 1 -
 a )^{0.1292} ] .
\end{equation}
Then, $f_{\rm QNM} \approx 0.083 / M_{\rm BH,f}$ for $Q=2$ and $\approx
0.076 / M_{\rm BH,f}$ for $Q=3$, respectively. Assuming that $C_e /
4\pi$ gives an approximate value of $M_{\rm BH,f}$ as described in
Sec.~\ref{subsec:simulation_diagnostics}, these values are in good
agreement with the ringdown part of gravitational waves for models
Bss-Q2M135 and B-Q3M135, for which the disk masses are negligibly small,
respectively. We note that this estimation is valid only when the
quasinormal modes of the BH are excited, and actually the tidal
disruption of the NS often suppresses the quasinormal-mode excitation as
can be seen in Figs.~\ref{fig:GW1} and \ref{fig:GW2}, in particular, for
the stiff EOS such as 2H.

\section{Summary} \label{sec:summary}

We performed numerical simulations for the merger of nonspinning BH-NS
binaries using an AMR code {\small SACRA} with eight piecewise
polytropic EOSs. In this work, we employed the EOSs with two free
parameters which determine the core EOSs. The crust EOS was fixed
whereas the core EOS was varied for a wide range, to investigate the
dependence of gravitational waveforms, merger process, and merger
remnant on the core EOS. We focused, in particular, on the case in which
the NS is tidally disrupted by the companion BH, choosing relatively low
values of mass ratio as $Q=2$ and 3 as well as low masses for the NS as
$M_{\rm NS} = 1.2$ and $1.35M_\odot$. By preparing the initial condition
with a distant orbit and a small eccentricity, we always tracked
$\gtrsim 5$ quasicircular orbits in the inspiral phase and studied the
merger phase with a realistic setting. We also evolved the merger
remnant (BH-disk system) until they settled to a quasistationary state.

A wide variety of simulations were systematically performed to
investigate the dependence of the tidal-disruption process and resulting
gravitational waveforms on the EOS. For the case in which the tidal
disruption occurs before the orbit reaches the ISCO, the
gravitational-wave amplitude decreases quickly at its onset and the
emission of ringdown gravitational waves associated with the quasinormal
mode of the remnant BH is suppressed. Only in the BH-NS binaries with
low values of mass ratio (for the nonspinning BH), the tidal effects
play an important role, and hence, the remarkable dependence of the
gravitational waveforms on the EOS is found only for such cases: With
stiffer EOSs, the radius of the NS becomes larger and the tidal effect
is more relevant than with softer EOSs. For given masses of the BH and
NS, the tidal disruption occurs in a lower frequency with stiffer EOSs
than with softer EOSs, and consequently, the emission of gravitational
waves terminates at a lower frequency in the inspiral phase. The
corresponding Fourier spectrum of gravitational waves is characterized
by a cutoff frequency, $f_{\rm cut}$, above which the spectrum amplitude
exponentially damps. From the analysis of the gravitational-wave
spectra, we find that the cutoff frequency $f_{\rm cut}$ depends
strongly on the mass ratio and the compactness ${\cal C}$ of the NS. For
a given small mass ratio such as $Q=2$, the value of $f_{\rm cut}$
increases monotonically and steeply with ${\cal C}$, depending weakly on
the adiabatic index, $\Gamma_2$, of the core EOS. We derive the relation
between ${\cal C}$ and $f_{\rm cut}$ for $Q=2$ and $\Gamma_2 = 3$ as
$f_{\rm cut} \propto {\cal C}^{3.9}$, in which the power index of ${\cal
C}$ is significantly larger than 1.5 which is expected from the analysis
of the mass-shedding limit. This implies that the dependence of $f_{\rm
cut}$ on ${\cal C}$ is stronger than that for $f_{\rm shed}$, and
indicates that the observation of $f_{\rm cut}$ will play a role for
constraining the value of ${\cal C}$. Varying the core EOS also modifies
the value of $f_{\rm cut}$, because the central density profile of the
NS depends on the stiffness of the core EOS and susceptibility to the
tidal force of its companion BH is modified. For the variation from
$\Gamma_2=3$ to 2.4, the value of $f_{\rm cut}$ is modified by $\sim
20\%$. This suggests that the details of the core EOS for $\rho \agt
10^{15}~{\rm g/cm^3}$ may play an important role for determining the
gravitational waveform from the BH-NS binaries composed of high-mass
NSs.

We also determined the mass of the disk surrounding the remnant BH. The
disk mass depends strongly on the EOS, because the EOS determines the
location at which the tidal disruption occurs through the compactness
${\cal C}$ of the NS. The disk mass is correlated strongly with the NS
compactness ${\cal C}$, and for $Q=2$, it can be $\gtrsim 0.01 M_\odot$
for a wide range of the EOSs and the NS masses $M_{\rm NS}$. However,
the disk mass is tiny for $Q=3$, unless the EOS is extremely stiff like
2H EOS or the NS mass is low. For the BH-NS binaries consisting of a
nonspinning BH, the disk mass can be $\agt 0.01M_{\odot}$ for $Q=3$,
only for the case ${\cal C} \alt 0.16$.

Using the quantities calculated on the apparent horizon, we estimated
the dimensionless spin parameter of the remnant BH. We find that this
spin parameter depends only weakly on the EOS for given masses of the BH
and NS, unlike the disk mass. The BH spin depends primarily on the mass
ratio $Q$ and becomes smaller for a binary with a larger value of $Q$:
$a \approx 0.66 \pm 0.03$ for $Q=2$ and $\approx 0.54 \pm 0.02$ for
$Q=3$.

Finally we list the issues for the future. The two-piece EOS employed in
this paper is not accurate enough to describe high-mass NSs for which
the inner core is composed of a high-density matter with $\rho \agt
10^{15}~{\rm g/cm}^3$. For the study of a BH-NS binary composed of a
high-mass NS with small values of $Q$ (i.e., for a binary in which the
tidal interaction plays a role), it is necessary to adopt piecewise
polytrope EOSs with three or four free parameters. It is also necessary
to take into account the BH spin for a systematic survey of the BH-NS
binary merger process, because the orbital frequency at the ISCO depends
strongly on the BH spin as well as the mass of the BH; e.g., for the
ISCO around Kerr BHs, the orbital angular frequency increases by a
factor of $6^{3/2}$ if the spin is changed from zero to unity. This
difference in the ISCO will be crucial for determining the criteria for
the onset of tidal disruption, the mass of the remnant disk, and
gravitational waveforms. Currently we are working on this subject and
will report the numerical results in the next paper.

\begin{acknowledgments}
We thank J.L. Friedman for the suggestion of parameter sets of the
piecewise polytrope employed in this paper, for careful reading of this
paper, and for helpful comments. Numerical computation of
quasiequilibrium states is performed using the free library LORENE
\cite{LORENE}. We thank members in the Meudon Relativity Group for
developing LORENE. This work was supported by Grant-in-Aid for
Scientific Research (21340051), by a Grant-in-Aid for Scientific
Research on Innovative Area (20105004) of Japanese MEXT, by a
Grant-in-Aid of JSPS, and by a Grant-in-Aid for the Global COE Program
``The Next Generation of Physics, Spun from Universality and Emergence''
of Japanese MEXT.
\end{acknowledgments}

\section*{Appendix: convergence} \label{sec:app_conv}

\begin{table}
 \caption{Several numerical results for models HB-Q2M135 and H-Q3M135
 with different grid resolutions, $N=50$, 42, and 36. All the quantities
 are defined in the body text. In this table, we compare the disk mass
 at $t-t_{\rm merger} \approx 10$ ms.}
\begin{tabular}{cccccc} \hline
 $N$ & $f_{\rm cut} m_0$ & $M_{r>r_{\rm AH}} [M_\odot] (10 {\rm ms})$ &
 $a$ & $\Delta E / M_0 (\%)$ & $\Delta J / J_0 (\%)$ \\ \hline
 & & HB-Q2M135 &&& \\ \hline
 50 & 0.0613 & 0.025 & 0.67 & 1.36 & 21.8 \\
 42 & 0.0621 & 0.022 & 0.67 & 1.34 & 21.4 \\
 36 & 0.0644 & 0.022 & 0.68 & 1.35 & 21.4 \\ \hline
 & & H-Q3M135 &&& \\ \hline
 50 & 0.0790 & 0.0027 & 0.55 & 1.39 & 21.8 \\
 42 & 0.0794 & 0.0021 & 0.56 & 1.36 & 21.3 \\
 36 & 0.0788 & 0.0022 & 0.56 & 1.33 & 20.7 \\ \hline
\end{tabular}
\label{table:conv}
\end{table}

\begin{figure}[tbp]
 \includegraphics[width=90mm,clip]{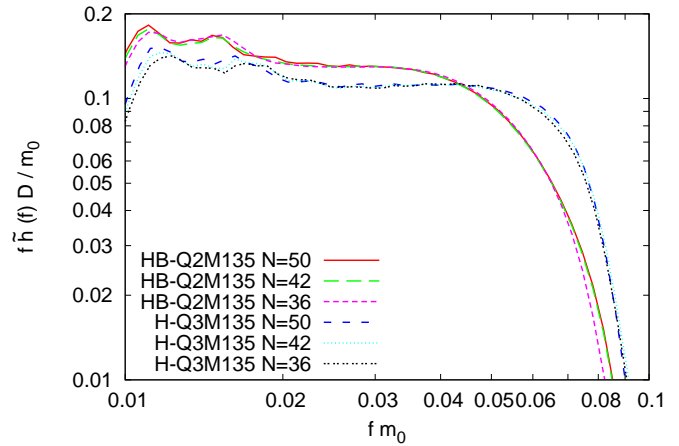} \caption{Comparison of
 gravitational-wave spectra for models HB-Q2M135 and H-Q3M135 with
 different grid resolutions.} \label{fig:spec_conv}
\end{figure}

\begin{figure}[tbp]
 \includegraphics[width=90mm,clip]{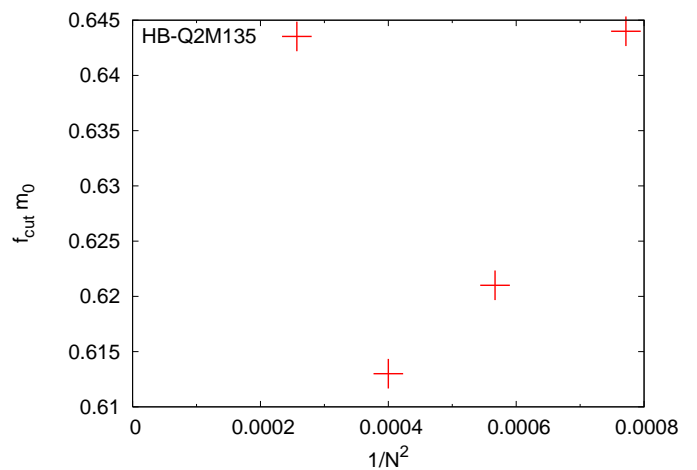} \caption{$f_{\rm cut} m_0$
 as a function of the inverse of a squared grid resolution $1/N^2$ for
 model HB-Q2M135.} \label{fig:fcut_conv}
\end{figure}

In this Appendix, we demonstrate that the convergence is approximately
achieved for the numerical results shown in Sec.~\ref{sec:result}. We
here compare numerical results obtained with different grid resolutions
for models HB-Q2M135 and H-Q3M135. Table \ref{table:conv} lists several
numerical results. This shows that the numerical results depend only
weakly on the grid resolutions, and thus, we conclude that the
convergence is approximately achieved in our simulation. Most
importantly, Fig.~\ref{fig:spec_conv} shows that the gravitational-wave
spectra approximately converge and $f_{\rm cut} m_0$ shown in Table
\ref{table:conv} does not vary by $\gtrsim 5\%$. Figure
\ref{fig:fcut_conv} plots $f_{\rm cut} m_0$ for model HB-Q2M135 as a
function of the inverse of a squared grid resolution $1/N^2$. This
figure shows that the value of $f_{\rm cut} m_0$ converges at better
than second order, and thus the values of $f_{\rm cut} m_0$ for $N = 50$
are obtained in $\lesssim 3\%$ error. For model H-Q3M135, the value of
$f_{\rm cut} m_0$ does not converge systematically and fluctuates with
the amplitude of $\sim 0.5 \%$. This fluctuation may be ascribed to the
variance associated with the fitting procedure using Eq.~(\ref{eq:fit}),
which involves a number of free parameters. We estimate roughly the
variance of $f_{\rm cut} m_0$ at $\sim 0.5\%$ within 95\% accuracy of
the fitting with respect to the norm defined by Eq.~(\ref{eq:norm}) for
model H-Q3M135. We note that the merger time $t_{\rm merger}$ depends on
the grid resolution; it is systematically larger for better grid
resolutions. However, the spectrum near $f = f_{\rm cut}$ depends weakly
on the grid resolution. $\Delta E$ and $\Delta J$ also approximately
converge. The errors are $\lesssim 0.1 \%$ for $\Delta E$ and $\lesssim
1 \%$ for $\Delta J$, respectively.

\begin{figure}[tbp]
 \includegraphics[width=90mm,clip]{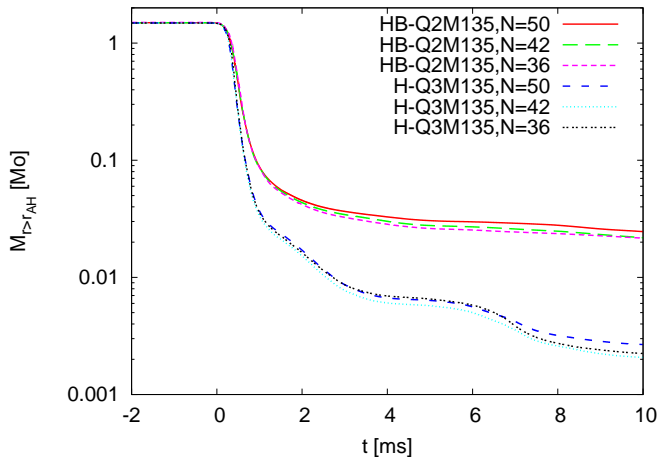} \caption{Comparison of the
 disk-mass evolution for models HB-Q2M135 and H-Q3M135 with different
 grid resolutions.} \label{fig:disk_conv}
\end{figure}

Among many quantities, the disk mass is most sensitive to the numerical
dissipation because the spurious dissipation of the angular momentum in
the disk enhances the accretion of the materials surrounding BH and
results in a lower disk mass. Hence, the values of disk mass described
in the body text should be regarded as the lower limit of the actual
mass of the remnant disk.

We plot the time evolution of the disk mass for different grid
resolutions in Fig.~\ref{fig:disk_conv}. Roughly speaking, the numerical
results for the disk mass increase with improving the grid resolution,
although systematic convergence property is not seen. The reason for
this unsystematic behavior is likely that the motion of the disk
material is affected slightly by the atmosphere (in particular for
low-mass disks of relatively low densities), and thus, convergence
property should not be expected. Assuming most conservatively that the
convergence is achieved only at first order for the results of $N = 42$
and 50, the error of the results for $N=50$ may be a factor of 2 for the
low disk mass case $M_{r>r_{\rm AH}} \lesssim 0.01 M_\odot$. However, we
expect that systematic quantitative relations between the disk mass and
the compactness of the NS, and between the disk mass and the maximum
density shown in Figs.~\ref{fig:mtoc} and \ref{fig:rhotom} are not
drastically changed.

\section{erratum} \label{sec:erratum}

(The original version of this article was submitted on Aug 9, 2010. This
erratum is added on Aug 5, 2011.)

In previous sections, we performed numerical simulations for the merger
of a nonspinning black hole (BH) and a neutron star (NS), and explored
gravitational waves emitted and the final outcome formed after the
merger. We recently noticed that we systematically {\it underestimated}
disk masses in previous sections. The reason is that we evolved
hydrodynamic variables and estimated disk masses only in domains of the
size $\sim 200^3 \; {\rm km}^3$, although Einstein's field equation was
solved in domains of the size $\sim 800^3 \; {\rm km}^3$. A small
domain size for hydrodynamics is insufficient for the estimation of the
disk mass, because, if tidal disruption occurs at a distant orbit,
especially for the case in which the NS radius is large ($\sim 15$~km),
tidal disrupted material extends far away from the central region. For
this reason, we performed again the same simulations as in previous
sections, enlarging the computational domain of hydrodynamics. To
estimate disk mass more accurately, in addition, we enlarged the size of
the computational domain to the size $1500^3$--$2000^3 \; {\rm km}^3$.
This is done by increasing a coarse domain by one more level in the
adaptive mesh refinement algorithm (AMR). Specifically, the number of
coarser domains is increased from three to four.
Table~\ref{table:grid_e} (new version of Table III) summarizes the
parameters of the new grid structure. In these simulations, the total
rest mass of the atmosphere is always less than $10^{-4}
M_\odot$. Results for gravitational waves do not change within the level
of numerical accuracy in our simulations.

\begin{table}[tbp]
 \caption{Setup of the grid structure for the computation with our AMR
 algorithm. $\Delta x = h_7 = L / (2^7 N)$ is the grid spacing at the
 finest-resolution domain with $L$ being the location of the outer
 boundaries for each axis. $R_{\rm diam}/\Delta x$ denotes the grid
 number assigned inside the semimajor diameter of the NS. $\lambda_0$ is
 the gravitational wavelength of the initial configuration.}
 \begin{tabular}{cccc} \hline
 Model & $\Delta x / M_0$ & $R_{\rm diam} / \Delta x$ & $L /
 \lambda_0$ \\ \hline \hline
 2H-Q2M135 & 0.0471 & 90.8 & 2.377 \\
 H-Q2M135 & 0.0377 & 86.2 & 2.130 \\
 HB-Q2M135 & 0.0347 & 87.0 & 1.963 \\
 HBs-Q2M135 & 0.0353 & 85.2 & 1.996 \\
 HBss-Q2M135 & 0.0353 & 84.0 & 1.996 \\
 B-Q2M135 & 0.0330 & 85.1 & 1.863 \\
 Bs-Q2M135 & 0.0324 & 84.4 & 1.830 \\
 Bss-Q2M135 & 0.0270 & 95.4 & 1.650 \\ \hline
 2H-Q3M135 & 0.0353 & 89.0 & 1.996 \\
 H-Q3M135 & 0.0282 & 84.7 & 1.711 \\
 HB-Q3M135 & 0.0269 & 82.7 & 1.631 \\
 B-Q3M135 & 0.0247 & 83.8 & 1.497 \\ \hline
 2H-Q2M12 & 0.0565 & 86.9 & 2.510 \\
 H-Q2M12 & 0.0453 & 83.1 & 2.563 \\
 HB-Q2M12 & 0.0420 & 83.6 & 2.377 \\
 B-Q2M12 & 0.0392 & 83.4 & 2.218 \\ \hline
 HB-Q3M12 & 0.0306 & 84.6 & 1.713 \\
 B-Q3M12 & 0.0278 & 86.9 & 1.572 \\ \hline
 \end{tabular}
 \label{table:grid_e}
\end{table}

\begin{figure*}[th]
 \begin{tabular}{cc}
 \includegraphics[width=85mm,clip]{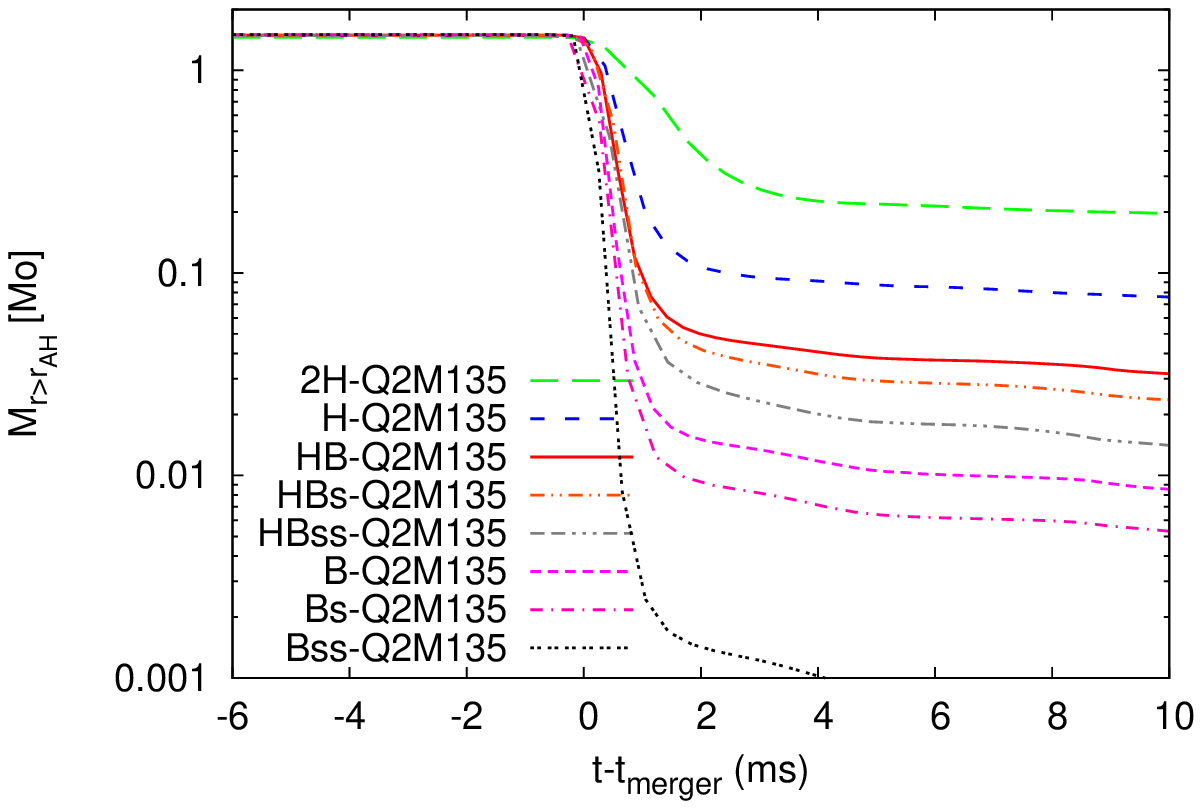} &
 \includegraphics[width=85mm,clip]{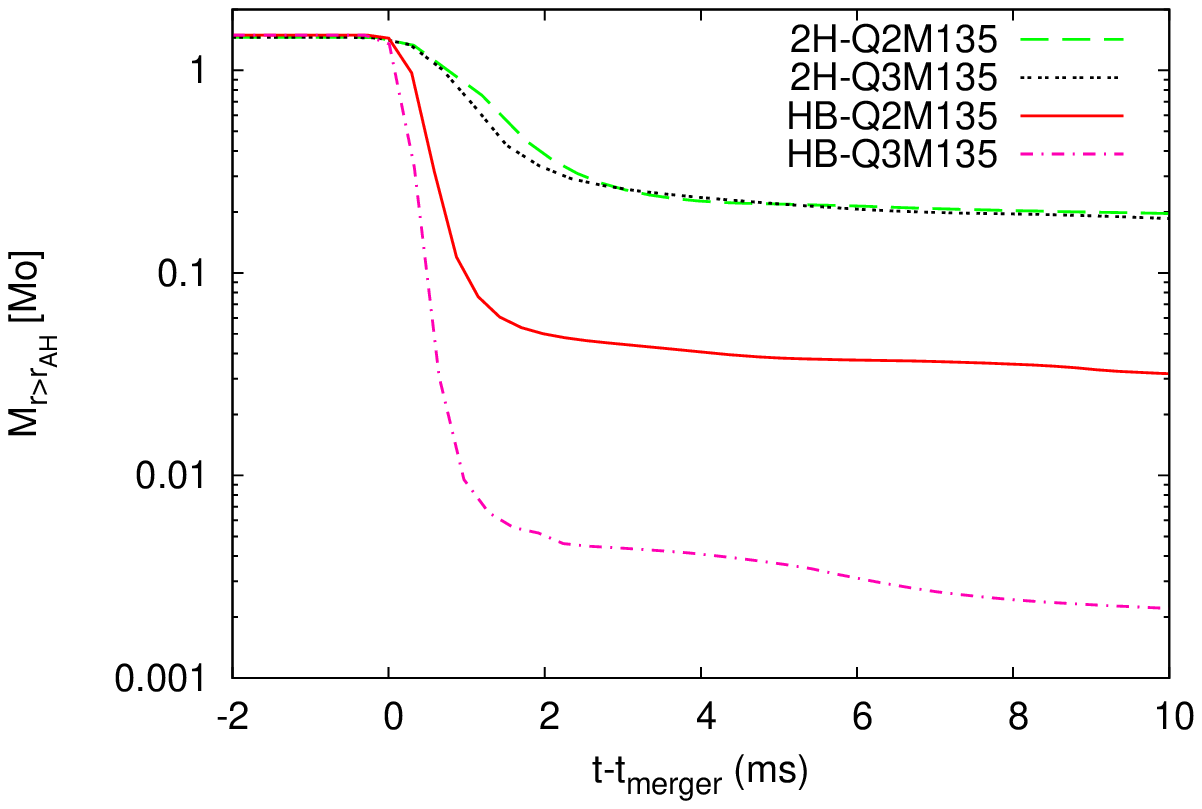} \\
 \includegraphics[width=85mm,clip]{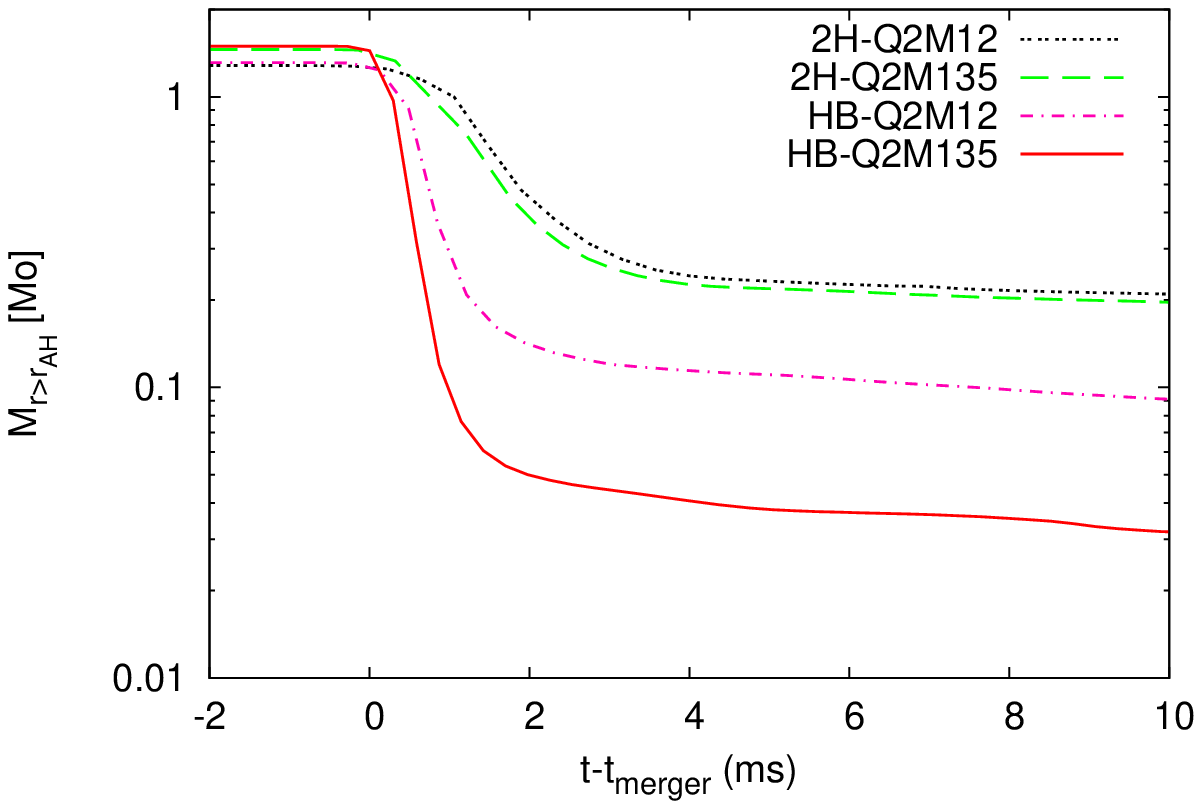} &
 \includegraphics[width=85mm,clip]{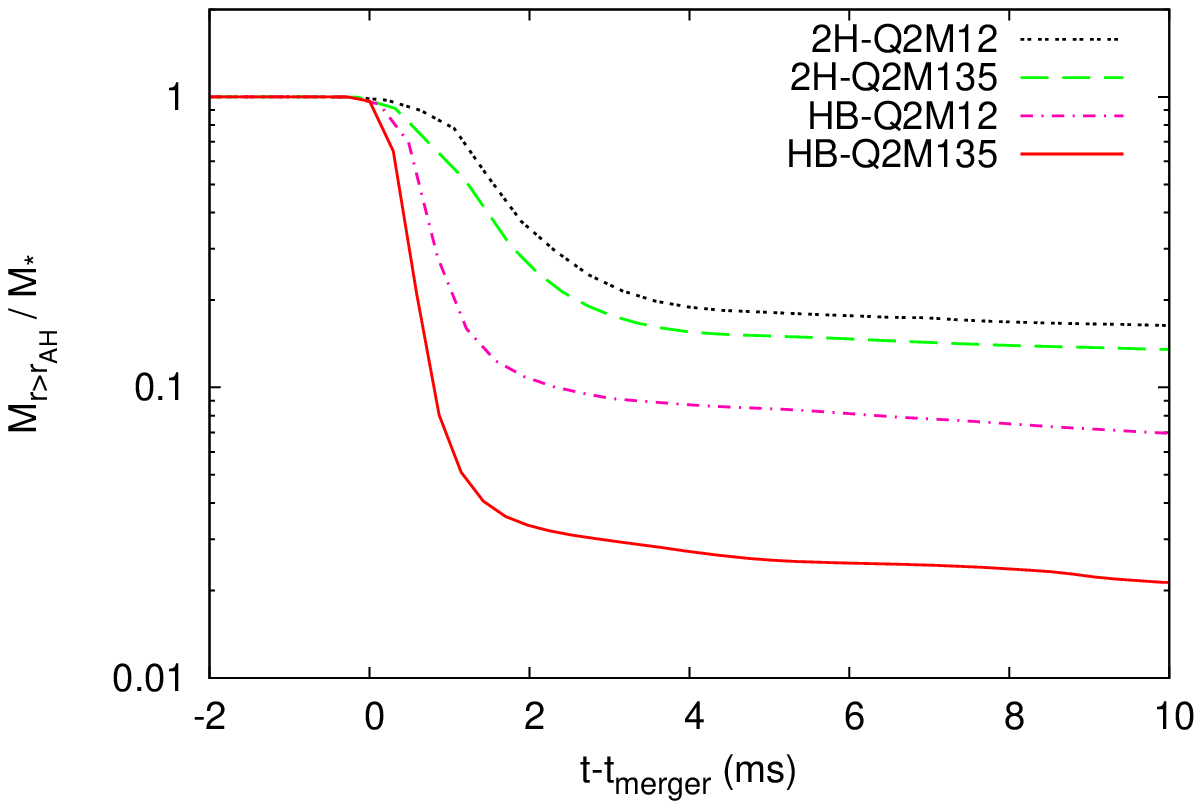}
 \end{tabular}
 \caption{Evolution of the rest mass of the material located outside the
 apparent horizon, $M_{r > r_{\rm AH}}$, with an appropriate time shift;
 in these plots, the time at the onset of the merger is taken as the
 time origin. The top-left panel shows the results for models with $Q=2$
 and $M_{\rm NS} = 1.35 M_\odot$ for all the EOSs employed in this paper
 The top-right panel shows the results for selected models with $M_{\rm
 NS} = 1.35 M_\odot$ but with different values of $Q$. The bottom-left
 panel shows the results for selected models with $Q=2$ but with the
 different NS mass $M_{\rm NS}$. The bottom-right panel is the same as
 the bottom-left panel except for the normalization of the mass, with
 respect to the initial rest mass $M_*$.} \label{fig:disk_e}
\end{figure*}

\begin{figure}[th]
 \includegraphics[width=85mm,clip]{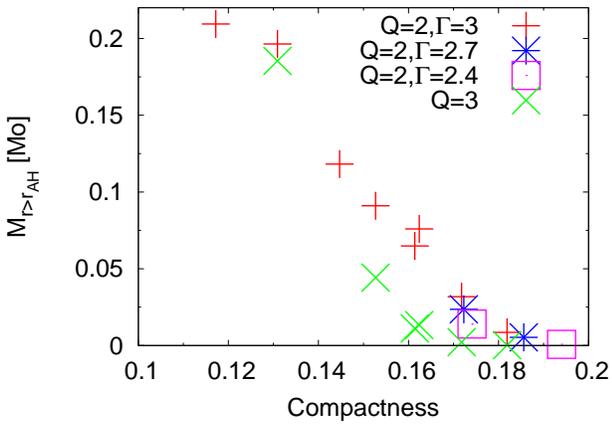} \caption{Disk mass $M_{r >
 r_{\rm AH}}$ at $t-t_{\rm merger} \approx 10$ ms as a function of the
 NS compactness ${\cal C}$.}
 \label{fig:mtoc_e}
\end{figure}

\begin{figure}[th]
 \includegraphics[width=85mm,clip]{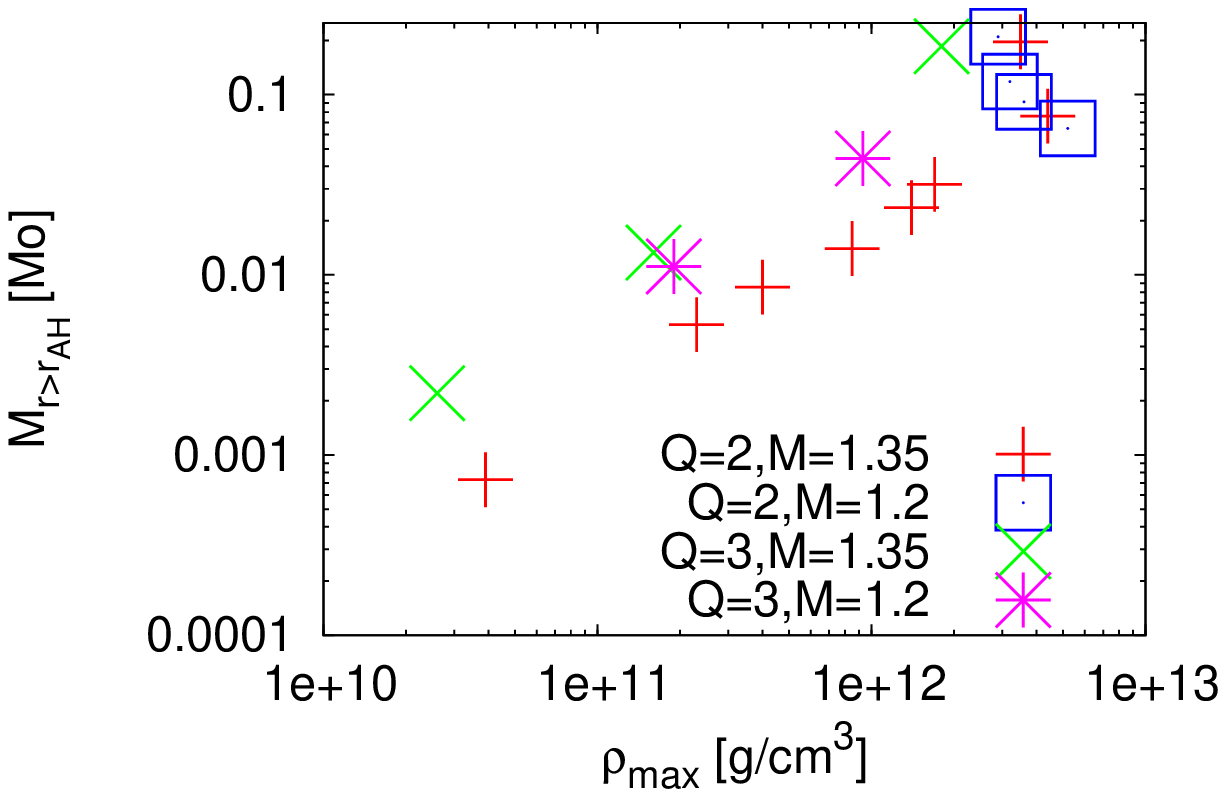} \caption{Relation between
 disk mass $M_{r > r_{\rm AH}}$ and the maximum density, $\rho_{\rm
 max}$, estimated at $t-t_{\rm merger} \approx 10$ ms. The maximum
 density oscillates with time even in the quasistationary phase, and we
 here plot a value averaged in one oscillation period.}
 \label{fig:rhotom_e}
\end{figure}

Table~\ref{table:remnant_e} (corrected version of Table V) lists
corrected values for quantities associated with the merger remnants. We
estimated all the values at the end of the simulations in previous
sections. In this section, we present the values evaluated at $\approx
10$ ms after the merger to perform more systematic
comparisons. Quantities associated with the remnant BH do not change
appreciably. Taking into account the change in the time at which the
disk mass is estimated, the mass of the remnant disk becomes larger by a
factor of $\sim 2$--3 for $Q=2$, and by a factor of $\sim 5$ for
$Q=3$. Figure \ref{fig:disk_e} (corrected version of Fig.~12) plots the
time evolution of $M_{r>r_{\rm AH}}$. Although qualitative behavior is
not altered, $M_{r>r_{\rm AH}}$ is systematically larger for the new
computations. In particular, the sudden decrease of $M_{r>r_{\rm AH}}$
at $\sim 3$--5 ms after the merger seen in Fig.~12 now
disappears. Approximate accretion time scale $t_{\rm d}$ becomes longer
by a factor of $\lesssim 2$ for many cases. Figure \ref{fig:mtoc_e}
(corrected version of Fig.~13) plots the values of $M_{r>r_{\rm AH}}$ at
$\approx 10$ ms after the merger as a function of ${\cal C}$. Although
we again see the systematic increase of $M_{r>r_{\rm AH}}$, the
conclusion that the disk mass is much smaller than $0.01 M_\odot$ for
BH-NS binaries with the typical NS mass of $M_{\rm NS} =
1.2$--$1.35M_\odot$ and ${\cal C} \gtrsim 0.16$ does not
change. Figure~\ref{fig:rhotom_e} (corrected version of Fig.~14) plots
the relation between $M_{r>r_{\rm AH}}$ and the maximum rest mass
density $\rho_{\rm max}$ of the remnant disk. Approximately speaking,
the relations between them are not changed qualitatively and
quantitatively.

\begin{table*}[th]
 \caption{Several key quantities for the merger remnants. All the
 quantities are estimated at $t - t_{\rm merger} \approx 10$ ms, where
 $t_{\rm merger}$ denotes the time of the merger. $M_{r>r_{\rm AH}}$ is
 the rest mass of the disk surrounding the BH; because the accretion is
 still ongoing at the end of simulations due to the hydrodynamic angular
 momentum transport process, the values listed give only an approximate
 mass of the long-lived accretion disk, which survives for a time scale
 longer than the dynamical time scale $\sim 10$ ms. $t_{\rm d}$ is the
 approximate accretion time scale estimated around $\approx 10$ ms after
 the merger, which we show only for the case $M_{r > r_{\rm AH}} \gtrsim
 0.001 M_\odot$. $C_e$ and $C_p$ are the circumferential radii of the
 apparent horizon along the equatorial plane and meridional plane,
 respectively, and $C_e/4\pi$ is the approximate mass of the remnant BH.
 $M_{\rm irr}$ is the irreducible mass of the remnant BH. $a$ is the
 nondimensional spin parameter of the remnant BH estimated from
 $C_p/C_e$.}
 \begin{tabular}{c|cccccc} \hline
 Model & $M_{r>r_{\rm AH}} [M_\odot]$ & $t_{\rm d}$ (ms) & $C_e / 4 \pi
 M_0$ & $M_{\rm irr} / M_0$ & $C_p / C_e$ & $a$ \\ \hline
 \hline
 2H-Q2M135 & 0.20 & 57 & 0.942 & 0.886 & 0.913 & 0.64 \\
 H-Q2M135 & 0.076 & 32 & 0.969 & 0.905 & 0.903 & 0.67 \\
 HB-Q2M135 & 0.032 & 24 & 0.978 & 0.912 & 0.902 & 0.67 \\
 HBs-Q2M135 & 0.024 & 22 & 0.980 & 0.914 & 0.902 & 0.67 \\
 HBss-Q2M135 & 0.014 & 21 & 0.980 & 0.915 & 0.902 & 0.67 \\
 B-Q2M135 & 0.0085 & 18 & 0.980 & 0.916 & 0.904 & 0.67 \\
 Bs-Q2M135 & 0.0053 & 23 & 0.980 & 0.917 & 0.906 & 0.66 \\
 Bss-Q2M135 & $7 \times 10^{-4}$ & $\cdots$ & 0.977 & 0.917 & 0.910 &
                         0.65 \\
 \hline
 2H-Q3M135 & 0.19 & 26 & 0.958 & 0.923 & 0.945 & 0.52 \\
 H-Q3M135 & 0.013 & 26 & 0.982 & 0.940 & 0.936 & 0.56 \\
 HB-Q3M135 & 0.0022 & 25 & 0.983 & 0.941 & 0.936 & 0.56 \\
 B-Q3M135 & $2 \times 10^{-4}$ & $\cdots$ & 0.982 & 0.941 & 0.938 &
                         0.55 \\
 \hline
 2H-Q2M12 & 0.21 & 66 & 0.937 & 0.885 & 0.918 & 0.62 \\
 H-Q2M12 & 0.12 & 28 & 0.958 & 0.900 & 0.907 & 0.66 \\
 HB-Q2M12 & 0.091 & 31 & 0.965 & 0.902 & 0.905 & 0.66 \\
 B-Q2M12 & 0.065 & 27 & 0.970 & 0.906 & 0.903 & 0.67 \\
 \hline
 HB-Q3M12 & 0.044 & 30 & 0.977 & 0.936 & 0.937 & 0.55 \\
 B-Q3M12 & 0.011 & 28 & 0.982 & 0.939 & 0.935 & 0.56 \\
 \hline
 \end{tabular}
 \label{table:remnant_e}
\end{table*}

\begin{figure}[th]
 \includegraphics[width=85mm,clip]{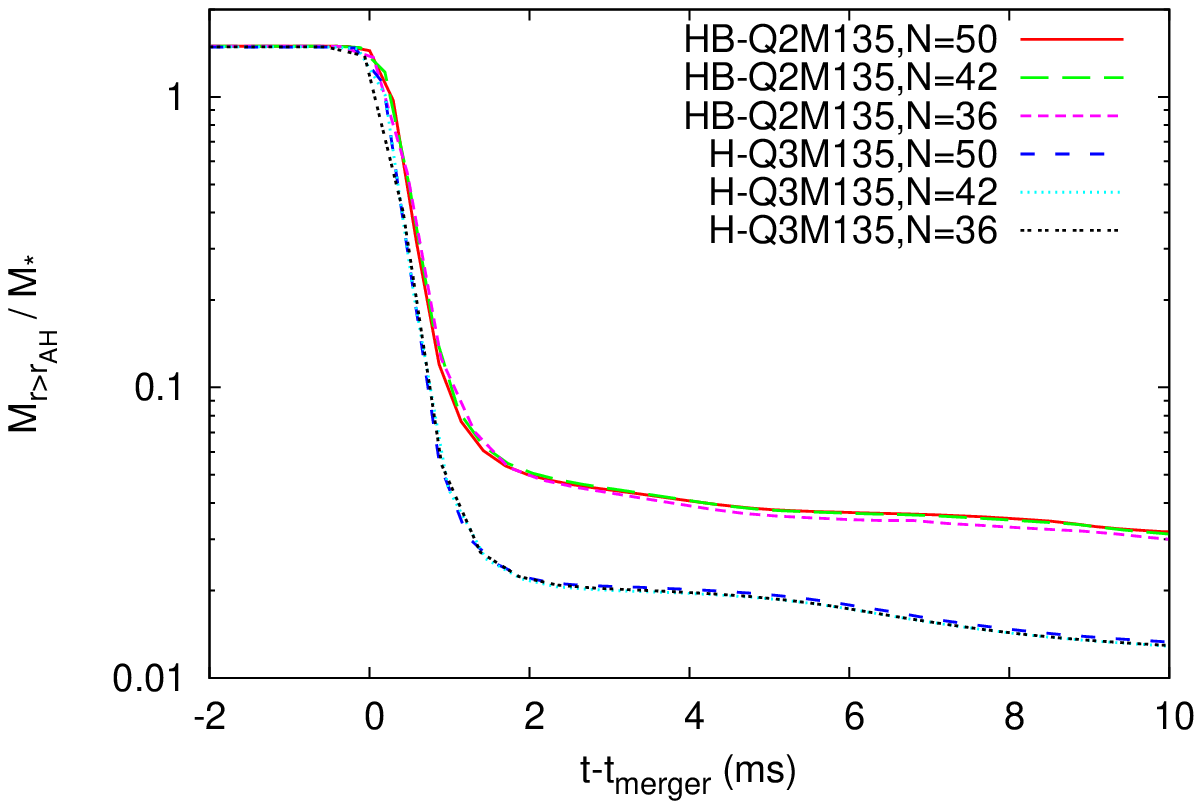} \caption{Comparison of
 the disk-mass evolution for models HB-Q2M135 and H-Q3M135 with
 different grid resolutions.} \label{fig:disk_conv_e}
\end{figure}

Table~\ref{table:conv_e} (corrected version of Table VI) lists several
numerical results for the merger remnants, and
Fig.~\ref{fig:disk_conv_e} (corrected version of Fig.~17) plots the time
evolution of $M_{r>r_{\rm AH}}$ for different grid resolutions. The
convergence of the remnant disk mass becomes slightly better than that
in previous sections.

\begin{table}[th]
 \caption{The disk masses at $t - t_{\rm merger} \approx 10$ ms and
 nondimensional spin parameters of the remnant BHs for models HB-Q2M135
 and H-Q3M135 with different grid resolutions, $N=50$, 42, and 36.}
\begin{tabular}{cccccc} \hline
 $N$ & $M_{r>r_{\rm AH}} [M_\odot] (10 {\rm ms})$ & $a$ \\
 \hline
 & HB-Q2M135 & \\
 \hline
 50 & 0.032 & 0.67 \\
 42 & 0.031 & 0.67 \\
 36 & 0.030 & 0.67 \\
 \hline
 & H-Q3M135 & \\
 \hline
 50 & 0.013 & 0.56 \\
 42 & 0.013 & 0.56 \\
 36 & 0.013 & 0.56 \\
 \hline
\end{tabular}
\label{table:conv_e}
\end{table}

%

\end{document}